\documentclass{emulateapj}

\usepackage{graphicx}
\usepackage[]{natbib}
\usepackage{placeins}
\usepackage{lineno}
\usepackage{ulem}

\pdfoutput=1

\newcommand{\ltsima} {$\; \buildrel < \over \sim \;$}
\newcommand{\gtsima} {$\; \buildrel > \over \sim \;$}
\newcommand{\lta} {\lower.5ex\hbox{\ltsima}}
\newcommand{\gta} {\lower.5ex\hbox{\gtsima}}
\newcommand{\grbnos} {GRB~$120323$A}
\newcommand{\grb} {GRB~$120323$A~}

\shorttitle{Possible Photospheric Emission Scenario in the Prompt Spectra of~\grbnos}
\shortauthors{Guiriec et al.}

\begin{document}

\title{Photospheric Emission in the Joint GBM and Konus Prompt Spectra of~\grb}

\author{S. Guiriec\altaffilmark{1,2,3,4}, N. Gehrels\altaffilmark{2}, J. McEnery\altaffilmark{2}, C. Kouveliotou\altaffilmark{1}, and D. H. Hartmann\altaffilmark{5}}


\altaffiltext{1}{Department of Physics, The George Washington University, 725 21st Street NW, Washington, DC 20052, USA}
\altaffiltext{2}{NASA Goddard Space Flight Center, Greenbelt, MD 20771, USA}
\altaffiltext{3}{Department of Astronomy, University of Maryland, College Park, MD 20742, USA}
\altaffiltext{4}{Center for Research and Exploration in Space Science and Technology (CRESST)}
\altaffiltext{5}{Department of Physics and Astronomy, Clemson University, Kinard Lab of Physics}




\email{sylvain.guiriec@nasa.gov}

\begin{abstract}

\grb is a very intense short Gamma Ray Burst (GRB) detected simultaneously during its prompt $\gamma$-ray emission phase with the Gamma-ray Burst Monitor (GBM) on board the {\it Fermi Gamma-ray Space Telescope} and the Konus experiment on board the {\it Wind} satellite. GBM and Konus operate in the keV--MeV regime, however, the GBM range is broader both toward the low and the high parts of the $\gamma$-ray spectrum. Analysis of such bright events provide a unique opportunity to check the consistency of the data analysis as well as cross-calibrate the two instruments.

We performed time-integrated and coarse time-resolved spectral analysis of~\grb prompt emission. We conclude that the analyses of GBM and Konus data are only consistent when using a double-hump spectral shape for both data sets; in contrast, the single-hump of the empirical Band function, traditionally used to fit GRB prompt emission spectra, leads to significant discrepancies between GBM and Konus analysis results. Our two-hump model is a combination of a thermal-like and a non-thermal component. We interpret the first component as a natural manifestation of the jet photospheric emission.

\end{abstract}


\section{Introduction}

Recent results on gamma-ray burst (GRB) prompt emission analysis revealed the existence of multiple simultaneous emission components~\citep[e.g.,][]{Guiriec:2010,Guiriec:2011,Guiriec:2013,Guiriec:2015a,Guiriec:2015b,Guiriec:2016a,Guiriec:2016b}. Using data collected with the {\it Fermi Gamma-ray Space Telescope} (hereafter, {\it Fermi}) for both short and long GRBs,~\citet{Guiriec:2015a} proposed a three-component model (C$_{nTh1}$+C$_{Th}$+C$_{nTh2}$) that adequately describes their $\gamma$-ray prompt emission: two components with non-thermal shapes, C$_{nTh1}$ and C$_{nTh2}$, and a thermal-like component, C$_{Th}$. While C$_{nTh1}$ and C$_{Th}$ seem to be present in most GRBs, C$_{nTh2}$ has only been detected in a few cases; it is not clear if this component is absent in most GRBs or if it is too weak to be detected. Reanalysis of GRB archival data collected with the Burst And Transient Source Experiment (BATSE) on board the {\it Compton Gamma-Ray Observatory} ({\it CGRO}) confirmed that C$_{nTh1}$+C$_{Th}$+C$_{nTh2}$ is a good model for $\gamma$-ray prompt emission~\citep{Guiriec:2016a}. Using optical, X- and $\gamma$-ray data collected with the three instruments on board {\it Swift} and {\it Suzaku}/Wide-band All-sky Monitor (WAM),~\cite{Guiriec:2016b} showed that C$_{nTh1}$+C$_{Th}$+C$_{nTh2}$ constitutes a unified model for the whole GRB broadband prompt emission, from the optical regime to higher energy $\gamma$-rays. Identification of the individual components of the prompt emission is a crucial step for disentangling the physical processes powering GRB jets.

Discovered in a long GRB~\citep{Guiriec:2011}, C$_{Th}$ has now been identified in an increasing number of both short and long GRBs. Although typically subdominant, compared to C$_{nTh1}$, C$_{Th}$ is particularly intense in GRB~120323A, a short and bright GRB, which triggered both the Gamma-ray Burst Monitor on board {\it Fermi} and the Konus instrument on board the {\it Wind} satellite\footnote{For a catalog of short GRBs observed with Konus, see \citet{Svinkin:2016}}.
This burst provides, therefore, a unique opportunity to compare and contrast the analysis of the data collected with these two instruments, which operate in similar energy bands.

\begin{figure*}
\begin{center}
\includegraphics[totalheight=0.55\textheight]{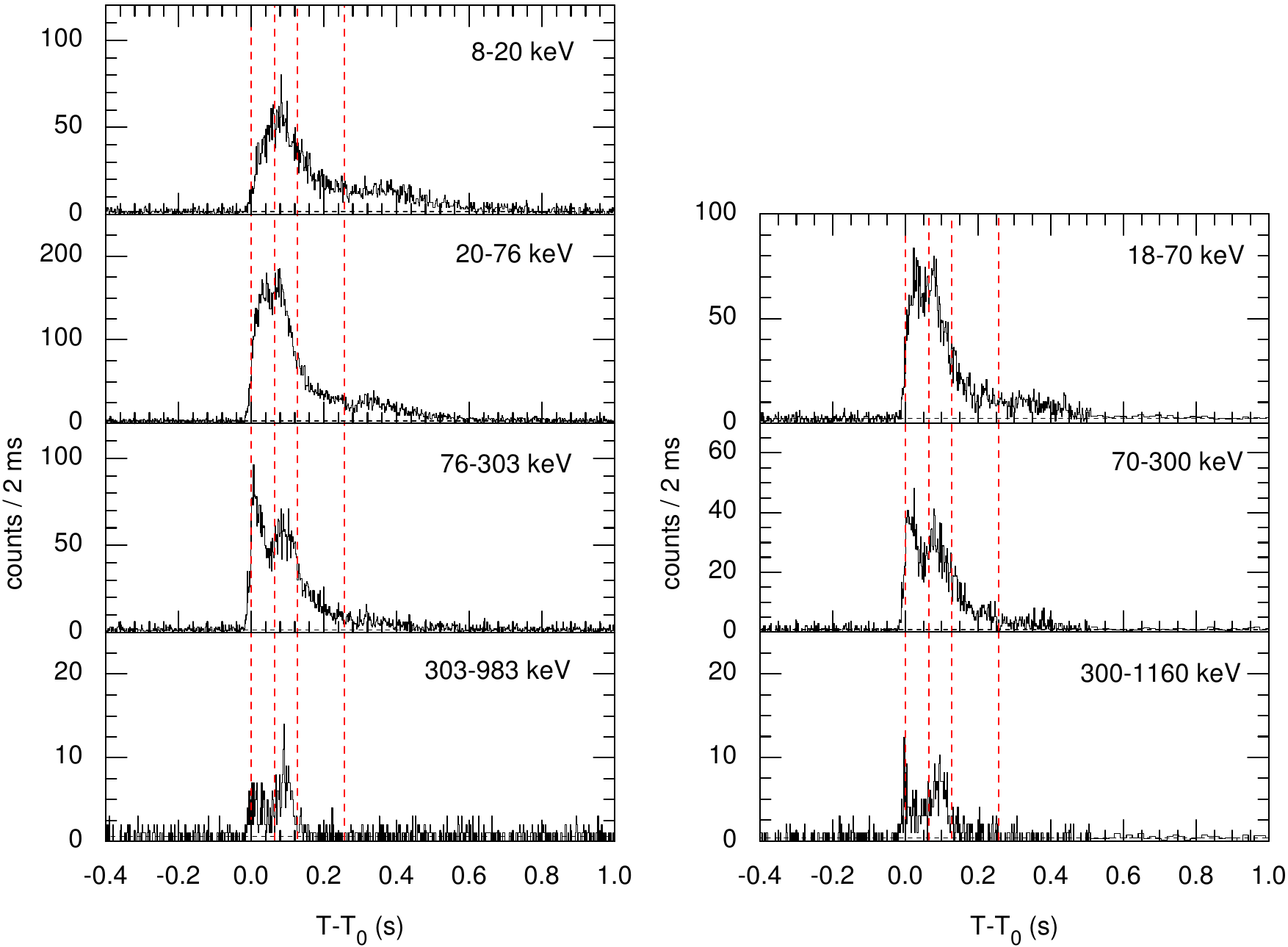}

\caption{\label{fig1}GBM and Konus light curves of~\grb with 2 ms time resolution (left and right respectively) after accounting for light propagation between the two spacecrafts; Konus trigger time, T$_0$, is used as the time reference. Because of the 18 keV low energy threshold of Konus, no light curve is available for this instrument below this value. For clarity, we did not plotted the light curves beyond 1 MeV. Indeed, while GBM and Konus high energy thresholds are 40 MeV and 7 MeV, respectively, few data are available beyond 1 MeV for~\grbnos. The vertical dashed red lines mark the three time intervals used for the spectral analysis: sp1--4 (i.e., T$_\mathrm{0}$ to T$_\mathrm{0}$+0.256 s), sp1 (i.e., T$_\mathrm{0}$ to T$_\mathrm{0}$+0.064 s) and sp2 (i.e., T$_\mathrm{0}$+0.064 s to T$_\mathrm{0}$+0.128 s).}
\end{center}
\end{figure*}

\section{Observations}

\grb triggered GBM~\citep{Gruber:2012} and Konus~\citep{Golenetskii:2012a} on 2012 March 23 at 12:10:19.72 UT and 12:10:15.97 UT\footnote{Konus trigger time will be referred as T$_\mathrm{0}$ later on}, respectively. This short and intense GRB has the highest peak flux among all short events observed with GBM thus far. While some emission can be observed up to several MeV with both GBM and Konus, this event is rather soft with the bulk of the emission below 100 keV. Despite an Autonomous Repointing Request (ARR) of the Fermi spacecraft triggered by the intensity of the event, the {\it Fermi}/Large Area Telescope (LAT) did not detect any significant emission even in the Low LAT Energy (LLE) data (20 to 100 MeV). Figure~\ref{fig1} shows the 2-ms time resolution light curves (LCs) of~\grb as observed with GBM (left) and with Konus (right) in energy bands ranging from 8 keV to 1 MeV. Due to the 18 keV energy threshold of Konus, the energy band 8-20 keV is missing for this instrument. The LCs of the two instruments are very similar above 20 keV exhibiting a sharp increase at the trigger time, two peaks with a separation of about 0.1s followed by a decaying tail with a small rebrightening from $\sim$0.30 to $\sim$0.40~s. The GBM LC below 20 keV exhibits one single peak with similar rising and decaying time up to 0.16~s followed by a decaying tail as observed in the other energy bands. The GBM and Konus T$_\mathrm{90}$ durations in the 50-300 keV energy range~\citep{Kouveliotou:1993} reported for this event were 0.448$\pm$0.090~s~\citep{Gruber:2012} and $\sim$0.5~s~\citep{Golenetskii:2012b}, respectively.

The best location for this event was estimated with the Inter-Planetary Network (IPN) at RA=340.4$^\mathrm{o}$, Dec=29.7$^\mathrm{o}$, inside an irregular error box with a minimal and maximal dimension of 0.25$^\mathrm{o}$ and 0.75$^\mathrm{o}$, respectively~\citep{Golenetskii:2012a}.

\begin{table*}
\caption{\label{tab:spectra}Spectral parameters and their 1--$\sigma$ confidence intervals resulting from CPL, Band, B+BB and C+BB fits to \grb spectra in time intervals sp1--4, sp1 and sp2, using the GBM detectors n0, n1, n3, b0 and Konus data.}
\begin{center}
\begin{tabular}{|c|c|c|c|c|c|c|c|c|c|c|}
\hline
\multicolumn{1}{|c|}{Model} &\multicolumn{1}{|c|}{Data Sets} &\multicolumn{3}{|c|}{Band or CPL} & \multicolumn{1}{|c|}{BB}& \multicolumn{4}{|c|}{EAC} & \multicolumn{1}{|c|}{pgstat/dof} \\
\hline
\multicolumn{1}{|c|}{} & \multicolumn{1}{|c|}{Parameters} & \multicolumn{1}{|c|}{E$_{\rm peak}$} & \multicolumn{1}{|c|}{$\alpha$} & \multicolumn{1}{|c|}{$\beta$} & \multicolumn{1}{|c|}{kT} & \multicolumn{1}{|c|}{N0/B0} & \multicolumn{1}{|c|}{N1/B0} & \multicolumn{1}{|c|}{N3/B0} & \multicolumn{1}{|c|}{K/B0} & \multicolumn{1}{|c|}{} \\
\hline
\multicolumn{11}{|c|}{sp1-4} \\
\hline
\multicolumn{1}{|c|}{CPL} & \multicolumn{1}{|c|}{GBM} & \multicolumn{1}{|c|}{196$\pm$14} & \multicolumn{1}{|c|}{-1.36$\pm$0.03} & \multicolumn{1}{|c|}{--} & \multicolumn{1}{|c|}{--} & \multicolumn{1}{|c|}{0.79$\pm$0.07} & \multicolumn{1}{|c|}{0.87$\pm$0.07} & \multicolumn{1}{|c|}{0.83$\pm$0.07} & \multicolumn{1}{|c|}{--} & \multicolumn{1}{|c|}{794/474} \\
\multicolumn{1}{|c|}{} & \multicolumn{1}{|c|}{GBM (20keV--7MeV)} & \multicolumn{1}{|c|}{231$\pm$23} & \multicolumn{1}{|c|}{-1.60$\pm$0.04} & \multicolumn{1}{|c|}{--} & \multicolumn{1}{|c|}{--} & \multicolumn{1}{|c|}{0.85$\pm$0.07} & \multicolumn{1}{|c|}{0.93$\pm$0.07} & \multicolumn{1}{|c|}{0.90$\pm$0.07} & \multicolumn{1}{|c|}{--} & \multicolumn{1}{|c|}{519/395} \\
\multicolumn{1}{|c|}{} & \multicolumn{1}{|c|}{Konus} & \multicolumn{1}{|c|}{316$_\mathrm{-40}^\mathrm{+55}$} & \multicolumn{1}{|c|}{-1.51$\pm$0.07} & \multicolumn{1}{|c|}{--} & \multicolumn{1}{|c|}{--} & \multicolumn{1}{|c|}{--} & \multicolumn{1}{|c|}{--} & \multicolumn{1}{|c|}{--} & \multicolumn{1}{|c|}{--} & \multicolumn{1}{|c|}{49/43} \\
\multicolumn{1}{|c|}{} & \multicolumn{1}{|c|}{GBM+Konus} & \multicolumn{1}{|c|}{216$\pm$11} & \multicolumn{1}{|c|}{-1.40$\pm$0.03} & \multicolumn{1}{|c|}{--} & \multicolumn{1}{|c|}{--} & \multicolumn{1}{|c|}{0.79(fix)} & \multicolumn{1}{|c|}{0.87(fix)} & \multicolumn{1}{|c|}{0.83(fix)} & \multicolumn{1}{|c|}{0.72$\pm$0.03} & \multicolumn{1}{|c|}{883/522} \\
\hline
\multicolumn{1}{|c|}{Band} & \multicolumn{1}{|c|}{GBM} & \multicolumn{1}{|c|}{72$\pm$6} & \multicolumn{1}{|c|}{-0.66$\pm$0.09} & \multicolumn{1}{|c|}{-2.09$\pm$0.04} & \multicolumn{1}{|c|}{--} & \multicolumn{1}{|c|}{0.85$\pm$0.07} & \multicolumn{1}{|c|}{0.93$\pm$0.08} & \multicolumn{1}{|c|}{0.90$\pm$0.08} & \multicolumn{1}{|c|}{--} & \multicolumn{1}{|c|}{579/473} \\
\multicolumn{1}{|c|}{} & \multicolumn{1}{|c|}{GBM (20keV--7MeV)} & \multicolumn{1}{|c|}{108$\pm$13} & \multicolumn{1}{|c|}{-1.21$\pm$0.12} & \multicolumn{1}{|c|}{-2.20$\pm$0.09} & \multicolumn{1}{|c|}{--} & \multicolumn{1}{|c|}{0.79$\pm$0.08} & \multicolumn{1}{|c|}{0.86$\pm$0.08} & \multicolumn{1}{|c|}{0.83$\pm$0.08} & \multicolumn{1}{|c|}{--} & \multicolumn{1}{|c|}{484/394} \\
\multicolumn{1}{|c|}{} & \multicolumn{1}{|c|}{Konus} & \multicolumn{1}{|c|}{290$_\mathrm{-184}^\mathrm{+62}$} & \multicolumn{1}{|c|}{-1.49$_\mathrm{-0.08}^\mathrm{+0.67}$} & \multicolumn{1}{|c|}{-2.78$_\mathrm{-3.47}^\mathrm{+0.69}$} & \multicolumn{1}{|c|}{--} & \multicolumn{1}{|c|}{--} & \multicolumn{1}{|c|}{--} & \multicolumn{1}{|c|}{--} & \multicolumn{1}{|c|}{--} & \multicolumn{1}{|c|}{46/42} \\
\multicolumn{1}{|c|}{} & \multicolumn{1}{|c|}{GBM+Konus} & \multicolumn{1}{|c|}{78$\pm$9} & \multicolumn{1}{|c|}{-0.75$\pm$0.05} & \multicolumn{1}{|c|}{-2.09$\pm$0.10} & \multicolumn{1}{|c|}{--} & \multicolumn{1}{|c|}{0.85(fix)} & \multicolumn{1}{|c|}{0.93(fix)} & \multicolumn{1}{|c|}{0.90(fix)} & \multicolumn{1}{|c|}{0.75$\pm$0.03} & \multicolumn{1}{|c|}{642/521} \\
\hline
\multicolumn{1}{|c|}{B+BB} & \multicolumn{1}{|c|}{GBM} & \multicolumn{1}{|c|}{326$\pm$36} & \multicolumn{1}{|c|}{-1.35$\pm$0.05} & \multicolumn{1}{|c|}{$<$-3.00} & \multicolumn{1}{|c|}{12.0$\pm$0.8} & \multicolumn{1}{|c|}{0.90$\pm$0.08} & \multicolumn{1}{|c|}{0.98$\pm$0.08} & \multicolumn{1}{|c|}{0.95$\pm$0.08} & \multicolumn{1}{|c|}{--} & \multicolumn{1}{|c|}{527/471} \\
\multicolumn{1}{|c|}{} & \multicolumn{1}{|c|}{GBM (20keV--7MeV)} & \multicolumn{1}{|c|}{325$_\mathrm{-37}^\mathrm{+48}$} & \multicolumn{1}{|c|}{-1.49$\pm$0.08} & \multicolumn{1}{|c|}{$<$-3.00} & \multicolumn{1}{|c|}{13.2$_\mathrm{-1.2}^\mathrm{+1.7}$} & \multicolumn{1}{|c|}{0.88$\pm$0.07} & \multicolumn{1}{|c|}{0.96$\pm$0.07} & \multicolumn{1}{|c|}{0.92$\pm$0.07} & \multicolumn{1}{|c|}{--} & \multicolumn{1}{|c|}{444/392} \\
\multicolumn{1}{|c|}{} & \multicolumn{1}{|c|}{Konus} & \multicolumn{1}{|c|}{386$_\mathrm{-76}^\mathrm{+102}$} & \multicolumn{1}{|c|}{-1.46$_\mathrm{-0.12}^\mathrm{+0.15}$} & \multicolumn{1}{|c|}{-2.99$\pm0.52$} & \multicolumn{1}{|c|}{18.2$_\mathrm{-3.6}^\mathrm{+5.0}$} & \multicolumn{1}{|c|}{--} & \multicolumn{1}{|c|}{--} & \multicolumn{1}{|c|}{--} & \multicolumn{1}{|c|}{--} & \multicolumn{1}{|c|}{31/40} \\

\multicolumn{1}{|c|}{} & \multicolumn{1}{|c|}{GBM+Konus} & \multicolumn{1}{|c|}{343$\pm$30} & \multicolumn{1}{|c|}{-1.37$\pm$0.05} & \multicolumn{1}{|c|}{$<$-3.91} & \multicolumn{1}{|c|}{12.4$\pm$0.8} & \multicolumn{1}{|c|}{0.90(fix)} & \multicolumn{1}{|c|}{0.98(fix)} & \multicolumn{1}{|c|}{0.95(fix)} & \multicolumn{1}{|c|}{0.79$\pm$0.03} & \multicolumn{1}{|c|}{577/519} \\
\hline
\multicolumn{1}{|c|}{C+BB} & \multicolumn{1}{|c|}{GBM} & \multicolumn{1}{|c|}{326$\pm$36} & \multicolumn{1}{|c|}{-1.34$\pm$0.05} & \multicolumn{1}{|c|}{--} & \multicolumn{1}{|c|}{12.0$\pm$0.8} & \multicolumn{1}{|c|}{0.90$\pm$0.08} & \multicolumn{1}{|c|}{0.98$\pm$0.08} & \multicolumn{1}{|c|}{0.95$\pm$0.08} & \multicolumn{1}{|c|}{--} & \multicolumn{1}{|c|}{527/472} \\
\multicolumn{1}{|c|}{} & \multicolumn{1}{|c|}{GBM (20keV-7MeV)} & \multicolumn{1}{|c|}{325$_\mathrm{-37}^\mathrm{+48}$} & \multicolumn{1}{|c|}{-1.49$\pm$0.08} & \multicolumn{1}{|c|}{--} & \multicolumn{1}{|c|}{13.2$_\mathrm{-1.2}^\mathrm{+1.7}$} & \multicolumn{1}{|c|}{0.88$\pm$0.07} & \multicolumn{1}{|c|}{0.96$\pm$0.07} & \multicolumn{1}{|c|}{0.92$\pm$0.07} & \multicolumn{1}{|c|}{--} & \multicolumn{1}{|c|}{444/393} \\
\multicolumn{1}{|c|}{} & \multicolumn{1}{|c|}{Konus} & \multicolumn{1}{|c|}{404$_\mathrm{-69}^\mathrm{+106}$} & \multicolumn{1}{|c|}{-1.48$_\mathrm{-0.11}^\mathrm{+0.14}$} & \multicolumn{1}{|c|}{--} & \multicolumn{1}{|c|}{18.5$_\mathrm{-3.6}^\mathrm{+5.0}$} & \multicolumn{1}{|c|}{--} & \multicolumn{1}{|c|}{--} & \multicolumn{1}{|c|}{--} & \multicolumn{1}{|c|}{--} & \multicolumn{1}{|c|}{32/41} \\
\multicolumn{1}{|c|}{} & \multicolumn{1}{|c|}{GBM+Konus} & \multicolumn{1}{|c|}{346$\pm$30} & \multicolumn{1}{|c|}{-1.37$\pm$0.04} & \multicolumn{1}{|c|}{--} & \multicolumn{1}{|c|}{12.5$\pm$0.8} & \multicolumn{1}{|c|}{0.90(fix)} & \multicolumn{1}{|c|}{0.98(fix)} & \multicolumn{1}{|c|}{0.95(fix)} & \multicolumn{1}{|c|}{0.79$\pm$0.03} & \multicolumn{1}{|c|}{577/520} \\

\hline
\multicolumn{11}{|c|}{sp1} \\
\hline
\multicolumn{1}{|c|}{CPL} & \multicolumn{1}{|c|}{GBM} & \multicolumn{1}{|c|}{146$\pm10$} & \multicolumn{1}{|c|}{-0.96$\pm$0.07} & \multicolumn{1}{|c|}{--} & \multicolumn{1}{|c|}{--} & \multicolumn{1}{|c|}{0.77$\pm$0.12} & \multicolumn{1}{|c|}{0.86$\pm$0.13} & \multicolumn{1}{|c|}{0.83$\pm$0.12} & \multicolumn{1}{|c|}{--} & \multicolumn{1}{|c|}{725/474} \\
\multicolumn{1}{|c|}{} & \multicolumn{1}{|c|}{Konus} & \multicolumn{1}{|c|}{249$_\mathrm{-38}^\mathrm{+58}$} & \multicolumn{1}{|c|}{-1.44$\pm$0.13} & \multicolumn{1}{|c|}{--} & \multicolumn{1}{|c|}{--} & \multicolumn{1}{|c|}{--} & \multicolumn{1}{|c|}{--} & \multicolumn{1}{|c|}{--} & \multicolumn{1}{|c|}{--} & \multicolumn{1}{|c|}{44/26} \\
\multicolumn{1}{|c|}{} & \multicolumn{1}{|c|}{GBM+Konus} & \multicolumn{1}{|c|}{161$\pm$9} & \multicolumn{1}{|c|}{-1.06$\pm$0.05} & \multicolumn{1}{|c|}{--} & \multicolumn{1}{|c|}{--} & \multicolumn{1}{|c|}{0.77(fix)} & \multicolumn{1}{|c|}{0.86(fix)} & \multicolumn{1}{|c|}{0.83(fix)} & \multicolumn{1}{|c|}{0.73$\pm$0.04} & \multicolumn{1}{|c|}{821/505} \\
\hline
\multicolumn{1}{|c|}{Band} & \multicolumn{1}{|c|}{GBM} & \multicolumn{1}{|c|}{70$\pm$7} & \multicolumn{1}{|c|}{+0.07$\pm$0.06} & \multicolumn{1}{|c|}{-2.21$\pm$0.10} & \multicolumn{1}{|c|}{--} & \multicolumn{1}{|c|}{0.98$\pm$0.14} & \multicolumn{1}{|c|}{1.08$\pm$0.15} & \multicolumn{1}{|c|}{1.06$\pm$0.15} & \multicolumn{1}{|c|}{--} & \multicolumn{1}{|c|}{489/473} \\
\multicolumn{1}{|c|}{} & \multicolumn{1}{|c|}{Konus} & \multicolumn{1}{|c|}{70$\pm$10} & \multicolumn{1}{|c|}{+0.58$\pm$0.50} & \multicolumn{1}{|c|}{-2.11$\pm$0.12} & \multicolumn{1}{|c|}{--} & \multicolumn{1}{|c|}{--} & \multicolumn{1}{|c|}{--} & \multicolumn{1}{|c|}{--} & \multicolumn{1}{|c|}{--} & \multicolumn{1}{|c|}{15/25} \\
\multicolumn{1}{|c|}{} & \multicolumn{1}{|c|}{GBM+Konus} & \multicolumn{1}{|c|}{69$\pm$6} & \multicolumn{1}{|c|}{+0.08$\pm$0.17} & \multicolumn{1}{|c|}{-2.19$\pm$0.05} & \multicolumn{1}{|c|}{--} & \multicolumn{1}{|c|}{0.98(fix)} & \multicolumn{1}{|c|}{1.08(fix)} & \multicolumn{1}{|c|}{1.06(fix)} & \multicolumn{1}{|c|}{0.87$\pm$0.05} & \multicolumn{1}{|c|}{499/504} \\
\hline
\multicolumn{1}{|c|}{B+BB} & \multicolumn{1}{|c|}{GBM} & \multicolumn{1}{|c|}{366$_\mathrm{-58}^\mathrm{+82}$} & \multicolumn{1}{|c|}{-1.18$\pm$0.10} & \multicolumn{1}{|c|}{$<$-3.00} & \multicolumn{1}{|c|}{15.3$\pm$1.0} & \multicolumn{1}{|c|}{1.08$\pm$0.12} & \multicolumn{1}{|c|}{1.20$\pm$0.14} & \multicolumn{1}{|c|}{1.17$\pm$0.14} & \multicolumn{1}{|c|}{--} & \multicolumn{1}{|c|}{489/471} \\
\multicolumn{1}{|c|}{} & \multicolumn{1}{|c|}{Konus} & \multicolumn{1}{|c|}{404$_\mathrm{-155}^\mathrm{+301}$} & \multicolumn{1}{|c|}{-1.17$\pm0.28$} & \multicolumn{1}{|c|}{-2.66$\pm$01.89} & \multicolumn{1}{|c|}{17.3$_\mathrm{-1.7}^\mathrm{+3.2}$} & \multicolumn{1}{|c|}{--} & \multicolumn{1}{|c|}{--} & \multicolumn{1}{|c|}{--} & \multicolumn{1}{|c|}{--} & \multicolumn{1}{|c|}{15/23} \\
\multicolumn{1}{|c|}{} & \multicolumn{1}{|c|}{GBM+Konus} & \multicolumn{1}{|c|}{385$_\mathrm{-50}^\mathrm{+68}$} & \multicolumn{1}{|c|}{-1.20$\pm$0.08} & \multicolumn{1}{|c|}{$<$-3.00} & \multicolumn{1}{|c|}{15.5$\pm$0.9} & \multicolumn{1}{|c|}{1.08(fix)} & \multicolumn{1}{|c|}{1.20(fix)} & \multicolumn{1}{|c|}{1.17(fix)} & \multicolumn{1}{|c|}{0.96$\pm$0.05} & \multicolumn{1}{|c|}{506/502} \\
\hline
\multicolumn{1}{|c|}{C+BB} & \multicolumn{1}{|c|}{GBM} & \multicolumn{1}{|c|}{366$_\mathrm{-58}^\mathrm{+83}$} & \multicolumn{1}{|c|}{-1.18$\pm$0.10} & \multicolumn{1}{|c|}{--} & \multicolumn{1}{|c|}{15.3$\pm$1.0} & \multicolumn{1}{|c|}{1.08$\pm$0.14} & \multicolumn{1}{|c|}{1.19$\pm$0.15} & \multicolumn{1}{|c|}{1.17$\pm$0.15} & \multicolumn{1}{|c|}{--} & \multicolumn{1}{|c|}{489/472} \\
\multicolumn{1}{|c|}{} & \multicolumn{1}{|c|}{Konus} & \multicolumn{1}{|c|}{428$_\mathrm{-101}^\mathrm{+279}$} & \multicolumn{1}{|c|}{-1.20$_\mathrm{-0.31}^\mathrm{+0.52}$} & \multicolumn{1}{|c|}{--} & \multicolumn{1}{|c|}{17.4$_\mathrm{-2.2}^\mathrm{+3.1}$} & \multicolumn{1}{|c|}{--} & \multicolumn{1}{|c|}{--} & \multicolumn{1}{|c|}{--} & \multicolumn{1}{|c|}{--} & \multicolumn{1}{|c|}{15/24} \\
\multicolumn{1}{|c|}{} & \multicolumn{1}{|c|}{GBM+Konus} & \multicolumn{1}{|c|}{385$_\mathrm{-50}^\mathrm{+68}$} & \multicolumn{1}{|c|}{-1.21$\pm$0.09} & \multicolumn{1}{|c|}{--} & \multicolumn{1}{|c|}{15.5$\pm$0.9} & \multicolumn{1}{|c|}{1.08(fix)} & \multicolumn{1}{|c|}{1.19(fix)} & \multicolumn{1}{|c|}{1.17(fix)} & \multicolumn{1}{|c|}{0.96$\pm$0.05} & \multicolumn{1}{|c|}{506/503} \\

\hline
\multicolumn{11}{|c|}{sp2} \\
\hline
\multicolumn{1}{|c|}{CPL} & \multicolumn{1}{|c|}{GBM} & \multicolumn{1}{|c|}{314$_\mathrm{-39}^\mathrm{+52}$} & \multicolumn{1}{|c|}{-1.50$\pm$0.04} & \multicolumn{1}{|c|}{--} & \multicolumn{1}{|c|}{--} & \multicolumn{1}{|c|}{0.77$\pm0.08$} & \multicolumn{1}{|c|}{0.82$\pm0.09$} & \multicolumn{1}{|c|}{0.80$\pm$0.09} & \multicolumn{1}{|c|}{--} & \multicolumn{1}{|c|}{678/474} \\
\multicolumn{1}{|c|}{} & \multicolumn{1}{|c|}{Konus} & \multicolumn{1}{|c|}{482$_\mathrm{-104}^\mathrm{+181}$} & \multicolumn{1}{|c|}{-1.52$\pm$0.10} & \multicolumn{1}{|c|}{--} & \multicolumn{1}{|c|}{--} & \multicolumn{1}{|c|}{--} & \multicolumn{1}{|c|}{--} & \multicolumn{1}{|c|}{--} & \multicolumn{1}{|c|}{--} & \multicolumn{1}{|c|}{24/26} \\
\multicolumn{1}{|c|}{} & \multicolumn{1}{|c|}{GBM+Konus} & \multicolumn{1}{|c|}{345$_\mathrm{-33}^\mathrm{+41}$} & \multicolumn{1}{|c|}{-1.52$\pm$0.03} & \multicolumn{1}{|c|}{--} & \multicolumn{1}{|c|}{--} & \multicolumn{1}{|c|}{0.77(fix)} & \multicolumn{1}{|c|}{0.82(fix)} & \multicolumn{1}{|c|}{0.80(fix)} & \multicolumn{1}{|c|}{0.62$\pm$0.04} & \multicolumn{1}{|c|}{710/505} \\
\hline
\multicolumn{1}{|c|}{Band} & \multicolumn{1}{|c|}{GBM (Option 1)} & \multicolumn{1}{|c|}{48$\pm$3} & \multicolumn{1}{|c|}{-0.14$\pm$0.30} & \multicolumn{1}{|c|}{-1.96$\pm$0.05} & \multicolumn{1}{|c|}{--} & \multicolumn{1}{|c|}{0.79$\pm$0.10} & \multicolumn{1}{|c|}{0.83$\pm$0.10} & \multicolumn{1}{|c|}{0.82$\pm$0.10} & \multicolumn{1}{|c|}{--} & \multicolumn{1}{|c|}{607/473} \\
\multicolumn{1}{|c|}{} & \multicolumn{1}{|c|}{'' (Option 2)} & \multicolumn{1}{|c|}{314$_\mathrm{-39}^\mathrm{+52}$} & \multicolumn{1}{|c|}{-1.50$\pm$0.03} & \multicolumn{1}{|c|}{$<$-6.92} & \multicolumn{1}{|c|}{--} & \multicolumn{1}{|c|}{0.77$\pm$0.08} & \multicolumn{1}{|c|}{0.82$\pm$0.09} & \multicolumn{1}{|c|}{0.80$\pm$0.09} & \multicolumn{1}{|c|}{--} & \multicolumn{1}{|c|}{678/473} \\
\multicolumn{1}{|c|}{} & \multicolumn{1}{|c|}{Konus} & \multicolumn{1}{|c|}{478$_\mathrm{-107}^\mathrm{+187}$} & \multicolumn{1}{|c|}{-1.52$\pm$0.11} & \multicolumn{1}{|c|}{$<$-3.00} & \multicolumn{1}{|c|}{--} & \multicolumn{1}{|c|}{--} & \multicolumn{1}{|c|}{--} & \multicolumn{1}{|c|}{--} & \multicolumn{1}{|c|}{--} & \multicolumn{1}{|c|}{24/25} \\
\multicolumn{1}{|c|}{} & \multicolumn{1}{|c|}{GBM+Konus} & \multicolumn{1}{|c|}{48$\pm$3} & \multicolumn{1}{|c|}{-0.17$\pm$0.10} & \multicolumn{1}{|c|}{-1.93$\pm$0.20} & \multicolumn{1}{|c|}{--} & \multicolumn{1}{|c|}{0.79(fix)} & \multicolumn{1}{|c|}{0.83(fix)} & \multicolumn{1}{|c|}{0.82(fix)} & \multicolumn{1}{|c|}{0.61$\pm0.04$} & \multicolumn{1}{|c|}{658/504} \\
\multicolumn{1}{|c|}{} & \multicolumn{1}{|c|}{''} & \multicolumn{1}{|c|}{345$_\mathrm{-33}^\mathrm{+41}$} & \multicolumn{1}{|c|}{-1.52$\pm$0.03} & \multicolumn{1}{|c|}{$<$-3.00} & \multicolumn{1}{|c|}{--} & \multicolumn{1}{|c|}{0.77(fix)} & \multicolumn{1}{|c|}{0.82(fix)} & \multicolumn{1}{|c|}{0.80(fix)} & \multicolumn{1}{|c|}{0.62$\pm$0.04} & \multicolumn{1}{|c|}{709/504} \\
\hline
\multicolumn{1}{|c|}{B+BB} & \multicolumn{1}{|c|}{GBM} & \multicolumn{1}{|c|}{456$_\mathrm{-59}^\mathrm{+78}$} & \multicolumn{1}{|c|}{-1.25$\pm$0.08} & \multicolumn{1}{|c|}{$<$-3.00} & \multicolumn{1}{|c|}{9.7$\pm$0.6} & \multicolumn{1}{|c|}{0.85$\pm$0.09} & \multicolumn{1}{|c|}{0.89$\pm$0.08} & \multicolumn{1}{|c|}{0.88$\pm$0.08} & \multicolumn{1}{|c|}{--} & \multicolumn{1}{|c|}{507/471} \\
\multicolumn{1}{|c|}{} & \multicolumn{1}{|c|}{Konus} & \multicolumn{1}{|c|}{482$_\mathrm{-60}^\mathrm{+72}$} & \multicolumn{1}{|c|}{-1.28$\pm0.10$} & \multicolumn{1}{|c|}{$<$-3.00} & \multicolumn{1}{|c|}{9.4$_\mathrm{-2.6}^\mathrm{+3.2}$} & \multicolumn{1}{|c|}{--} & \multicolumn{1}{|c|}{--} & \multicolumn{1}{|c|}{--} & \multicolumn{1}{|c|}{--} & \multicolumn{1}{|c|}{21/23} \\
\multicolumn{1}{|c|}{} & \multicolumn{1}{|c|}{GBM+Konus} & \multicolumn{1}{|c|}{469$_\mathrm{-47}^\mathrm{+59}$} & \multicolumn{1}{|c|}{-1.26$\pm$0.08} & \multicolumn{1}{|c|}{$<$-3.00} & \multicolumn{1}{|c|}{9.7$\pm$0.6} & \multicolumn{1}{|c|}{0.85(fix)} & \multicolumn{1}{|c|}{0.89(fix)} & \multicolumn{1}{|c|}{0.88(fix)} & \multicolumn{1}{|c|}{0.65$\pm$0.05} & \multicolumn{1}{|c|}{538/502} \\
\hline
\multicolumn{1}{|c|}{C+BB} & \multicolumn{1}{|c|}{GBM} & \multicolumn{1}{|c|}{456$_\mathrm{-59}^\mathrm{+78}$} & \multicolumn{1}{|c|}{-1.25$\pm$0.08} & \multicolumn{1}{|c|}{--} & \multicolumn{1}{|c|}{9.7$\pm$0.6} & \multicolumn{1}{|c|}{0.85$\pm$0.09} & \multicolumn{1}{|c|}{0.89$\pm$0.08} & \multicolumn{1}{|c|}{0.88$\pm$0.08} & \multicolumn{1}{|c|}{--} & \multicolumn{1}{|c|}{507/472} \\
\multicolumn{1}{|c|}{} & \multicolumn{1}{|c|}{Konus} & \multicolumn{1}{|c|}{484$_\mathrm{-90}^\mathrm{+143}$} & \multicolumn{1}{|c|}{-1.28$_\mathrm{-0.19}^\mathrm{+0.36}$} & \multicolumn{1}{|c|}{--} & \multicolumn{1}{|c|}{9.4$\pm$2.0} & \multicolumn{1}{|c|}{--} & \multicolumn{1}{|c|}{--} & \multicolumn{1}{|c|}{--} & \multicolumn{1}{|c|}{--} & \multicolumn{1}{|c|}{21/24} \\
\multicolumn{1}{|c|}{} & \multicolumn{1}{|c|}{GBM+Konus} & \multicolumn{1}{|c|}{469$_\mathrm{-47}^\mathrm{+59}$} & \multicolumn{1}{|c|}{-1.26$\pm$0.08} & \multicolumn{1}{|c|}{--} & \multicolumn{1}{|c|}{9.7$\pm$0.6} & \multicolumn{1}{|c|}{0.85(fix)} & \multicolumn{1}{|c|}{0.89(fix)} & \multicolumn{1}{|c|}{0.88(fix)} & \multicolumn{1}{|c|}{0.65$\pm$0.04} & \multicolumn{1}{|c|}{538/503} \\
\hline
\end{tabular}
\end{center}
\end{table*}

\section{Data Analysis}

For the analysis of GBM data, we use the same selection criteria (i.e., data file type, detectors and energy selection) as described in Section 3 of~\citet{Guiriec:2013} as well as the same response files generated using the IPN location and the same background estimate. Konus data are extracted from detector S2 in two overlapping energy ranges covering from 20 keV to 16 MeV: 20 keV--1.6 MeV (i.e., high-gain PHA1) and 450 keV--16 MeV (i.e., low-gain PHA2). We exclude PHA1 data beyond 450 keV from the analysis to avoid duplication in the PHA1-PHA2 joint spectral fit process. No statistically significant emission is detected in Konus spectral channels beyond 7~MeV, which justifies the 7~MeV high energy cut used later on in this article.

The background of PHA1 and PHA2 data for \grb is modeled as a constant using the total counts accumulated from T$_0$+25~s to T$_0$+254~s. Located at L1 Lagrange point, Konus data are not affected by the same intense flux of charged particles captured in Earth's magnetic field that strongly affect the GBM background (low Earth orbit). The Konus background is, therefore, usually more stable. However, the very short duration of~\grb strongly reduces possible systematic effects in the background estimate used for the analysis of GBM data.

The full energy resolution of the Konus data is only available with a 64 ms time resolution. Therefore, this time resolution is the most constraining factor when defining time intervals for the joint GBM and Konus time resolved spectral analysis. We extracted GBM spectral data in the same four time intervals as those imposed by the Konus time resolution in taking into account the light propagation time between the {\it Fermi} and {\it Wind} spacecrafts. We performed a time-integrated spectral analysis of~\grb over these four time intervals of Konus (later sp1--4) and a time resolved spectral analysis of the first two: 0--64 ms (later sp1) corresponding mostly to the first peak of the LC observed above 20 keV, and 64--128 ms (later sp2) corresponding mostly to the second peak (see Figure~\ref{fig1}). The last two intervals correspond to the decay phase of the burst; we did not perform spectral analysis in each of them individually because their signal is weak. The vertical dashed red lines on Figure~\ref{fig1} indicate the three time intervals used in this analysis: sp1--4, sp1 and sp2.


We fitted the C$_{nTh1}$+C$_{Th}$ model~\citep{Guiriec:2011,Guiriec:2013,Guiriec:2015a,Guiriec:2015b,Guiriec:2016a,Guiriec:2016b}---where C$_{nTh1}$ is a non-thermal component that we approximate with a Band function or a power law with an exponential cutoff (CPL), and C$_{Th}$ is a thermal-like component that we approximate with $\emptyset$ or a black body (BB)---to the data in each time interval using either Konus and GBM data separately or together in a joint fit.
In this analysis we only consider the best models reported in~\citet{Guiriec:2013} to compare Konus results to GBM ones. To perform the spectral analysis, we used the software package XSpec\footnote{http://heasarc.nasa.gov/docs/xanadu/xspec/} and the best fit parameters as well as their corresponding uncertainties were estimated by minimizing PGSTAT (i.e., Poisson-Gaussian Statistic). This statistic is particularly interesting when fitting data in different statistical count regimes. The energy channels are grouped in order to have at least 1 count per energy bin as recommended for optimal usage of PGSTAT; the energy bins in the Konus data have typically more than 20 counts per bin.


While the two data sets of Konus are calibrated prior to the spectral analysis, no specific calibration is performed between the NaI and BGO detectors of GBM. Therefore, when fitting GBM data alone, we apply an effective area correction (EAC) factor between each NaI detector and BGO 0 (BGO 0 being the reference). This gives three additional free parameters to the fit of GBM data alone. When fitting simultaneously GBM and Konus data (later GBM+Konus), the EAC factors between BGO and NaI detectors are fixed to the values obtained from the best fit to the GBM data alone using the respective model. Therefore, GBM detectors are considered as a single instrument such as Konus. An EAC factor is then applied in between BGO 0 and Konus data to measure possible discrepancies in the flux estimates between the two instruments.

\begin{figure*}
\begin{center}
\includegraphics[totalheight=0.185\textheight, clip, viewport=0 20 512 395]{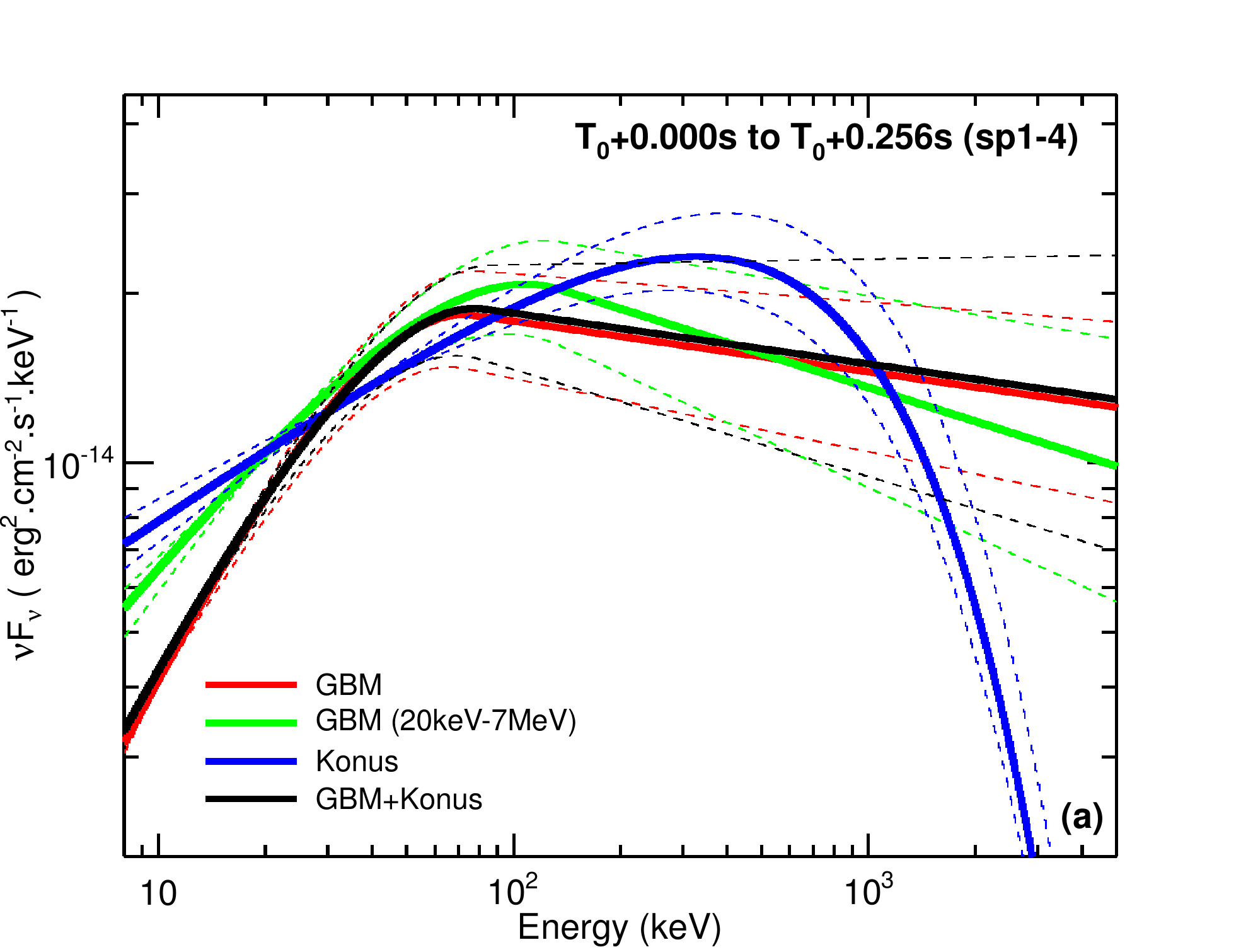}
\includegraphics[totalheight=0.185\textheight, clip, viewport=20 20 512 395]{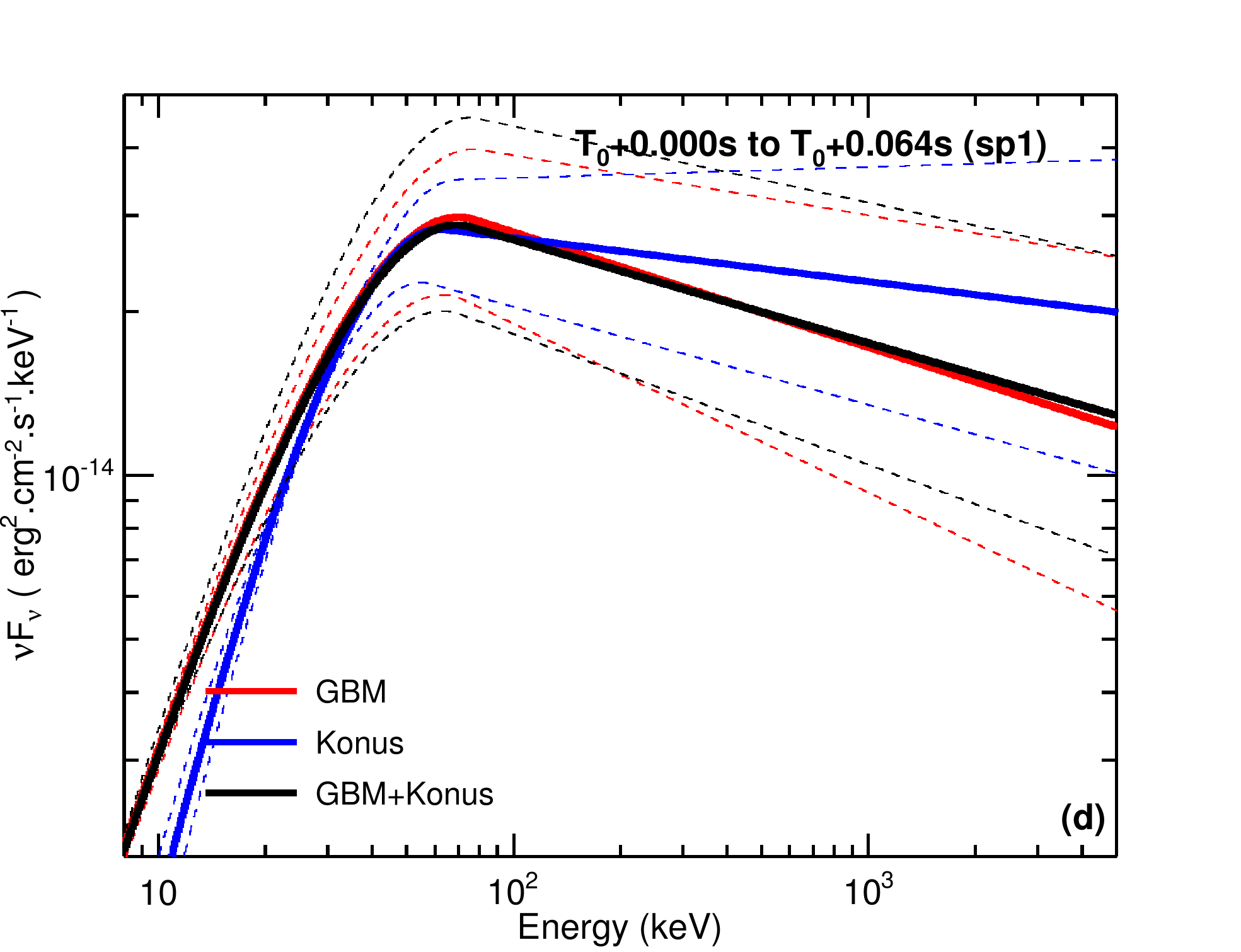}
\includegraphics[totalheight=0.185\textheight, clip, viewport=20 20 512 395]{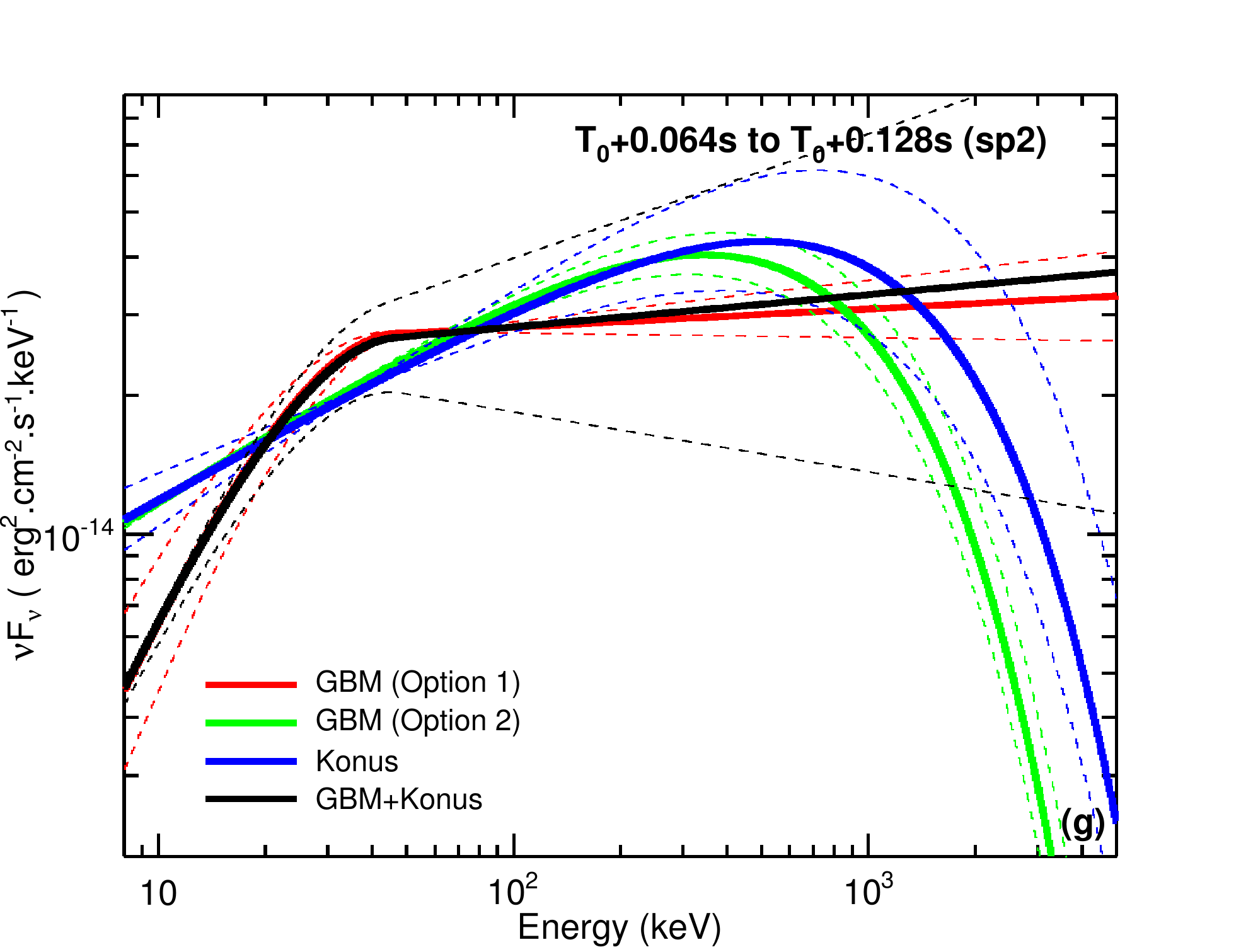}

\includegraphics[totalheight=0.185\textheight, clip, viewport=0 20 512 395]{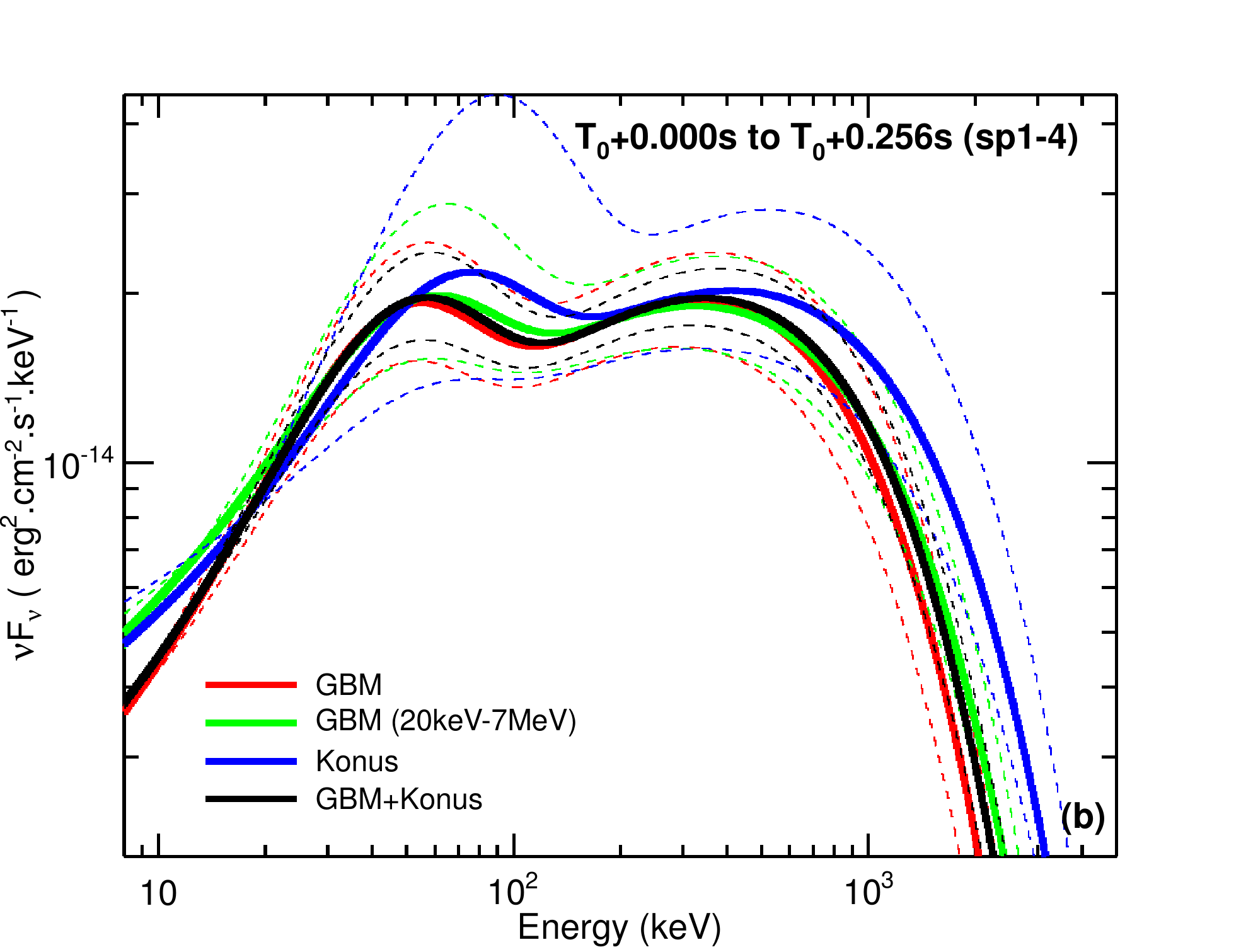}
\includegraphics[totalheight=0.185\textheight, clip, viewport=20 20 512 395]{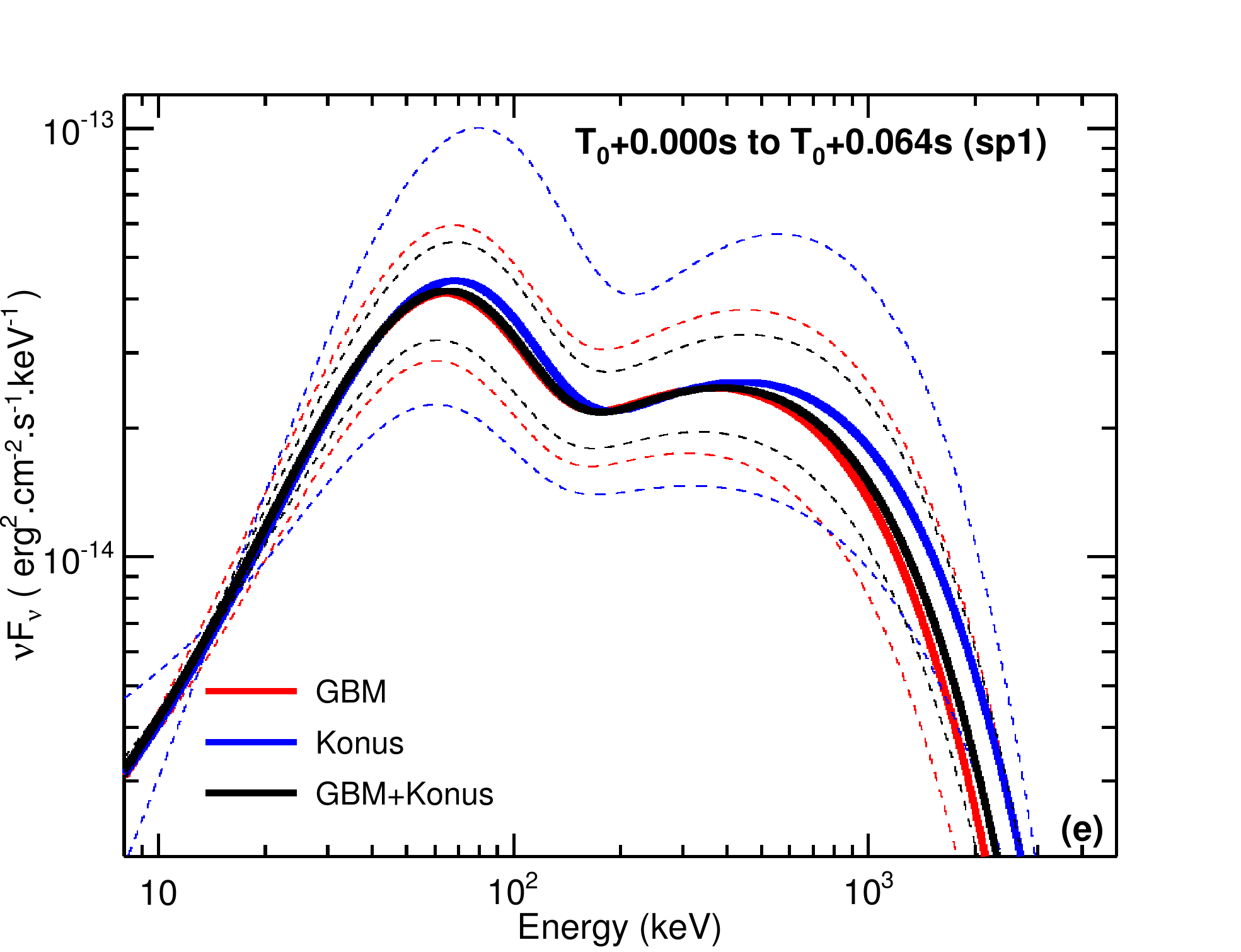}
\includegraphics[totalheight=0.185\textheight, clip, viewport=20 20 512 395]{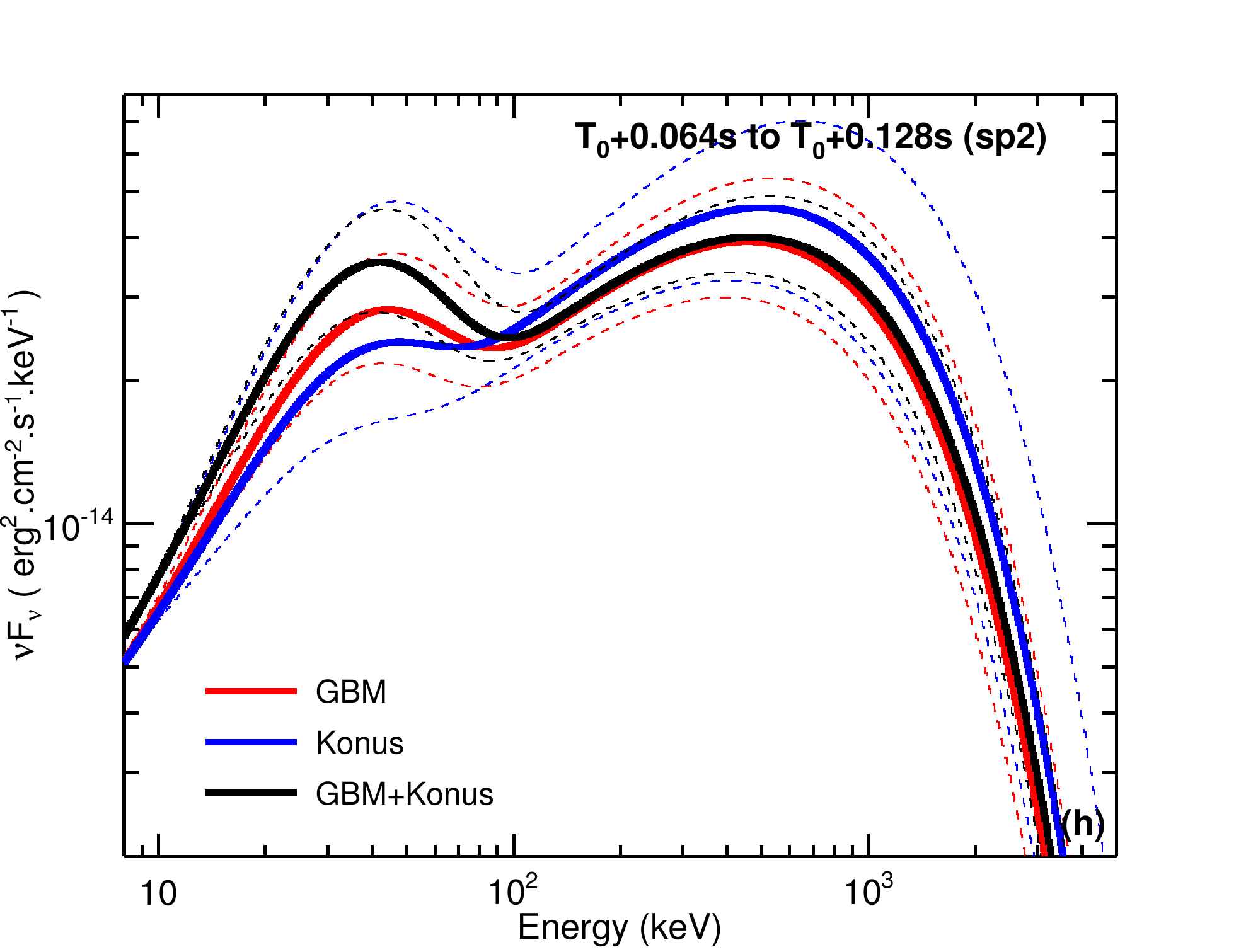}

\includegraphics[totalheight=0.194\textheight, clip, viewport=0 0 512 395]{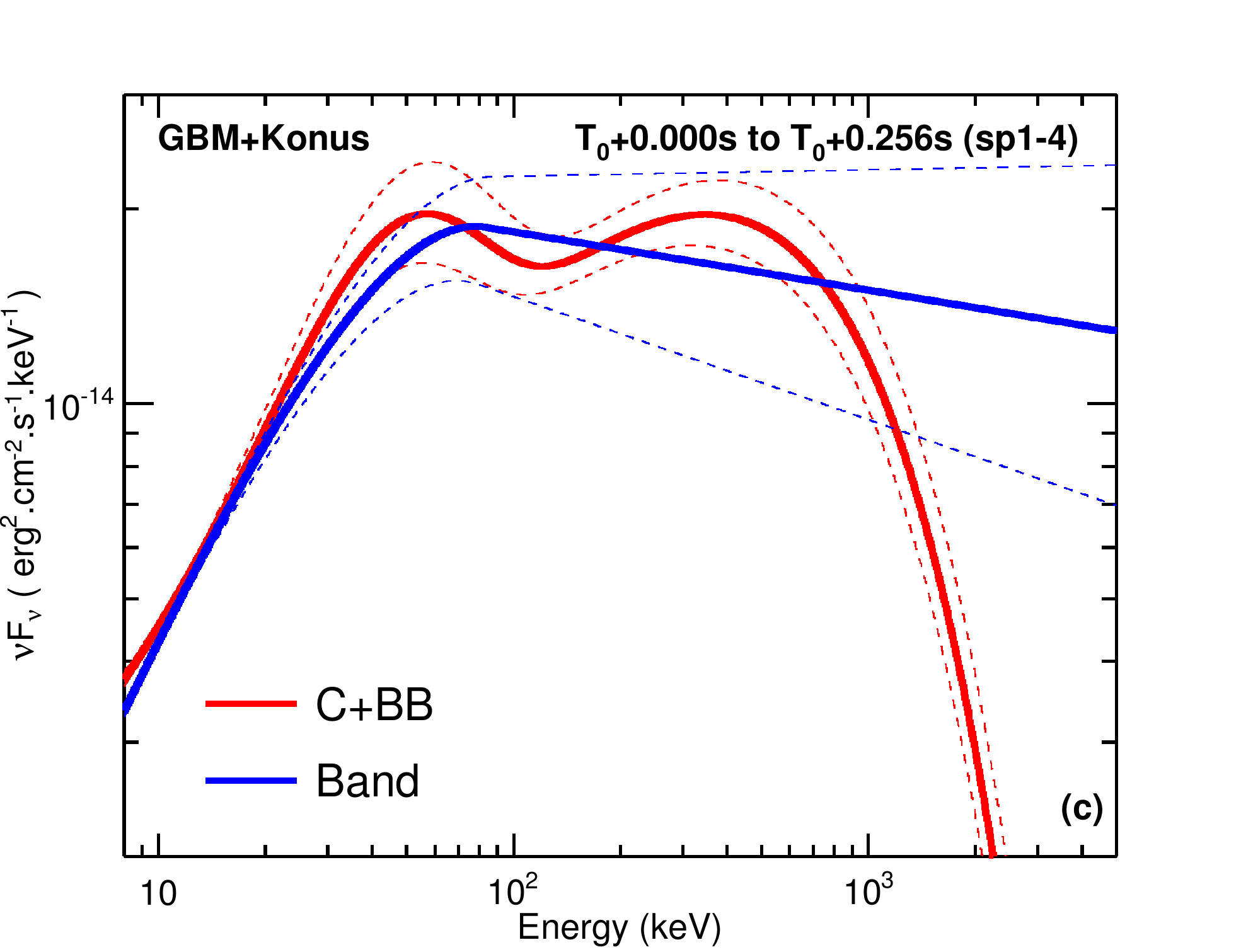}
\includegraphics[totalheight=0.194\textheight, clip, viewport=20 0 512 395]{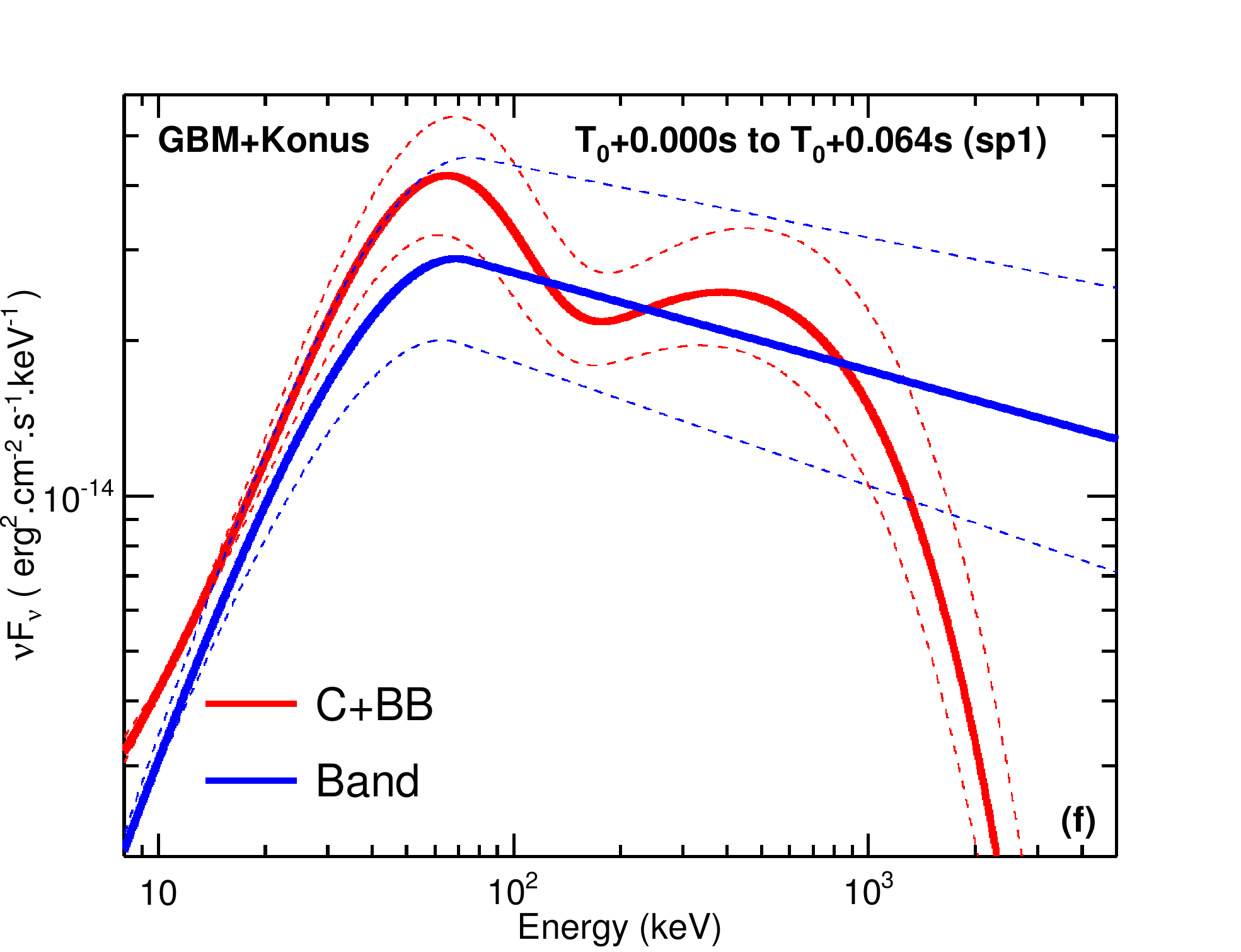}
\includegraphics[totalheight=0.194\textheight, clip, viewport=20 0 512 395]{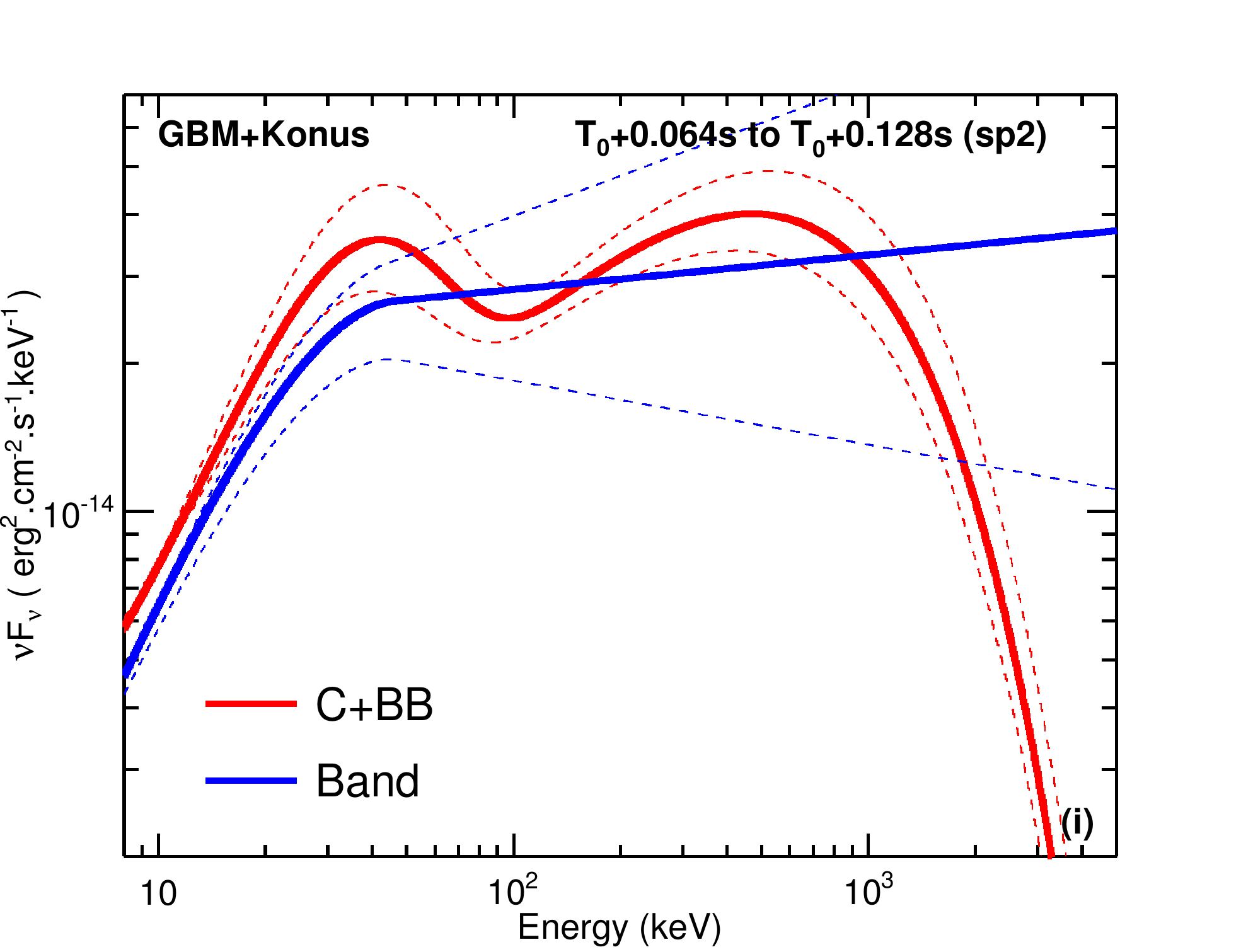}
\caption{\label{fig5}Band function (top line panels) and C+BB models (middle line panels) resulting from the fit of GBM, Konus and GBM+Konus data in time intervals sp1--4, sp1 and sp2 (left to right column, respectively). The solid lines correspond to the function resulting from the best fit parameters while the dashed lines correspond to the 1--$\sigma$ confidence region for the fit. The panels in the bottom line correspond to the Band functions overplotted with the C+BB models obtained when fitting jointly GBM and Konus data. The spectral shape resulting from the fit of the Konus data only are renormalized using the EAC correction factor estimated when fitting jointly GBM+Konus data with the same function and in the same time interval (see Table~\ref{tab:spectra}).}
\end{center}
\end{figure*}

The analysis results are reported in Table~\ref{tab:spectra}, Figure~\ref{fig5} and Figures~\ref{fig2} to~\ref{fig4}.

\section{Results}

\subsection{Time-Integrated Spectral Analysis}
\label{sec:tisa}

The Band function~\citep{Band:1993} fits to the GBM and Konus data separately in time interval sp1--4 result in two dramatically different spectral shapes (see red and blue lines in Figure~\ref{fig5}a, and Table~\ref{tab:spectra}). While the E$_{peak}$ value measured by GBM is very low for such a bright and short GRB ($\sim$80 keV), the Konus measurement is quite typical ($\sim$300 keV). Also, the value of the spectral index $\alpha$ of the Band function estimated with Konus data ($\alpha$$\sim$$-$1.50) is significantly lower than the one measured with GBM data ($\alpha$$\sim$$-$0.66).

We investigated the possible impact of the difference in the energy ranges covered by the two instruments to explain the observed discrepancies. To do so, we fitted the GBM data over the same energy range as Konus (i.e., from 20 keV to 7 MeV).
In reducing the GBM energy range we obtain a higher value for E$_\mathrm{peak}$ compared to the fit to GBM data over the whole energy range of the instrument, as well as lower values for $\alpha$ and $\beta$ (see green line in Figure~\ref{fig5}a and Table~\ref{tab:spectra}). This fit result is intermediate between the fit of GBM data over its full energy range and the fit of Konus data. This is an indication that the extended energy range covered by GBM can strongly impact the fit results, and it may explain, at least in part, the discrepancies observed between the two instruments. We note that the high-energy power law of the Band function systematically overshoots the BGO data at the highest energies showing that the Band function alone is unable to adequately capture the spectral shape of the GBM data over its whole energy range (see Figure~\ref{fig2}).

When fitting GBM and Konus data simultaneously, the resulting Band function is similar to the one obtained when fitting GBM data alone (see black line in Figure~\ref{fig5}a); this is because \grb is much more intense in GBM than in Konus. While EAC factors of only a few percent are required between NaI and BGO detectors when fitting a Band function to the GBM data alone, 20 to 30\% correction is necessary to align the GBM and Konus fluxes (see Table~\ref{tab:spectra}).

Conversely to the results reported for the Band only fits, the C$_{nTh1}$+C$_{Th}$ fits (with C$_{nTh1}$ being either Band or CPL, and C$_{Th}$ being BB) to the GBM and Konus data alone or together lead to consistent spectral shapes (see Table~\ref{tab:spectra} and Figure~\ref{fig5}b).

In the C$_{nTh1}$+C$_{Th}$ model, the shape of C$_{nTh1}$ is similar to the one obtained when fitting Band alone to the Konus data, and the shape of C$_{Th}$ mimics the Band function fit to the GBM (or GBM+Konus) data with E$_\mathrm{peak}$ matching with the BB spectrum peak energy\footnote{The maximum of the BB spectrum peaks at $\sim$3 times the value of the temperature kT.} (see Figure~\ref{fig5}c). We can also note that no correction is necessary between NaI and BGO data when fitting C$_{nTh1}$+C$_{Th}$  to the GBM data, and the correction between GBM and Konus is reduced by a few percent.

Moreover, the spectral shapes obtained with Band+BB and CPL+BB are identical (the parameter values and the fit statistics), so henceforth we will only consider CPL+BB when comparing models since it has one degree of freedom (later dof) less. The improvements in the PGSTAT values (later $\Delta$PGSTAT) between C$_{nTh1}$+C$_{Th}$ and Band fits are 52, 14 and 65 units for one additional dof. for GBM, Konus and GBM+Konus data, respectively. In the three cases, this is a statistically significant improvement making C$_{nTh1}$+C$_{Th}$ the best model with the lower number of free parameters.

\subsection{Time-Resolved Spectral Analysis}
\label{sec:trsa}

We performed a similar analysis as the one presented in Section~\ref{sec:tisa} for the time intervals sp1 and sp2. The data, the residuals of the fits and the fit results are presented in Table~\ref{tab:spectra}, Figure~\ref{fig5} and Figures~\ref{fig3} and~\ref{fig4} for sp1 and sp2, respectively.

The fits to Band functions alone to the GBM, Konus and GBM+Konus data in sp1 are globally consistent (see Figure~\ref{fig5}d and Table~\ref{tab:spectra}). The three fits are characterized by low E$_\mathrm{peak}$ values around 70 keV and positive values for $\alpha$. In addition, the value of $\alpha$ is significantly higher for Konus data than for GBM. Such high values of $\alpha$ point toward a thermal origin of the emission process. This is in good agreement with the results reported in~\citet{Guiriec:2013}.

The parameter values of C$_{nTh1}$ in C$_{nTh1}$+C$_{Th}$ are totally incompatible with those obtained when fitting Band alone to the data (see Figure~\ref{fig5}f and Table~\ref{tab:spectra}). Whatever data set is used---GBM, Konus or GBM+Konus---the values of E$_\mathrm{peak}^{nTh1}$ are much higher than the values of E$_\mathrm{peak}^{Band}$, and the values of $\alpha_{nTh1}$ ($\alpha_{nTh1}$$\sim$$-$1.2) are much lower than the values of $\alpha_{Band}$ ($>$0).

Similarly, the high-energy spectral slope of C$_{nTh1}$ is much steeper than the one resulting from the Band only fit (i.e., $\beta$) and cannot be differentiated from an exponential cutoff based on our data sets. Therefore, CPL+BB is equivalent to Band+BB (for both the parameters of the components and the statistical results), and since CPL+BB has one free parameter less than Band+BB, we will consider this model for statistical comparison. 

Figure~\ref{fig5}e shows the very good consistency of the C$_{nTh1}$+C$_{Th}$ fits for the three data sets with the confidence region progressively decreasing from the Konus fit alone to the GBM fit alone to the GBM+Konus fit. A C$_{Th}$ temperature of $\sim$16 keV is obtained with the three data sets. Once more, these results are consistent with the analysis presented in~\citet{Guiriec:2013}. 


For the three data sets (i.e., GBM, Konus and GBM+Konus), the Band and C$_{nTh1}$+C$_{Th}$ (i.e., Band+BB or CPL+BB) fits result in similar PGSTAT values. If the parameters of the Band function in the Band-only fits and of C$_{nTh1}$ in C$_{nTh1}$+C$_{Th}$ were identical, it would indicate that the additional parameters introduced with C$_{Th}$ are not required to improve the fits, and the intensity of this component should then be compatible with zero. Here, the shape of the Band functions and of C$_{nTh1}$ resulting from the Band-only and C$_{nTh1}$+C$_{Th}$ fits, respectively, are dramatically different, and in the C$_{nTh1}$+C$_{Th}$ scenario, C$_{Th}$ is an intense component which overpowers C$_{nTh1}$ between 20 and 150 keV. From this result, we conclude that given the quality of our data sets, a spectral shape with two humps (i.e., C$_{nTh1}$+C$_{Th}$) describes the data in sp1 as well as a spectral shape with a single hump (i.e., Band). However, the positive values of the low energy spectral indices in both scenarios require a thermal emission process. This is again in good agreement with the results reported in~\citet{Guiriec:2013}. Indeed, the authors reported no statistically significant improvement of C$_{nTh1}$+C$_{Th}$ over Band in the first peak of the LC, but strong changes in the parameters of the Band function pointing to a thermal emission origin for the two scenarios.

When Band or C$_{nTh1}$+C$_{Th}$ are fitted to the GBM data of sp1, no EAC is required between NaI and BGO detectors within the 1--$\sigma$ uncertainties. Similarly, the C$_{nTh1}$+C$_{Th}$ fits to the GBM+Konus data do not require any correction between GBM and Konus. However, a flux correction $>$12\% is required in the case of the Band-only fit. This shows once more that GBM and Konus data are fully consistent only when using C$_{nTh1}$+C$_{Th}$. 


\begin{figure}
\begin{center}
\includegraphics[totalheight=0.26\textheight, clip, viewport=0 20 512 395]{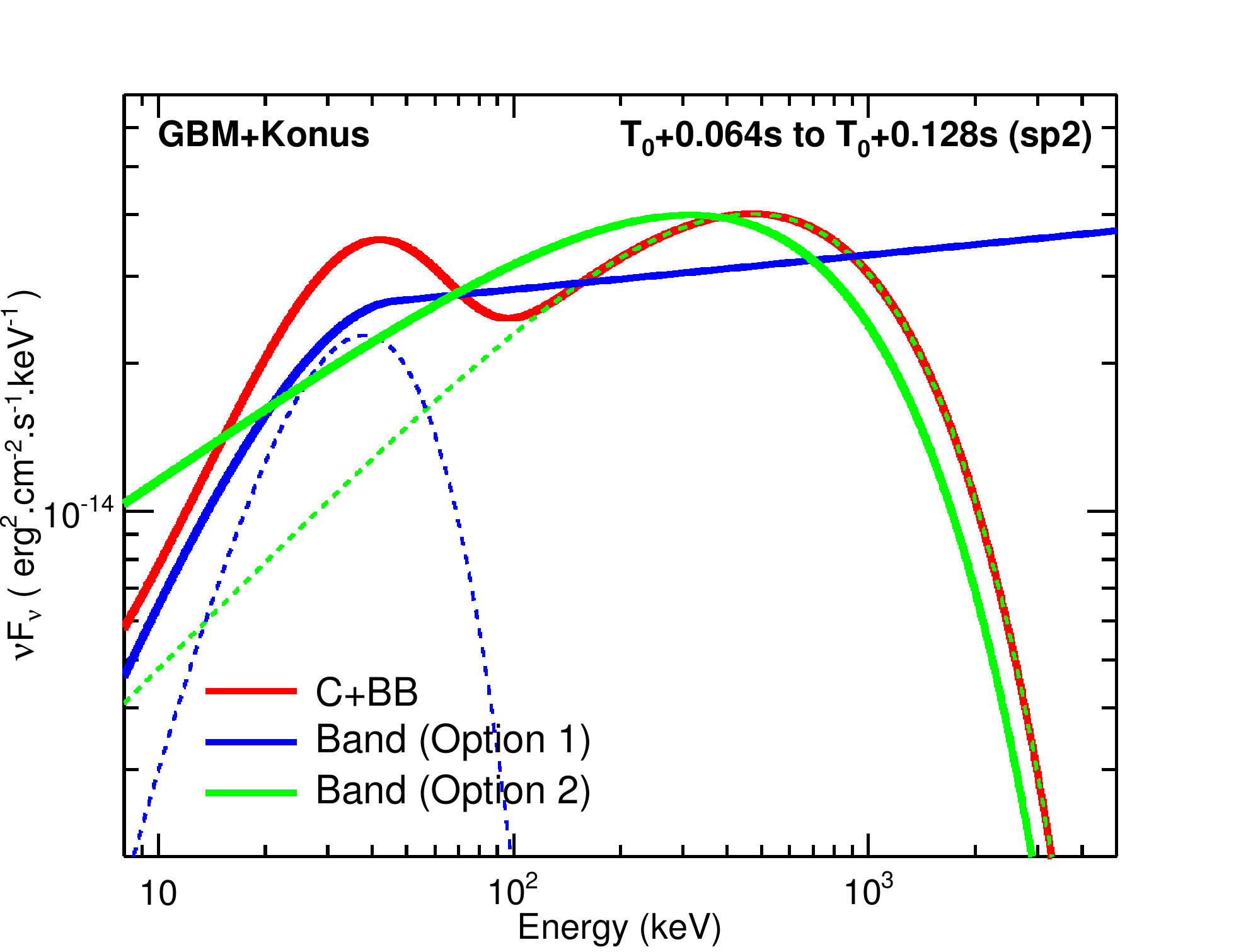}
\caption{\label{fig6}Best C+BB fit (red line) and the two options for the Band function fits (solid blue and green lines for options 1 and 2, respectively) to GBM+Konus data in the time interval sp2. The dashed blue and green lines correspond to the BB and CPL components of the C+BB model, respectively.}
\end{center}
\end{figure}

In time interval sp2, the C$_{nTh1}$+C$_{Th}$ fits to GBM, Konus and GBM+Konus data give consistent results with a temperature of $\sim$10 keV for the BB, E$_\mathrm{peak}$ around 400--500 keV, $\alpha$ around -1.2 and a very steep slope for the high energy PL making Band+BB similar to CPL+BB (see Table~\ref{tab:spectra} and Figure~\ref{fig5}h); this is good agreement with the results reported in~\citet{Guiriec:2013}. The spectral parameters obtained when fitting C$_{nTh1}$+C$_{Th}$ to the data in sp1 and sp2 are very similar. This is not the case when considering the Band-only fits for which the parameters are dramatically different between sp1 and sp2.

Interestingly, the minimization of a Band function alone to either GBM or GBM+Konus data converges towards two possible minima corresponding to dramatically different spectral parameters (see Table~\ref{tab:spectra} and Figures~\ref{fig5}g and~\ref{fig6}): the solution called Option 1 has a lower PGSTAT value than the solution called Option 2.
In Figure~\ref{fig6}, we overplot in solid lines the C$_{nTh1}$+C$_{Th}$ best fit (red) to GBM+Konus data as well as the two options for the Band function (blue and green for options 1 and 2, respectively). The dashed blue and green lines correspond to the C$_{Th}$ and C$_{nTh1}$ components of the C$_{nTh1}$+C$_{Th}$ model, respectively. We note the strong similarities of the solid and dashed blue lines up to $\sim$35 keV as well as the similarities of the solid and dashed green lines especially above $\sim$300 keV. Therefore, Option 1 mimics clearly the low-energy hump of the C$_{nTh1}$+C$_{Th}$ model (i.e., C$_{Th}$) while Option 2  mimics the high-energy hump of C$_{nTh1}$+C$_{Th}$ (i.e., C$_{nTh1}$).

A $\Delta$PGSTAT of 100, 3 and 120 is obtained between Band and C$_{nTh1}$+C$_{Th}$ with GBM, Konus and GBM+Konus data, respectively, showing that C$_{nTh1}$+C$_{Th}$ is statistically significantly better than Band in the time interval sp2 only with GBM data. The limited improvement obtained with Konus can be explained by a weaker signal detected with Konus as well as with the Konus energy band pass starting at 20 keV while the peak energy of the BB is at $\sim$30 keV (see Figure~\ref{fig6}). Once more, the results obtained in sp2 using GBM, Konus and GBM+Konus data globally support those reported in~\citet{Guiriec:2013}.

While a limited correction of order of a few percent may be needed to reconcile NaI and BGO detector fluxes in sp2, a $\sim$30\% correction is needed in between GBM and Konus. In sp2, fluxes obtained with BGO seem to be systematically higher than those obtained with NaI detectors and Konus. As in sp1, this correction factor required to re-normalize the fluxes for the joint analysis in sp2 does not impact the very good consistency of the parameters resulting from the C$_{nTh1}$+C$_{Th}$ fits when analyzing GBM and Konus data either separately or jointly.

\section{Conclusion}

Through the time-integrated and the coarse time-resolved spectral analysis of~\grb using GBM and Konus data fitted either jointly or separately, we found consistency of the data sets from these two different instruments. However, although the spectral parameters are similar when fitting the data from one instrument or the other, a normalization correction factor which can reach up to 30\% must be applied to calibrate the fluxes, Konus flux estimation being lower than GBM one. A few \% correction may also be applied in some cases to renormalize fluxes between NaI and BGO detectors. BGO 0 seems to systematically predict higher fluxes than the selected sample of NaI detectors and Konus. While the Band function fit to the data of GBM and Konus separately can lead to strong discrepancies in the resulting spectral parameters, C$_{nTh1}$+C$_{Th}$  results in fits that are perfectly compatible for the two instruments.

Overall, the fits obtained in this analysis are very similar to those presented in~\citet{Guiriec:2013} who fitted coarse time intervals. Unfortunately, it is not possible to perform time resolved spectral analysis of Konus data using as fine time intervals as in~\citet{Guiriec:2013} due to the limited time resolution of Konus compared to GBM. Such fine time-resolved spectroscopy was crucial to track the evolution of the components with time, which is an important argument supporting the existence of these two components. Discontinuities were indeed found in the temporal evolution of the parameters of the Band function when Band-alone was fitted to high time resolution data. Such discontinuities are difficult to interpret on physical ground. On another hand, spectral fits using two humps (i.e., C$_{nTh1}$+C$_{Th}$) did not show any discontinuities and the values of the parameters of C$_{nTh1}$ were very much compatible with synchrotron emission from electrons in fast-cooling regime. We observed similar results when comparing time interval sp1 and sp2 in Section~\ref{sec:trsa}. However, with only two time intervals, this result does not appear as clear as in~\citet{Guiriec:2013}.

In addition, fine time-resolved spectral analysis also revealed a relation between the instantaneous luminosity of C$_{nTh1}$, L$_{i}^{nTh1}$, and the corresponding instantaneous $\nu$F$_\nu$ peak energy, E$_\mathrm{peak,i}^{nTh1}$ (i.e., L$_{i}^{nTh1}$--E$_\mathrm{peak,i}^{nTh1}$ relation) only when fitting C$_{nTh1}$+C$_{Th}$ to the data, these two quantities being uncorrelated when fitting Band alone. Even with the coarse resolution of the Konus data our results confirm the analysis performed in~\citet{Guiriec:2013}, establishing two humps in the prompt emission spectra of~\grb. We interpret this double hump structure as resulting from two separate components (i.e., C$_{nTh1}$+C$_{Th}$), one with a thermal-like shape, C$_{Th}$ (most-likely of photospheric origin) and the other one with a non-thermal shape, C$_{nTh1}$ (possibly of synchrotron origin from electrons beyond the photosphere).

Our results show the importance of using the correct spectral model to intercalibrate data sets from different instruments.
Despite the very good consistency of the spectral shapes resulting from the C$_{nTh1}$+C$_{Th}$ fits to the GBM and Konus data, our results reveal flux normalization discrepancies between the two instruments.
To efficiently intercalibrate Konus and GBM, we would need to analyze a larger data set of common events.

\section{Acknowledgments}

To Neil Gehrels; you are an exceptional scientist, mentor, leader, friend and human being. Your support has always been instrumental to the success of this project. You are loved and missed.

We thank the Konus team and in particular D. Frederiks, V. Pal'shin and D. Svinkin for extracting the Konus data used in this study, for providing the time correction corresponding to the light propagation between {\it Fermi} and {\it Wind}, and for their useful comments and suggestions that significantly improved the quality of the article. To complete this work, SG was supported by the NASA Postdoctoral Program (NPP) at the NASA/Goddard Space Flight Center, administered by Oak Ridge Associated Universities through a contract with NASA. SG acknowledges financial support through the Cycles-5,7 and 9 NASA Fermi Guest Investigator program (NNH11ZDA001N \#51279, \#71377 and \#91264) to complete this work as part of a larger project.

\newpage

\setcounter{figure}{0}
\renewcommand{\thefigure}{A\arabic{figure}}

\begin{figure*}
\begin{center}
\includegraphics[totalheight=0.26\textheight, clip, viewport=40 38 523 717,angle=270]{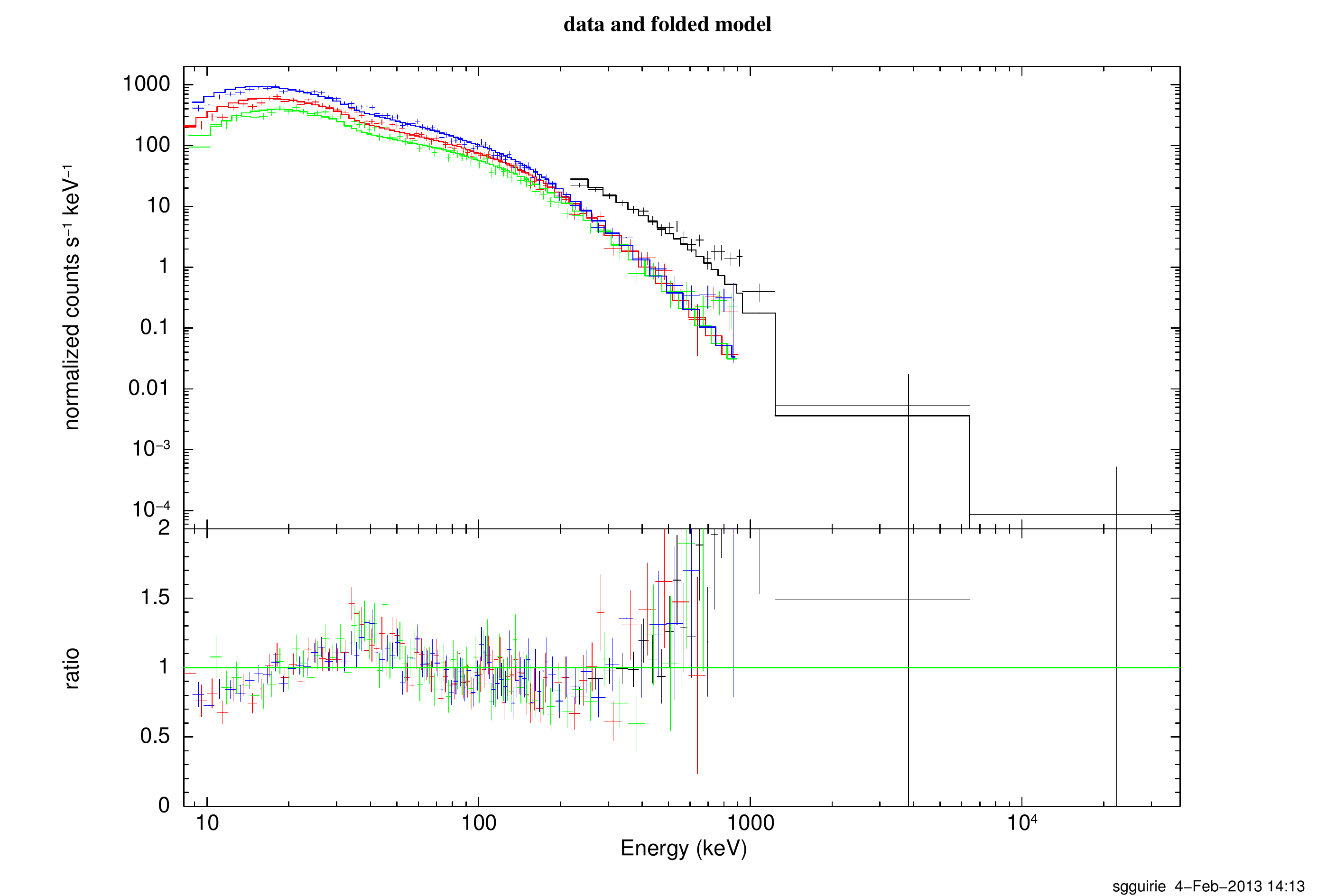}
\includegraphics[totalheight=0.245\textheight, clip, viewport=40 75 523 717,angle=270]{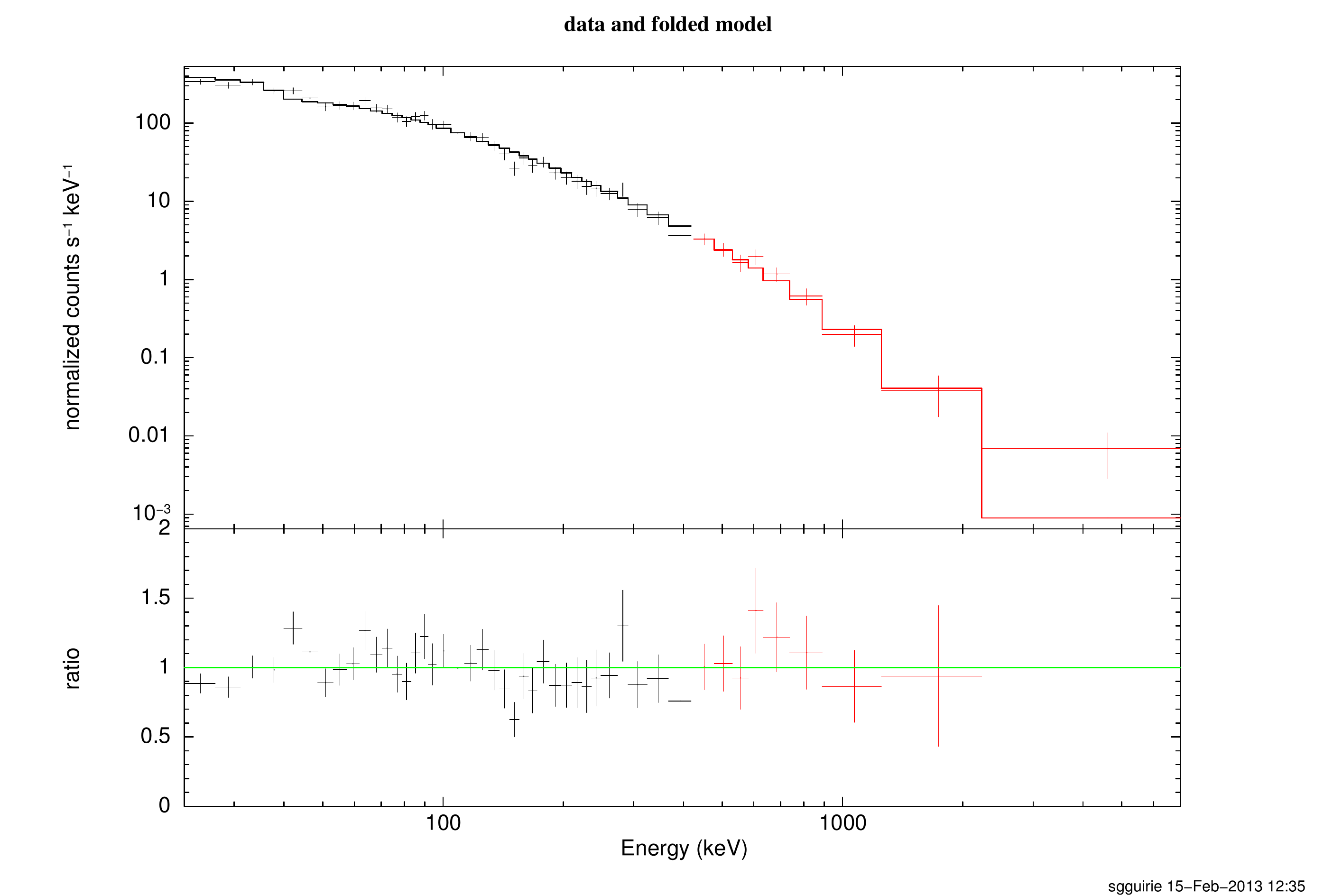}
\includegraphics[totalheight=0.245\textheight, clip, viewport=40 75 523 717,angle=270]{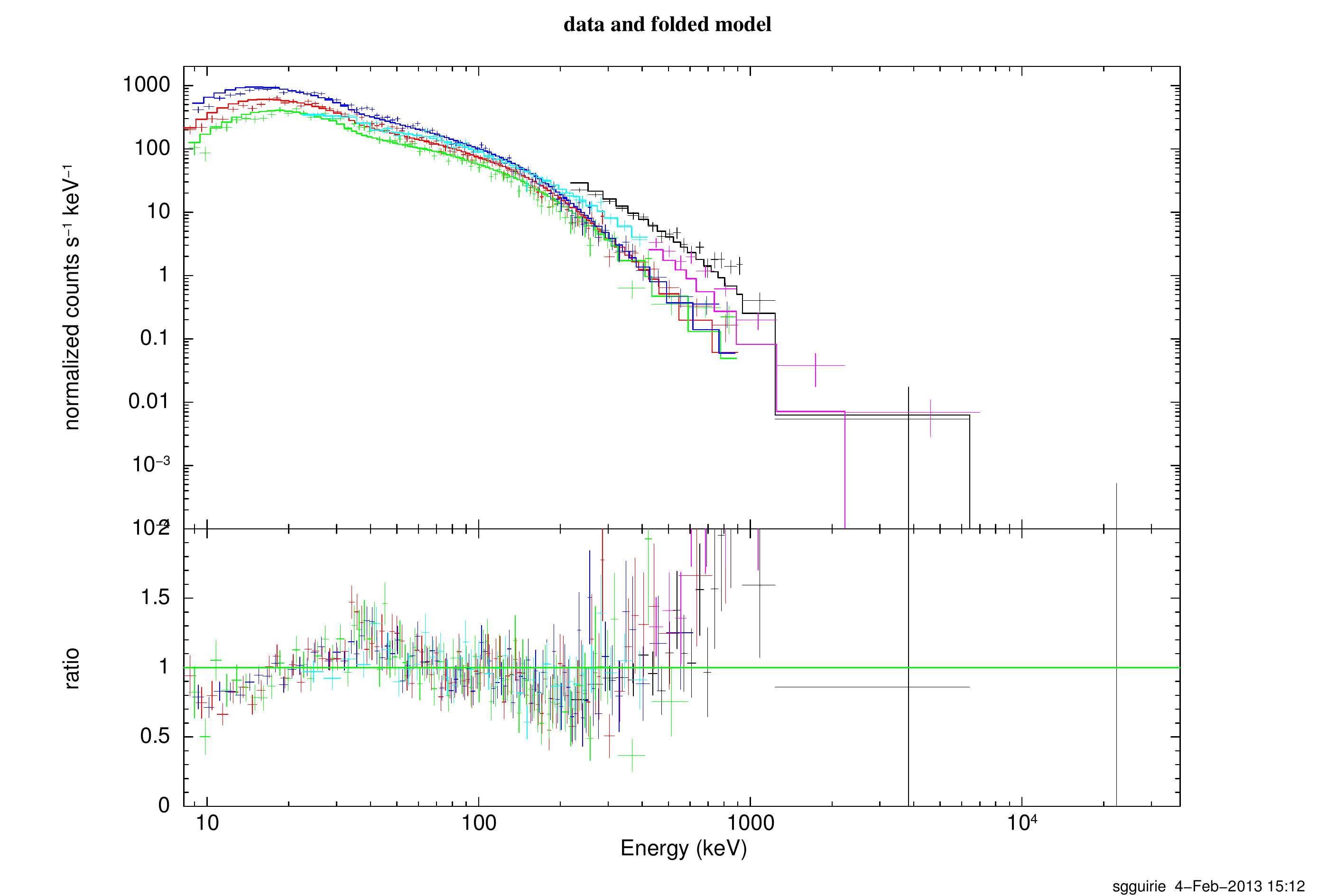}

\includegraphics[totalheight=0.26\textheight, clip, viewport=40 38 523 717,angle=270]{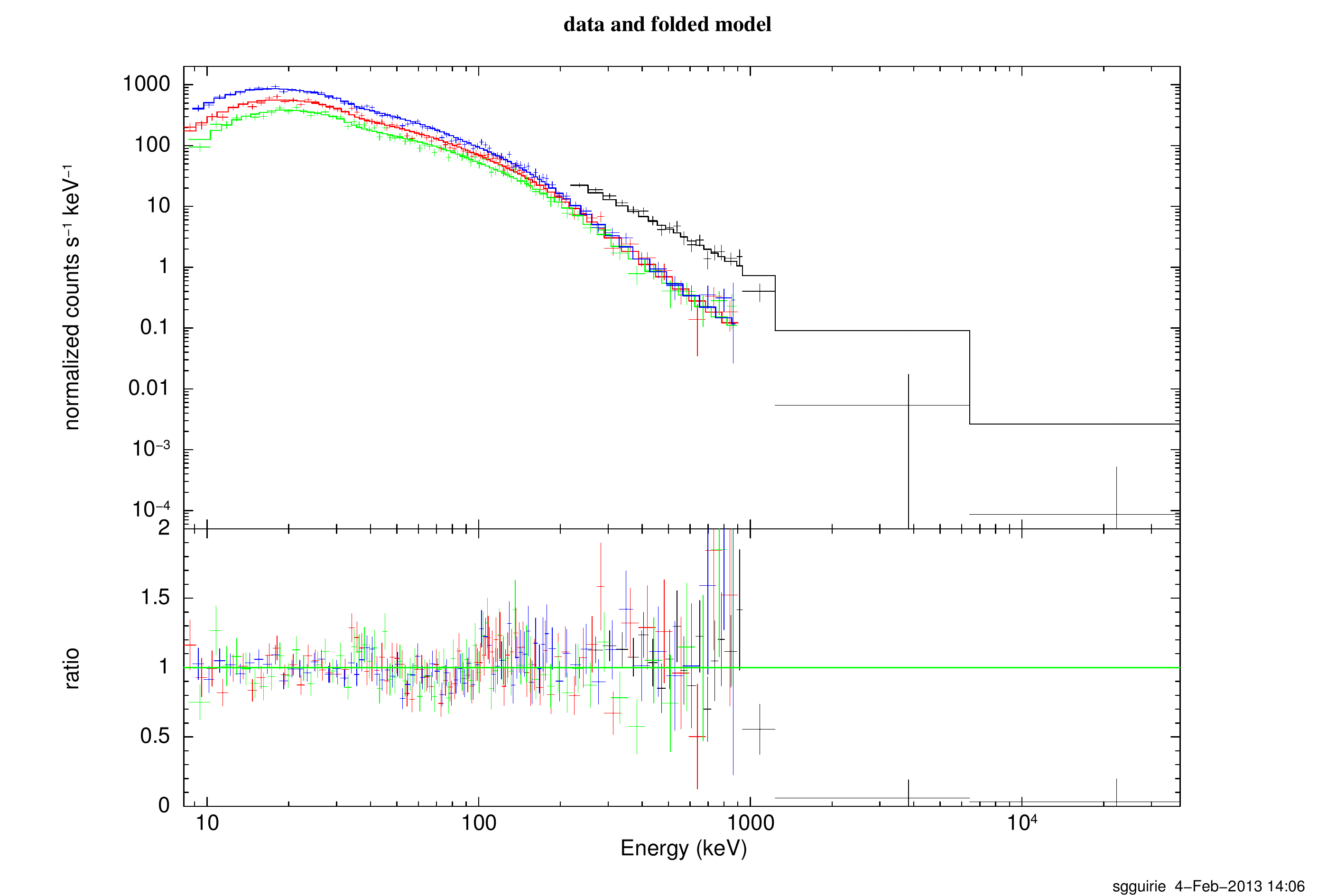}
\includegraphics[totalheight=0.245\textheight, clip, viewport=40 75 523 717,angle=270]{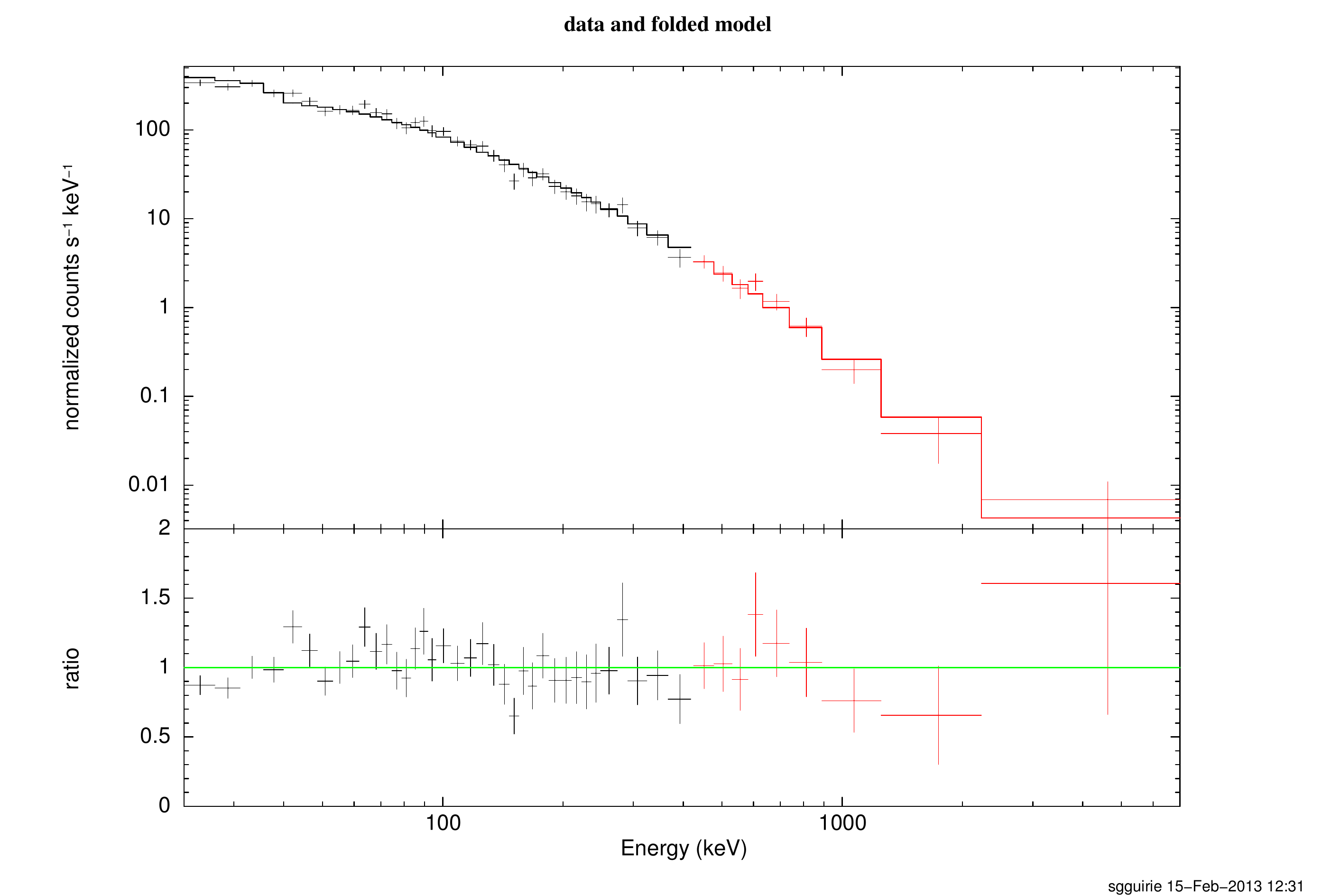}
\includegraphics[totalheight=0.245\textheight, clip, viewport=40 75 523 717,angle=270]{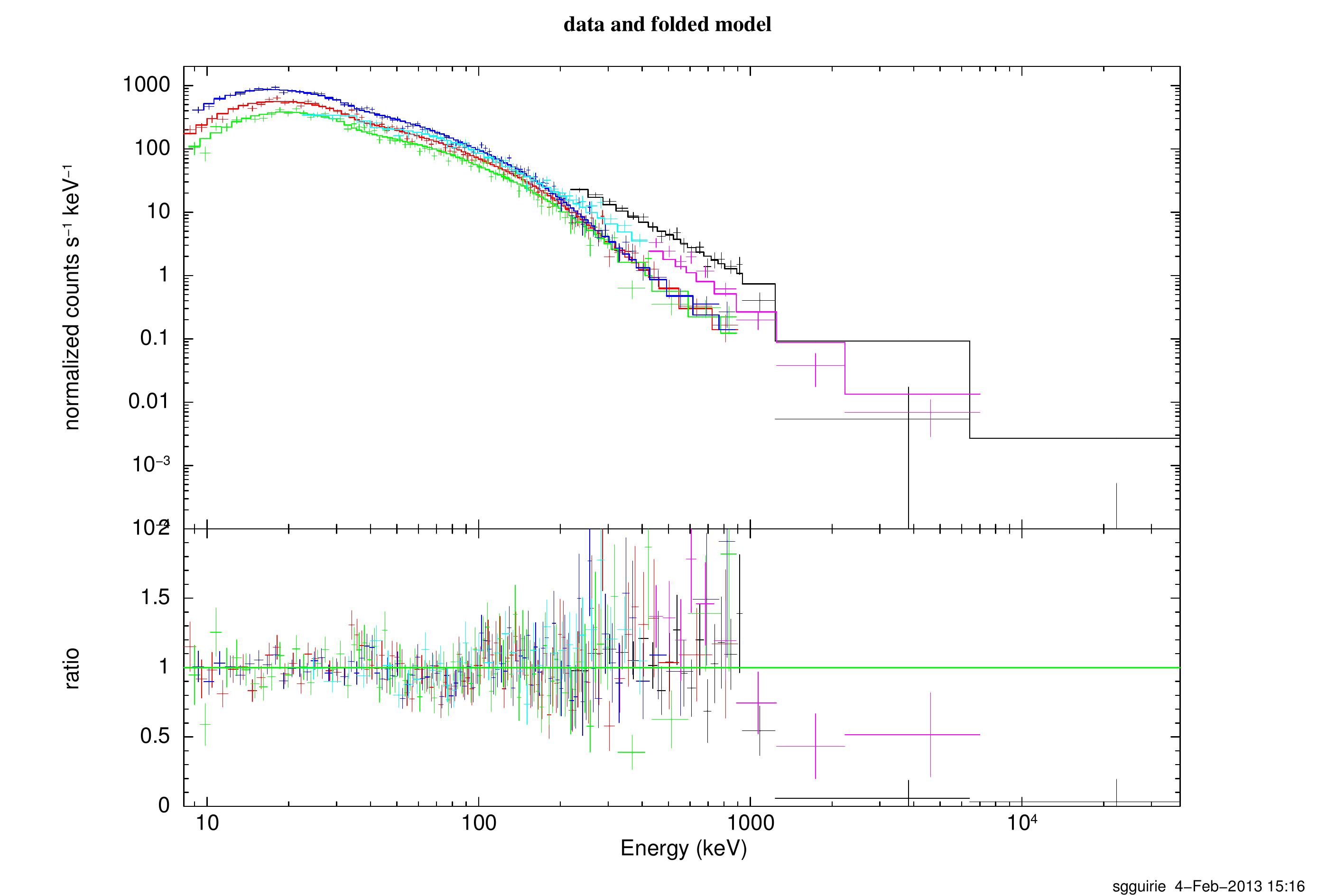}

\includegraphics[totalheight=0.26\textheight, clip, viewport=40 38 523 717,angle=270]{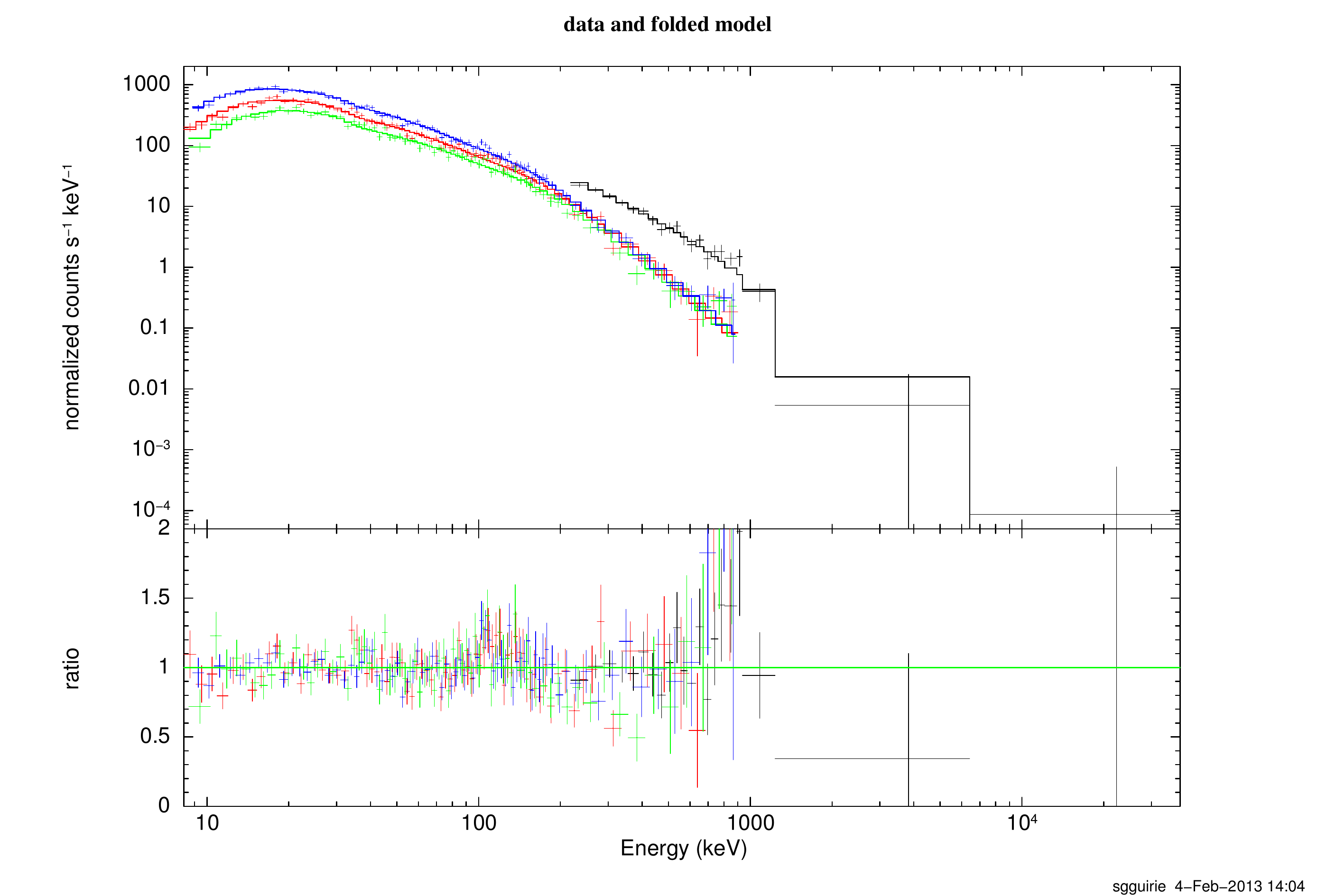}
\includegraphics[totalheight=0.245\textheight, clip, viewport=40 75 523 717,angle=270]{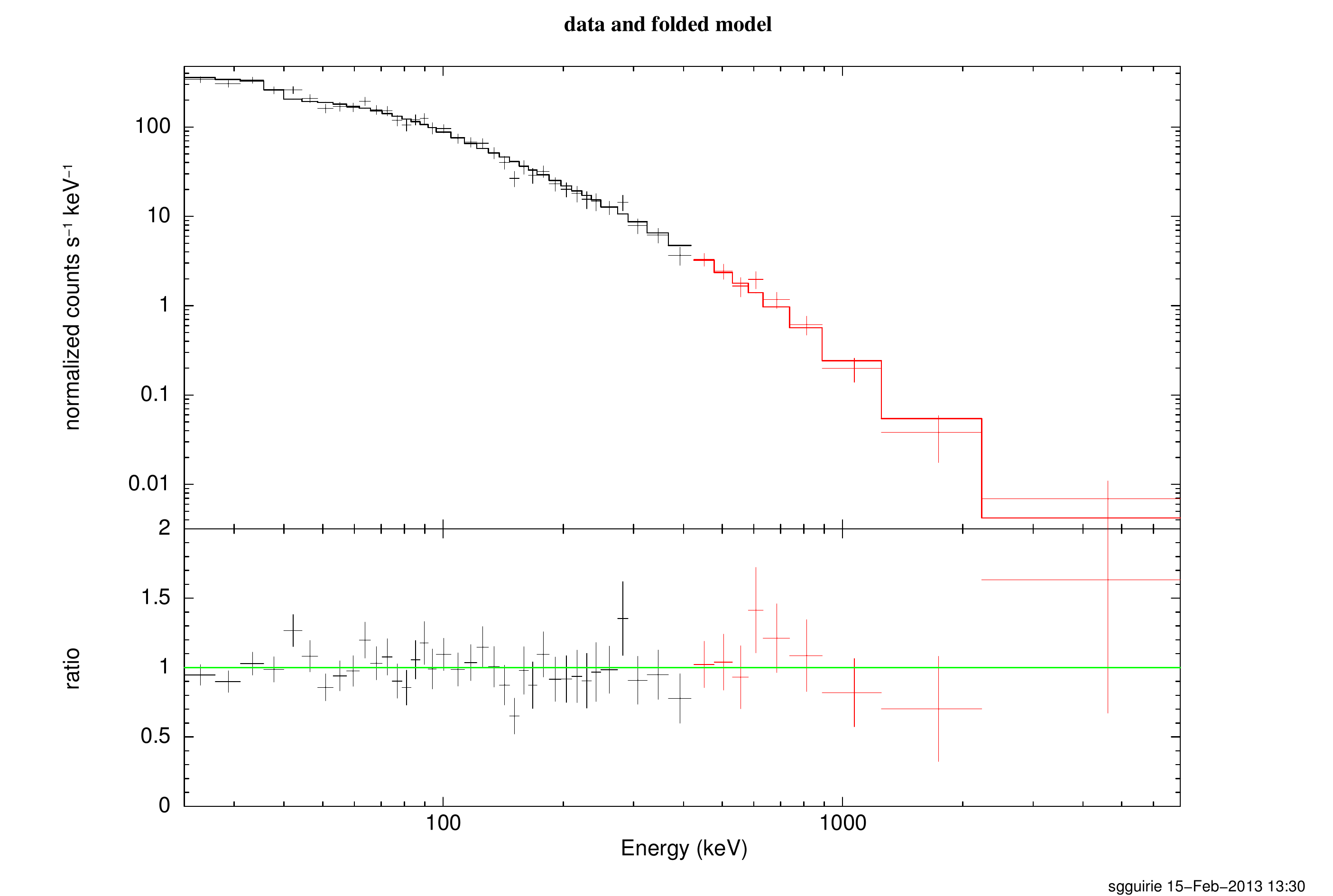}
\includegraphics[totalheight=0.245\textheight, clip, viewport=40 75 523 717,angle=270]{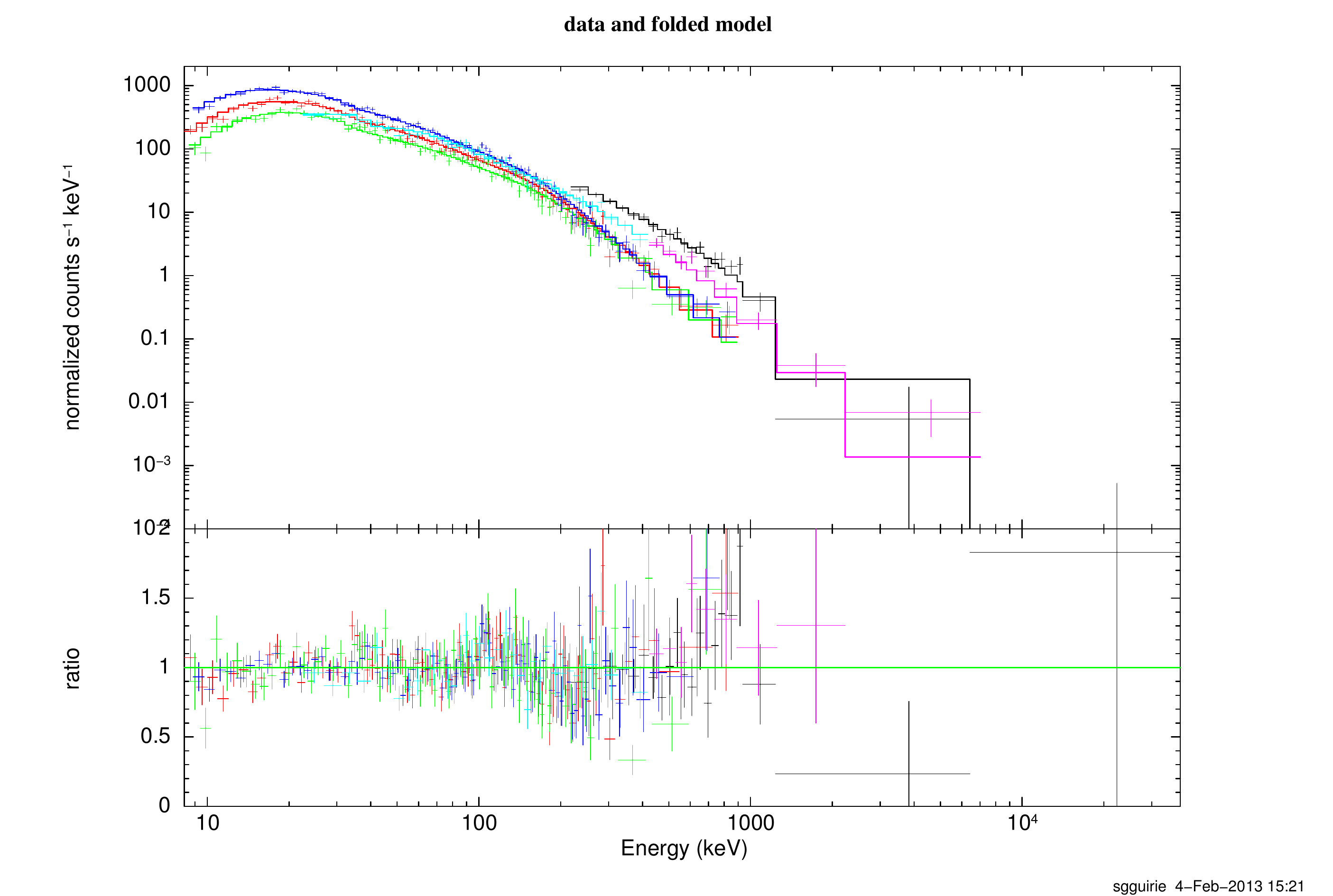}

\includegraphics[totalheight=0.26\textheight, clip, viewport=40 38 523 717,angle=270]{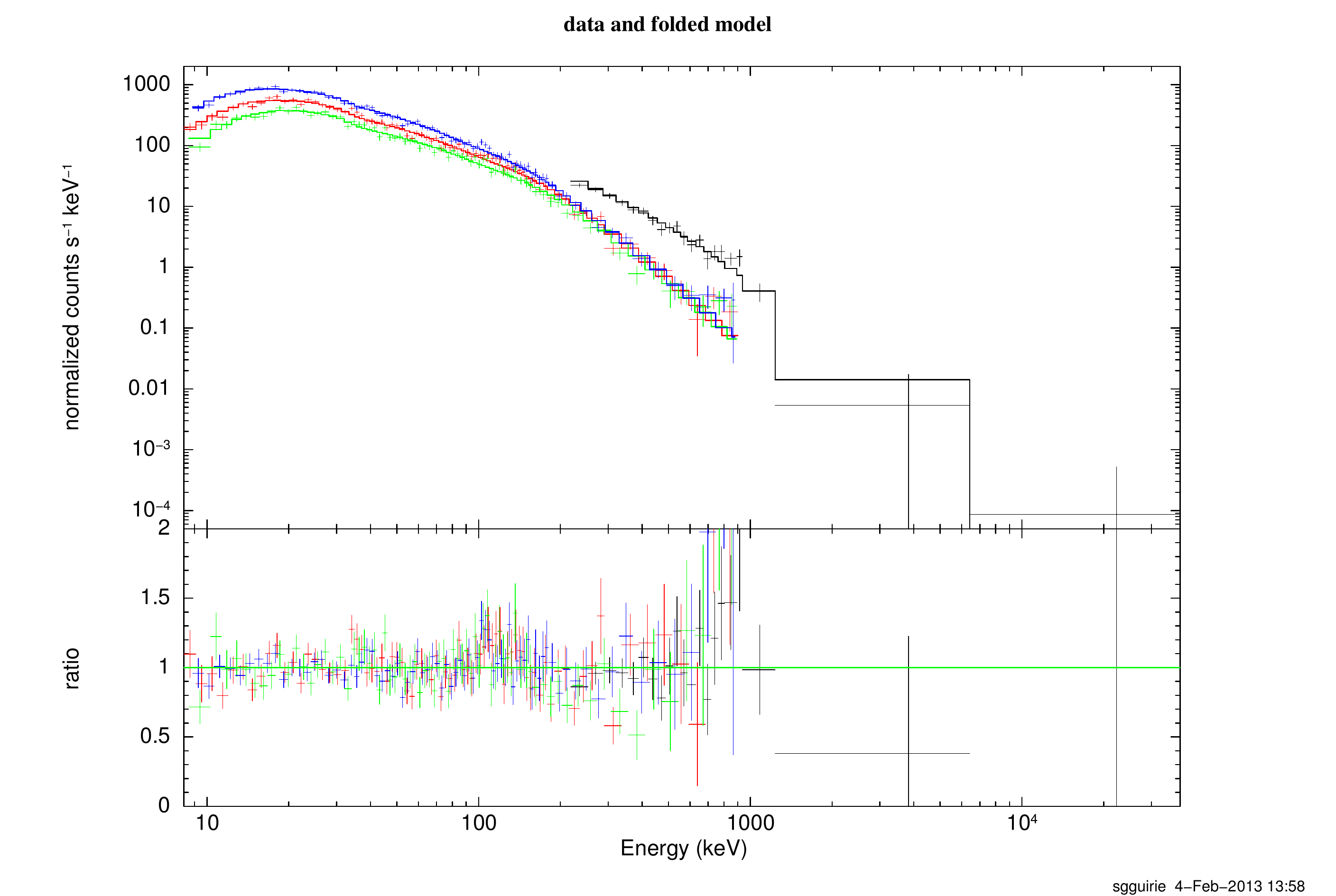}
\includegraphics[totalheight=0.245\textheight, clip, viewport=40 75 523 717,angle=270]{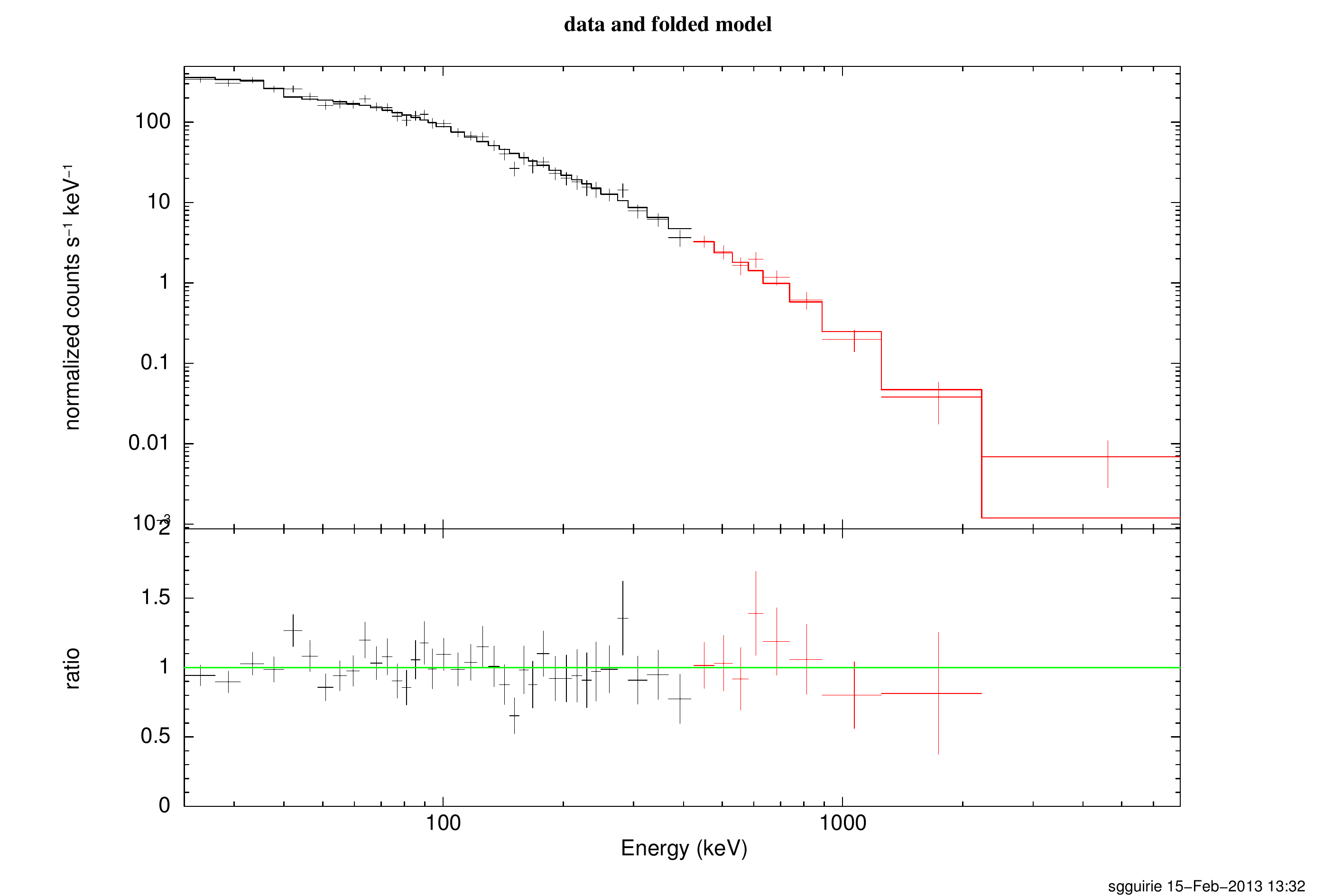}
\includegraphics[totalheight=0.245\textheight, clip, viewport=40 75 523 717,angle=270]{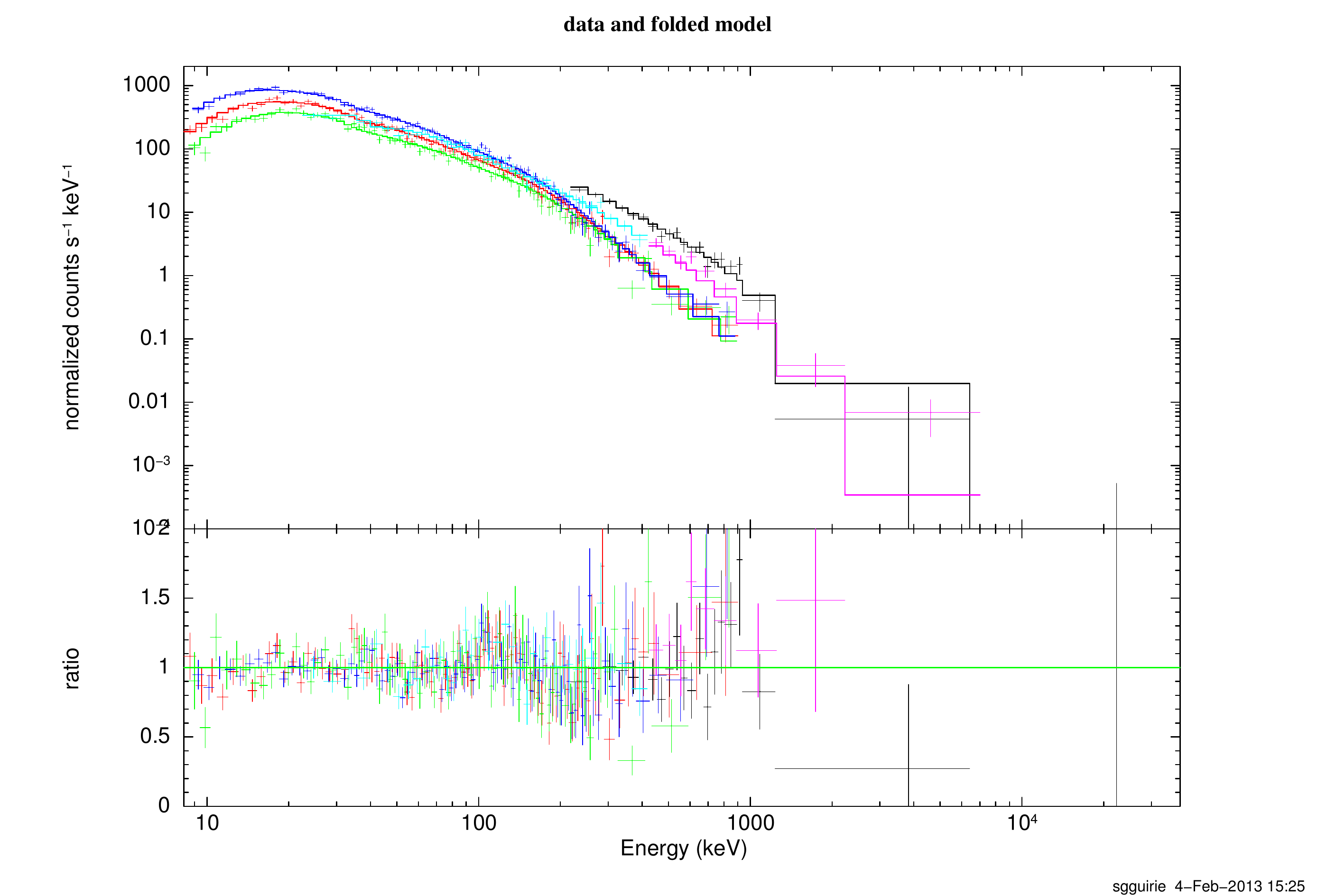}
\caption{\label{fig2}GBM, Konus and GBM+Konus count-rate spectra (left, center and right columns, respectively) in time interval sp1--4 (T$_\mathrm{0}$ to T$_\mathrm{0}$+0.256 s) with the residuals resulting from the CPL, Band, B+BB and C+BB fits (lines 1 to 4 from top to bottom, respectively). GBM energy channels have been grouped in larger energy bins for display purpose only; this grouping does not affect the fitting process that uses the best energy resolution of the instrument.}
\end{center}
\end{figure*}

\newpage

\begin{figure*}
\begin{center}
\includegraphics[totalheight=0.26\textheight, clip, viewport=40 38 523 717,angle=270]{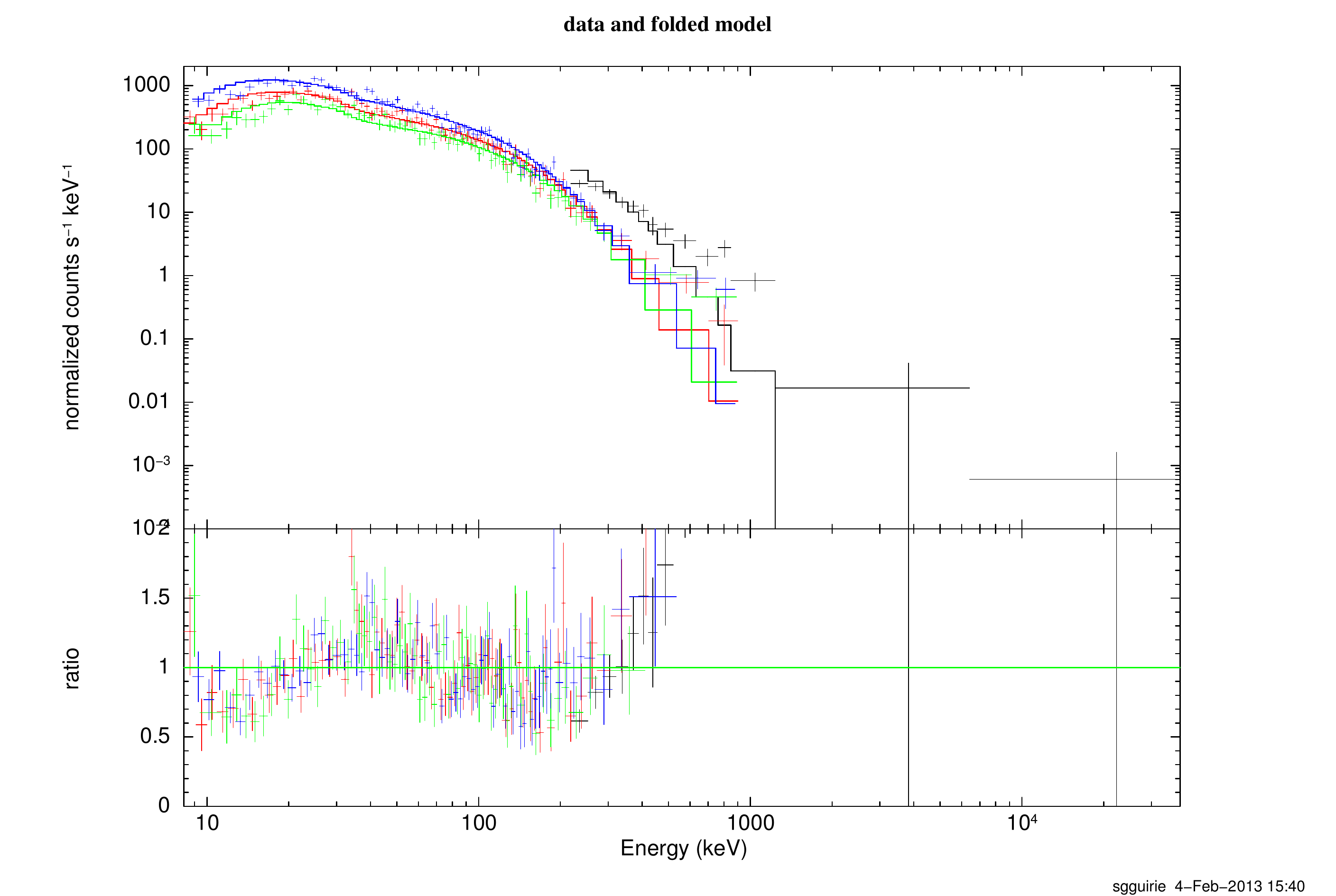}
\includegraphics[totalheight=0.245\textheight, clip, viewport=40 75 523 717,angle=270]{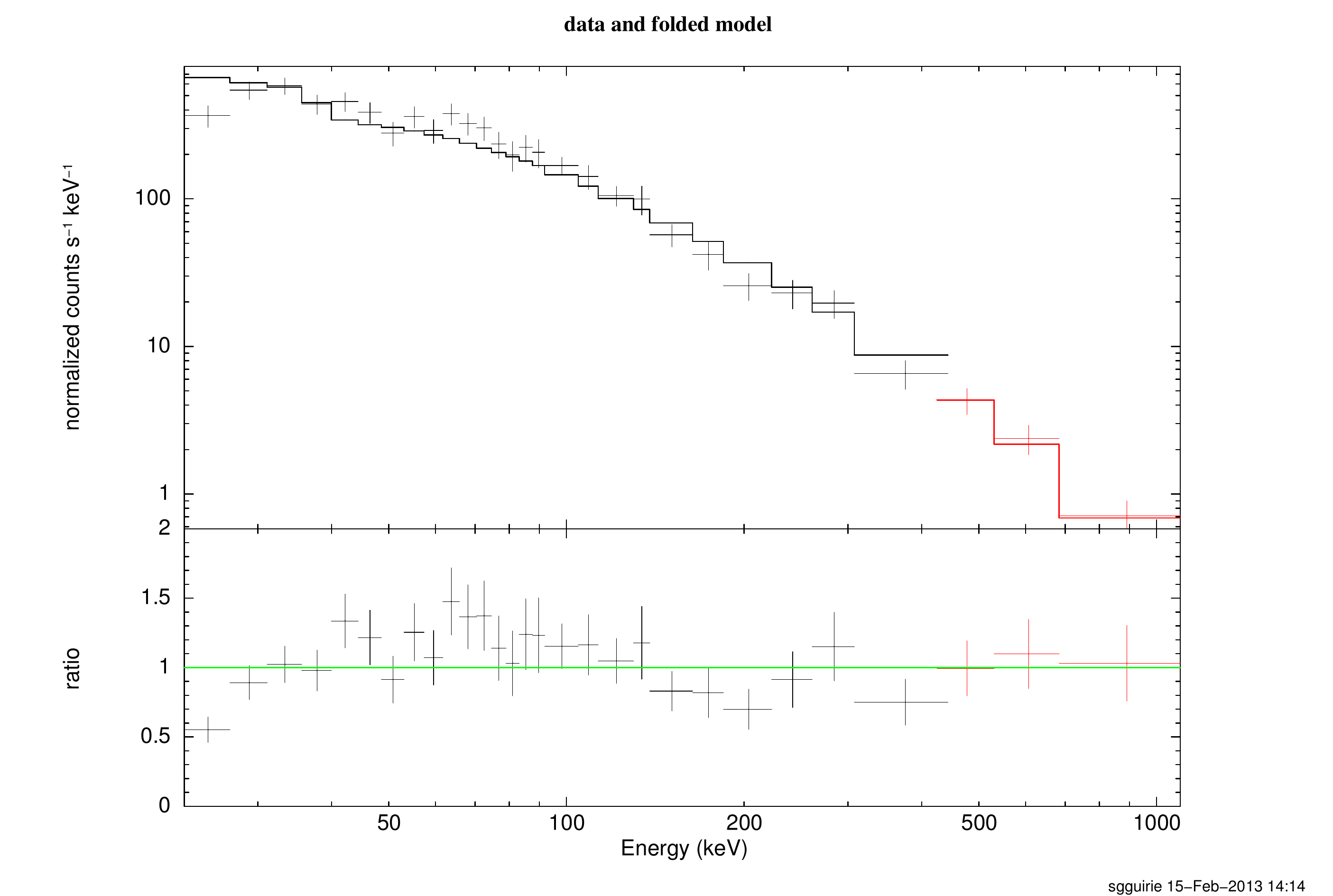}
\includegraphics[totalheight=0.245\textheight, clip, viewport=40 75 523 717,angle=270]{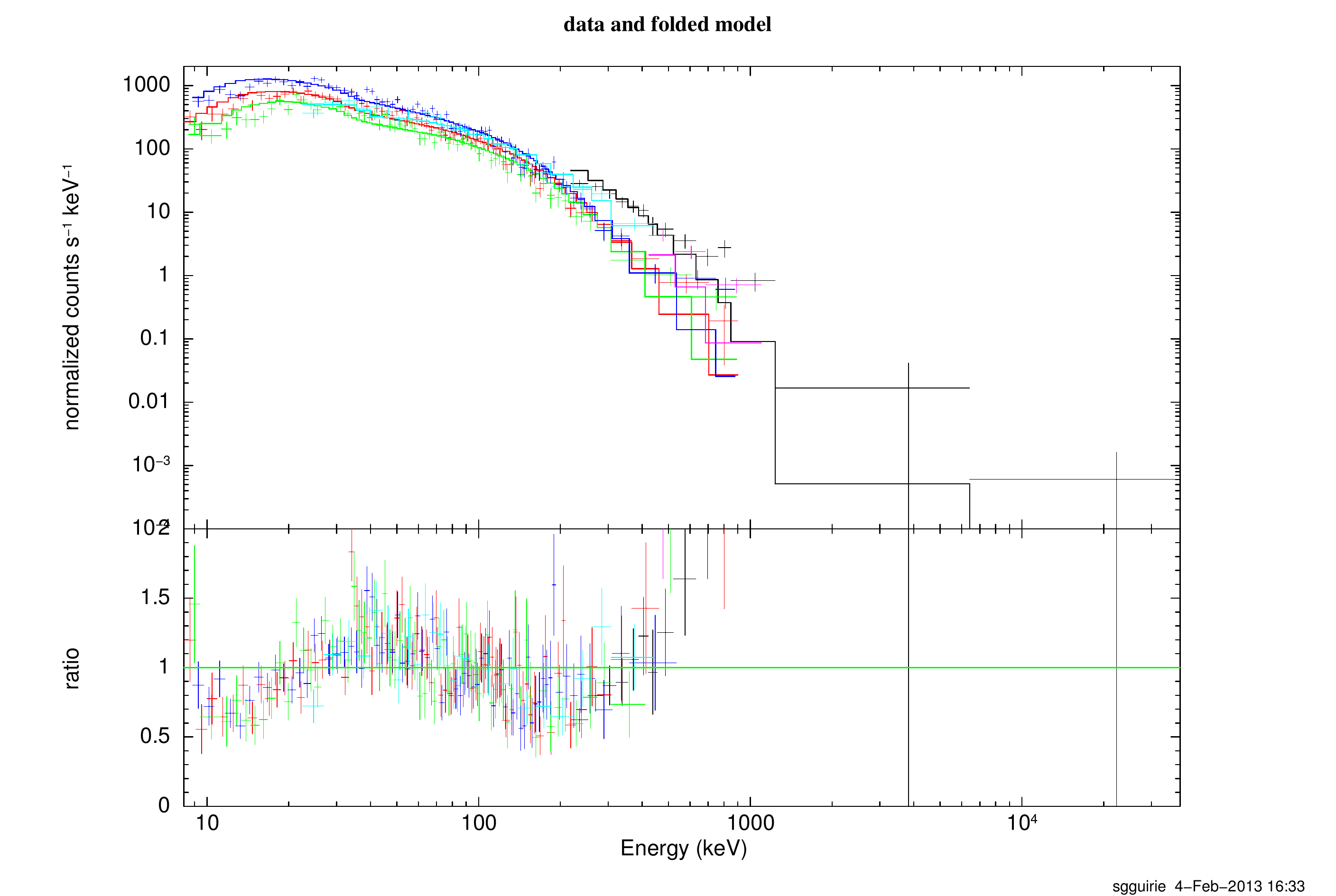}

\includegraphics[totalheight=0.26\textheight, clip, viewport=40 38 523 717,angle=270]{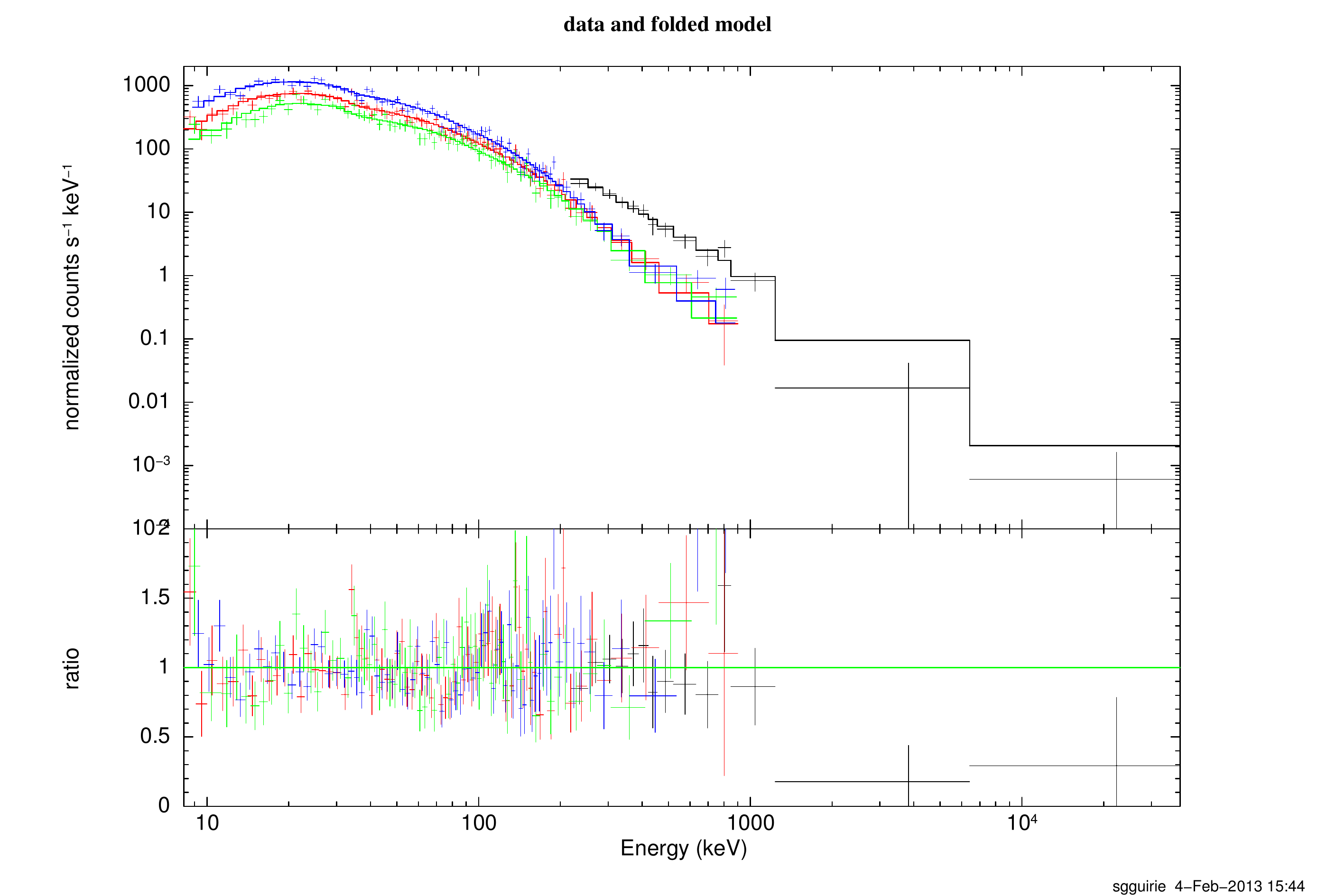}
\includegraphics[totalheight=0.245\textheight, clip, viewport=40 75 523 717,angle=270]{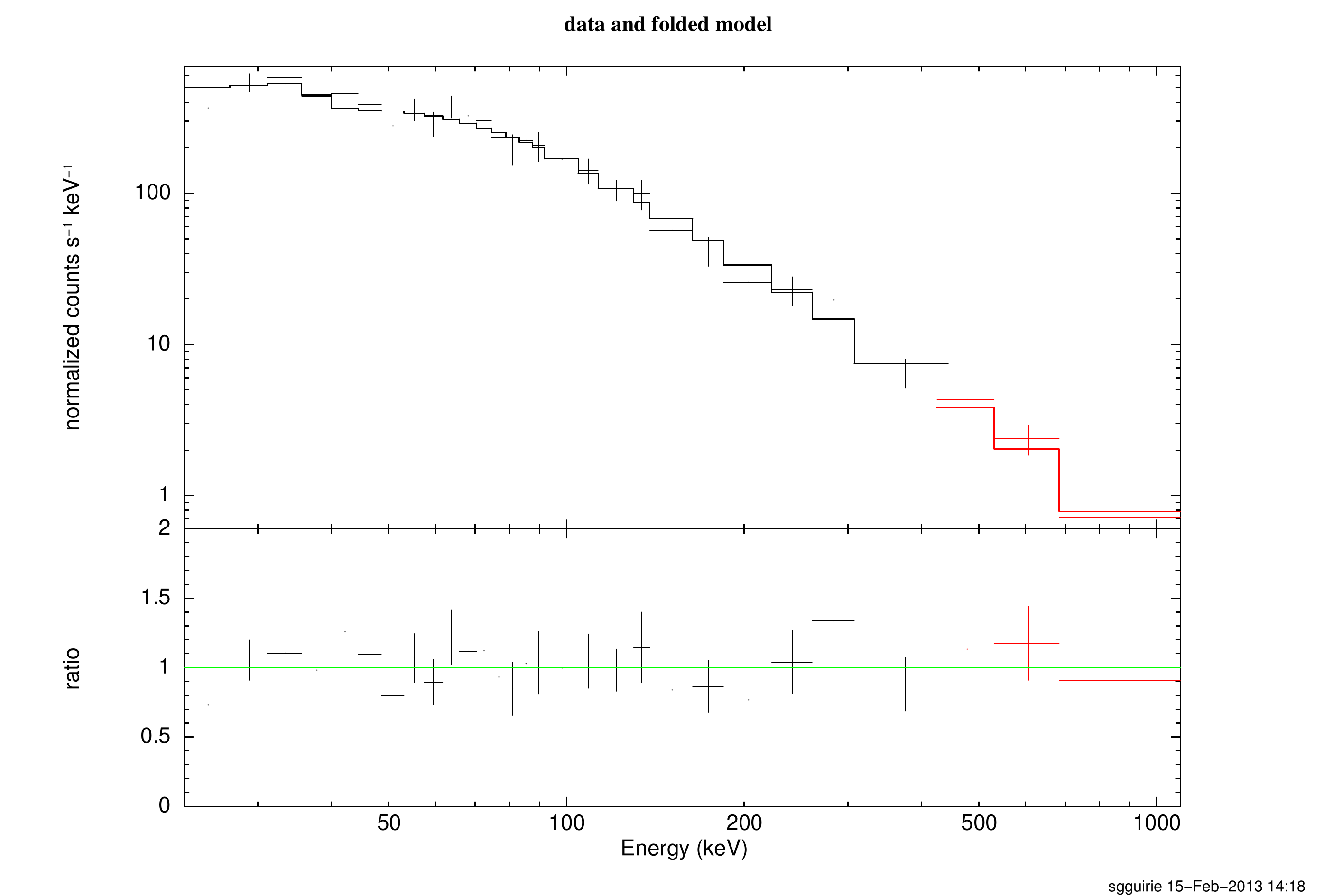}
\includegraphics[totalheight=0.245\textheight, clip, viewport=40 75 523 717,angle=270]{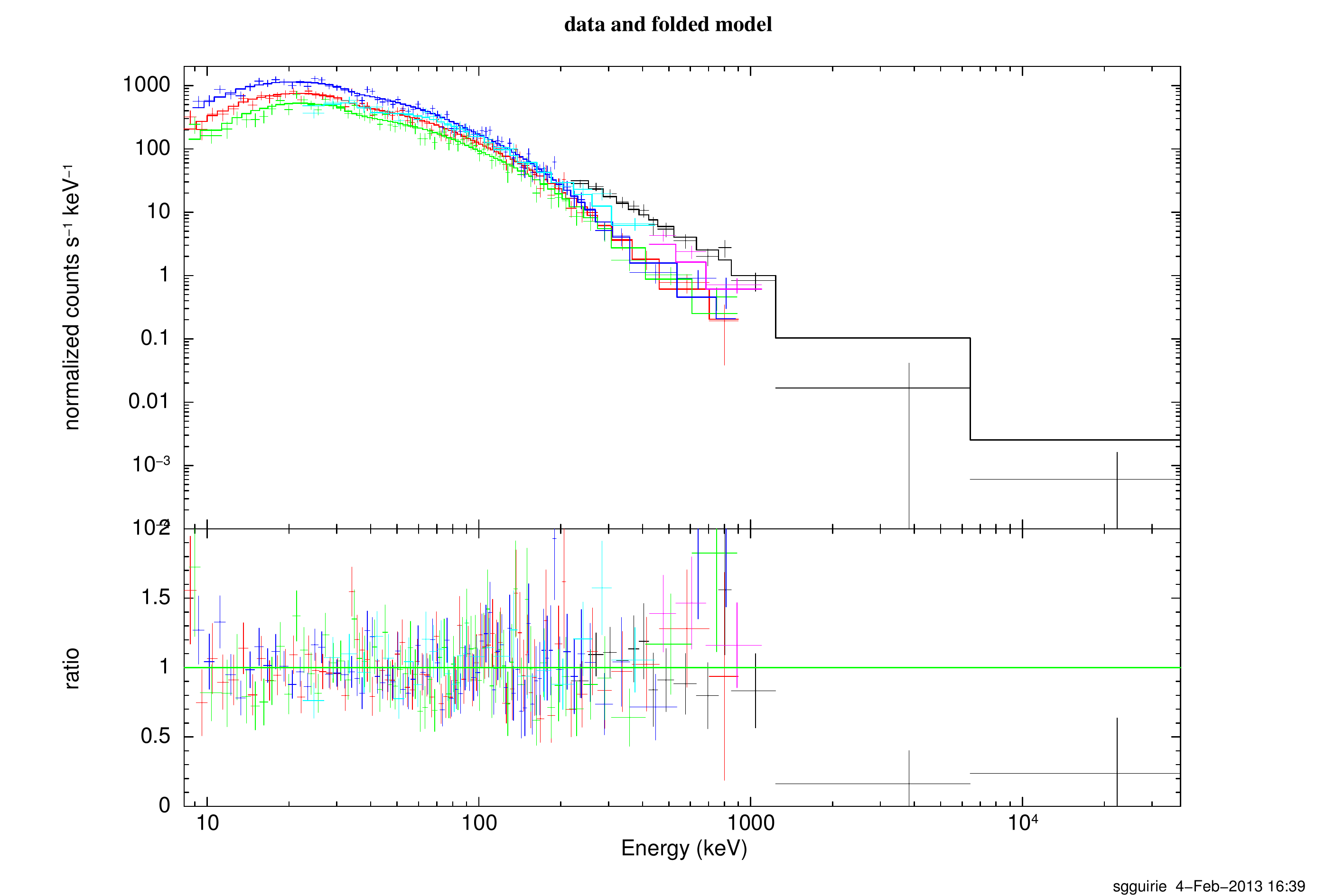}

\includegraphics[totalheight=0.26\textheight, clip, viewport=40 38 523 717,angle=270]{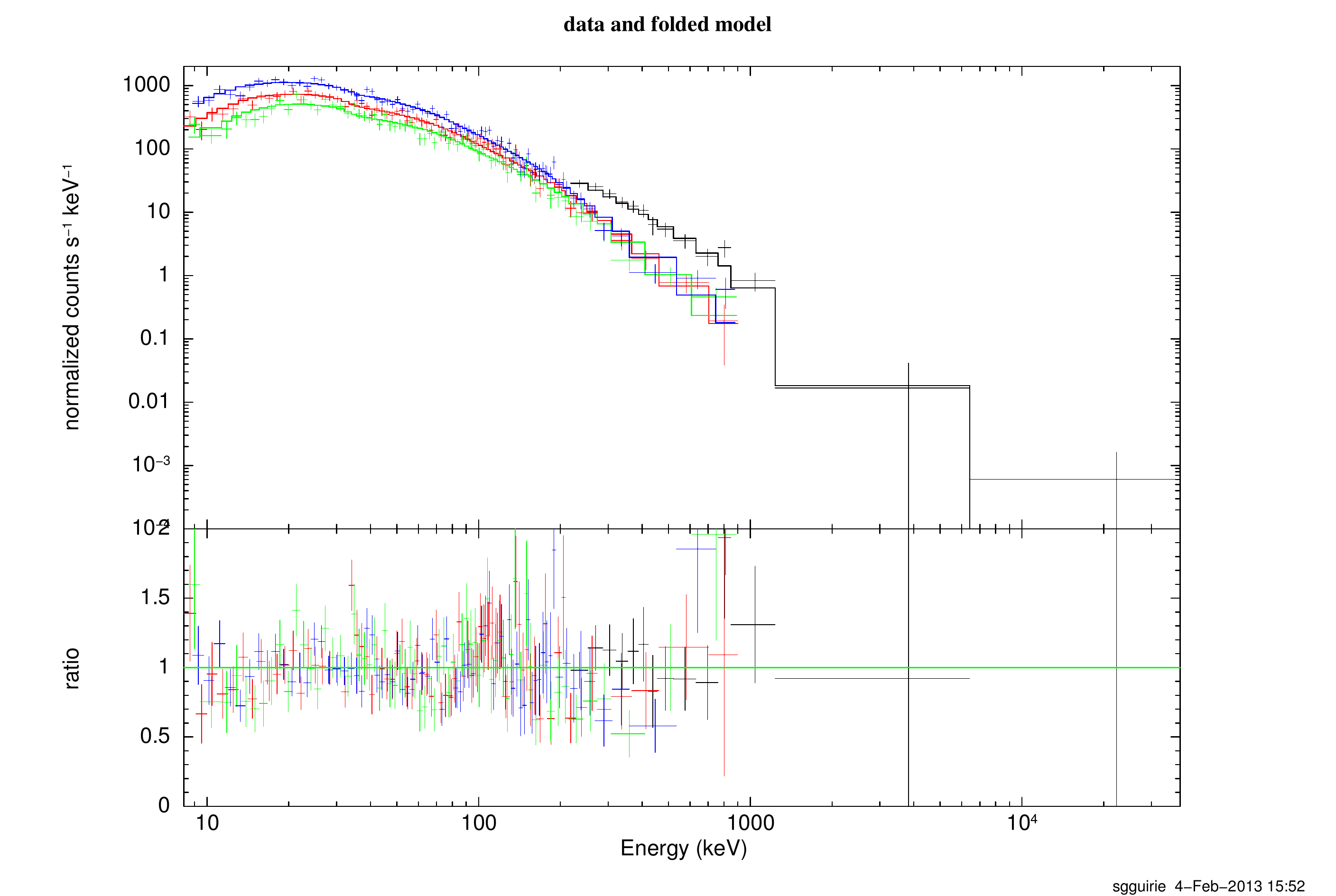}
\includegraphics[totalheight=0.245\textheight, clip, viewport=40 75 523 717,angle=270]{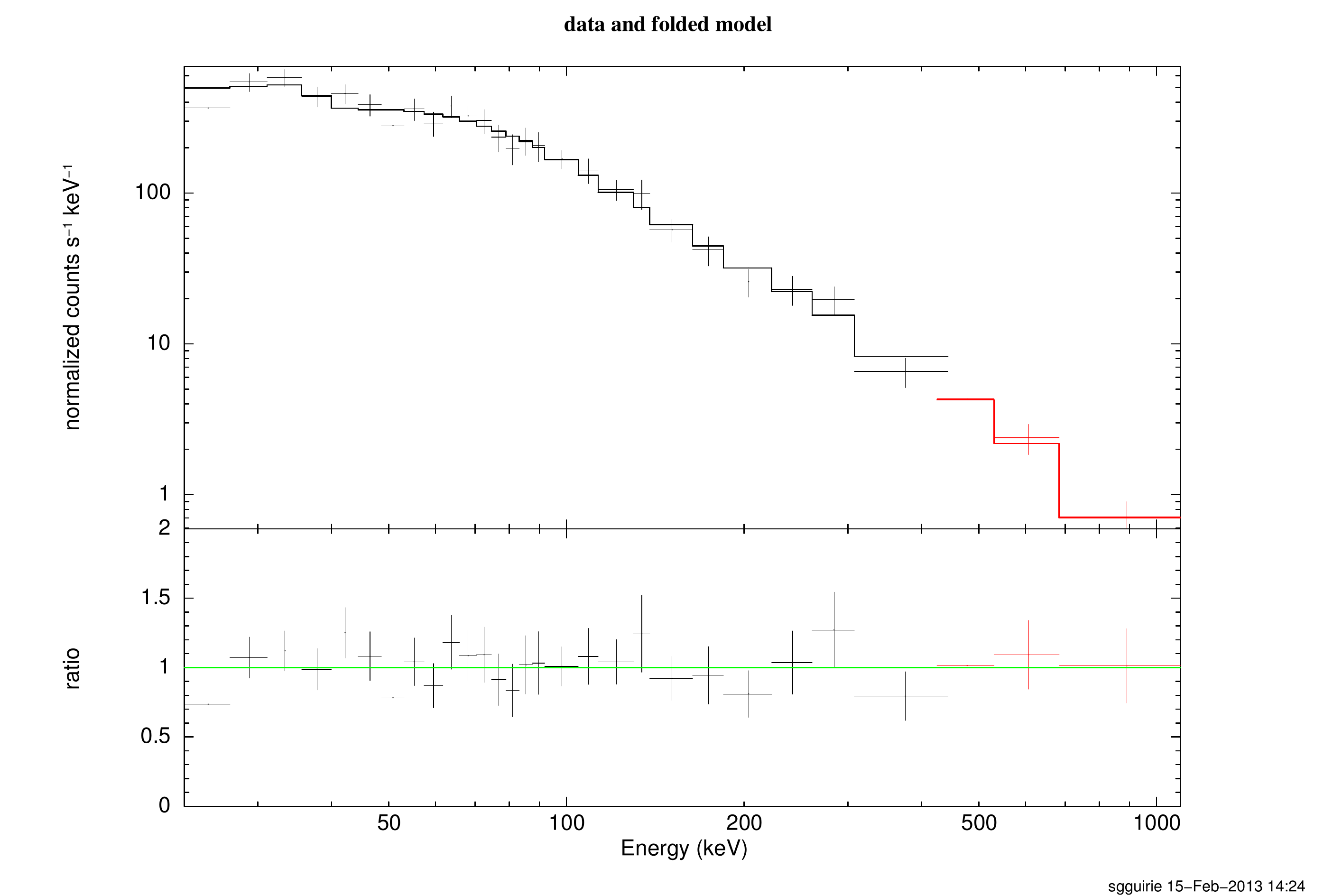}
\includegraphics[totalheight=0.245\textheight, clip, viewport=40 75 523 717,angle=270]{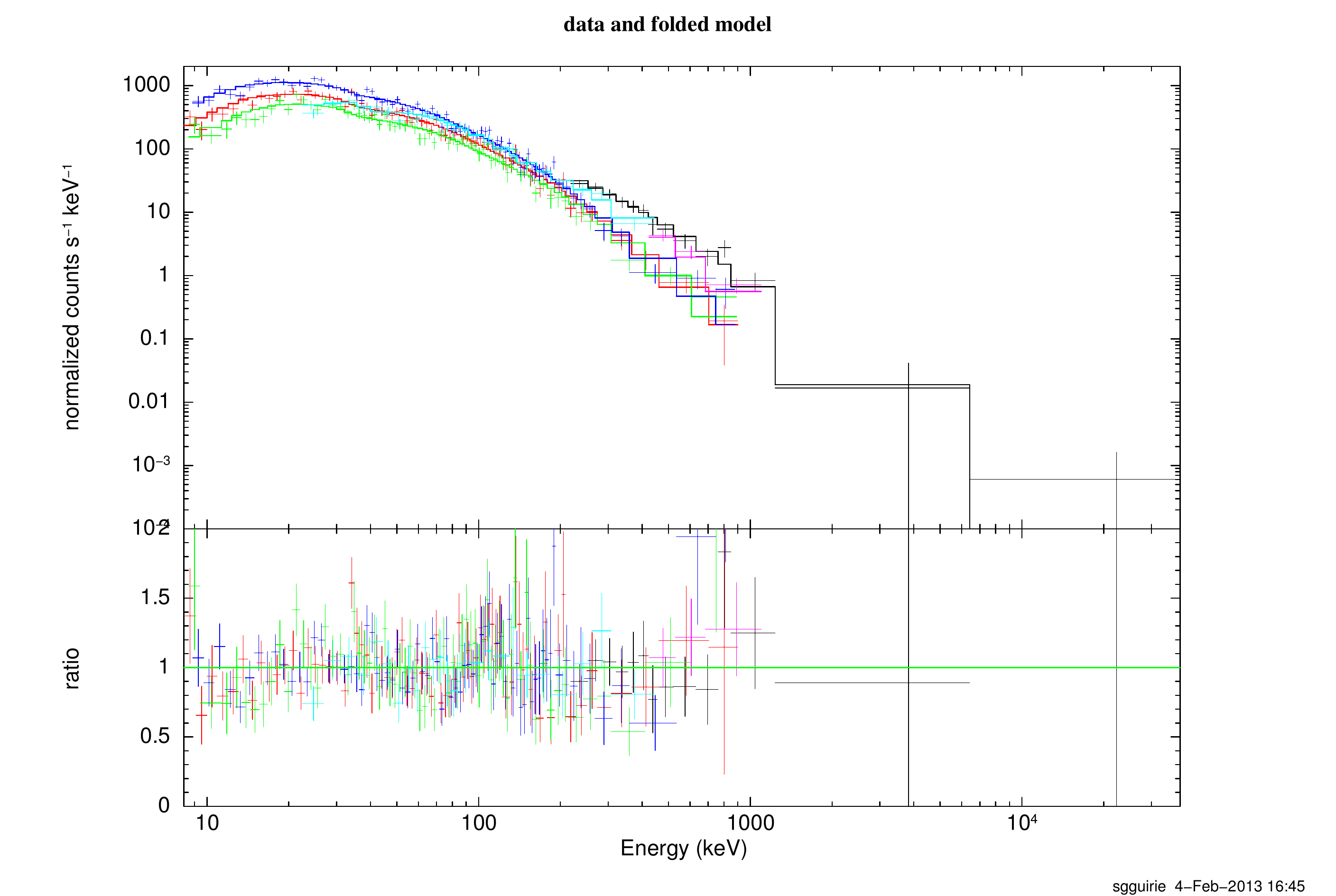}

\hspace{-11.95cm}
\includegraphics[totalheight=0.26\textheight, clip, viewport=40 38 523 717,angle=270]{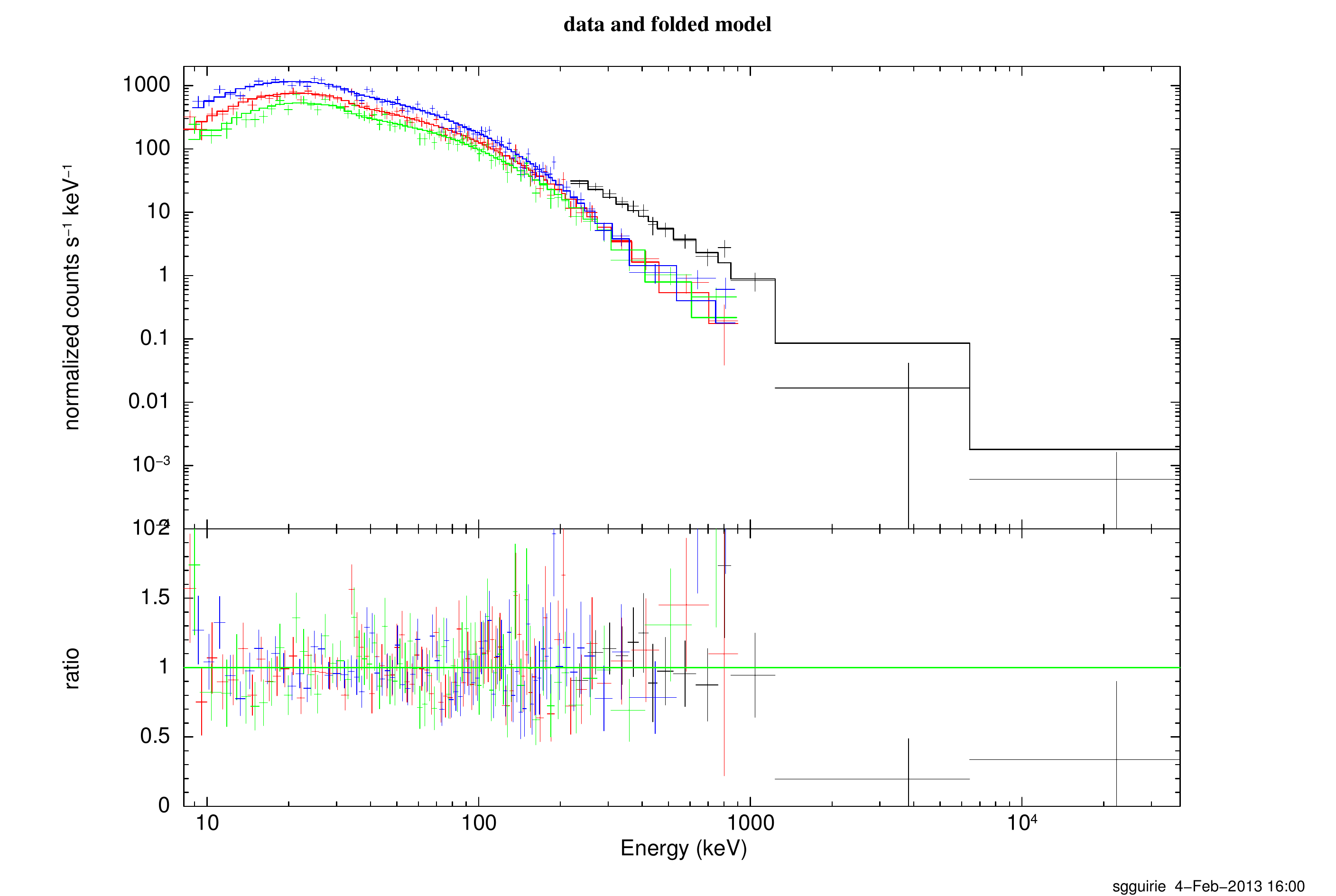}

\includegraphics[totalheight=0.26\textheight, clip, viewport=40 38 523 717,angle=270]{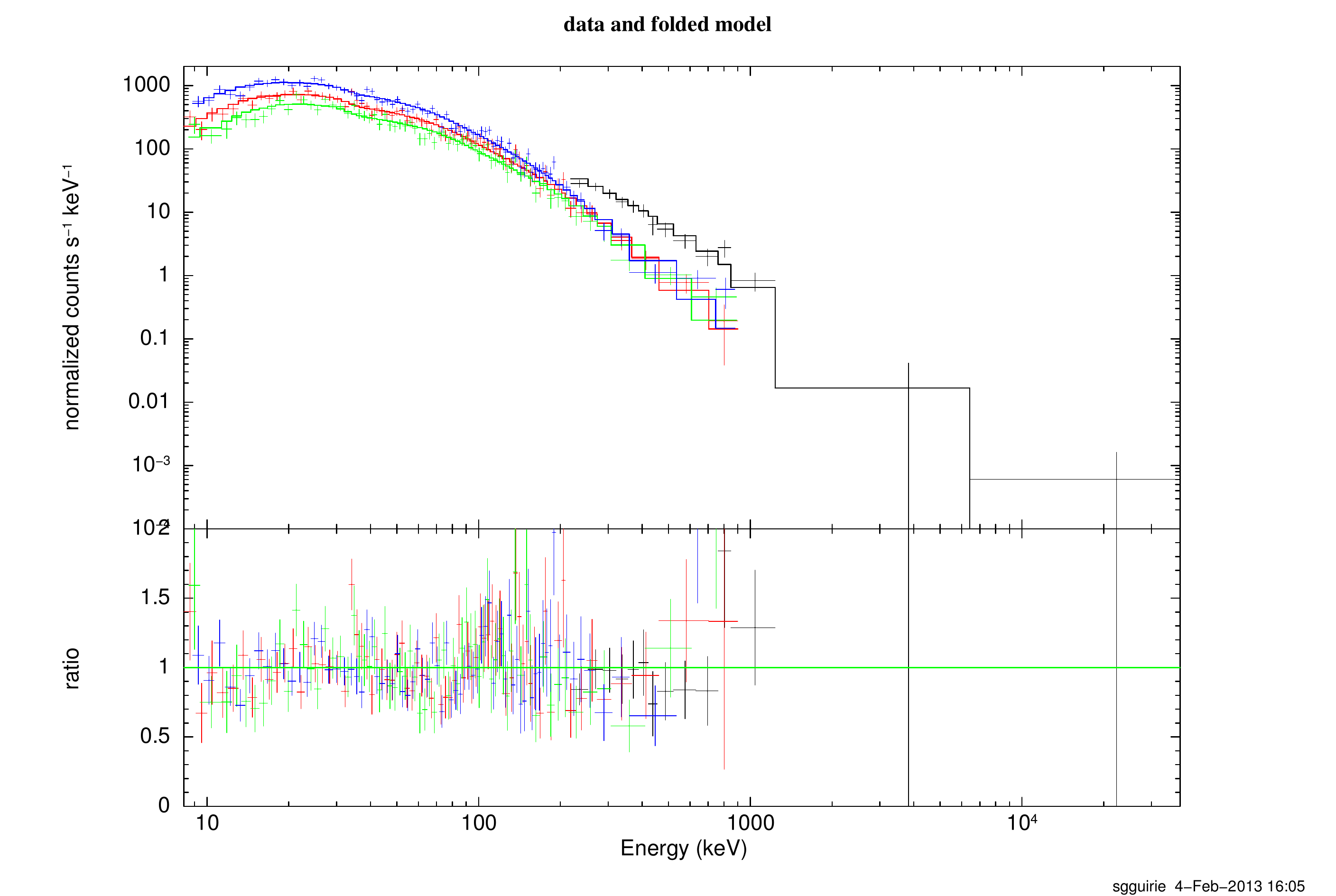}
\includegraphics[totalheight=0.245\textheight, clip, viewport=40 75 523 717,angle=270]{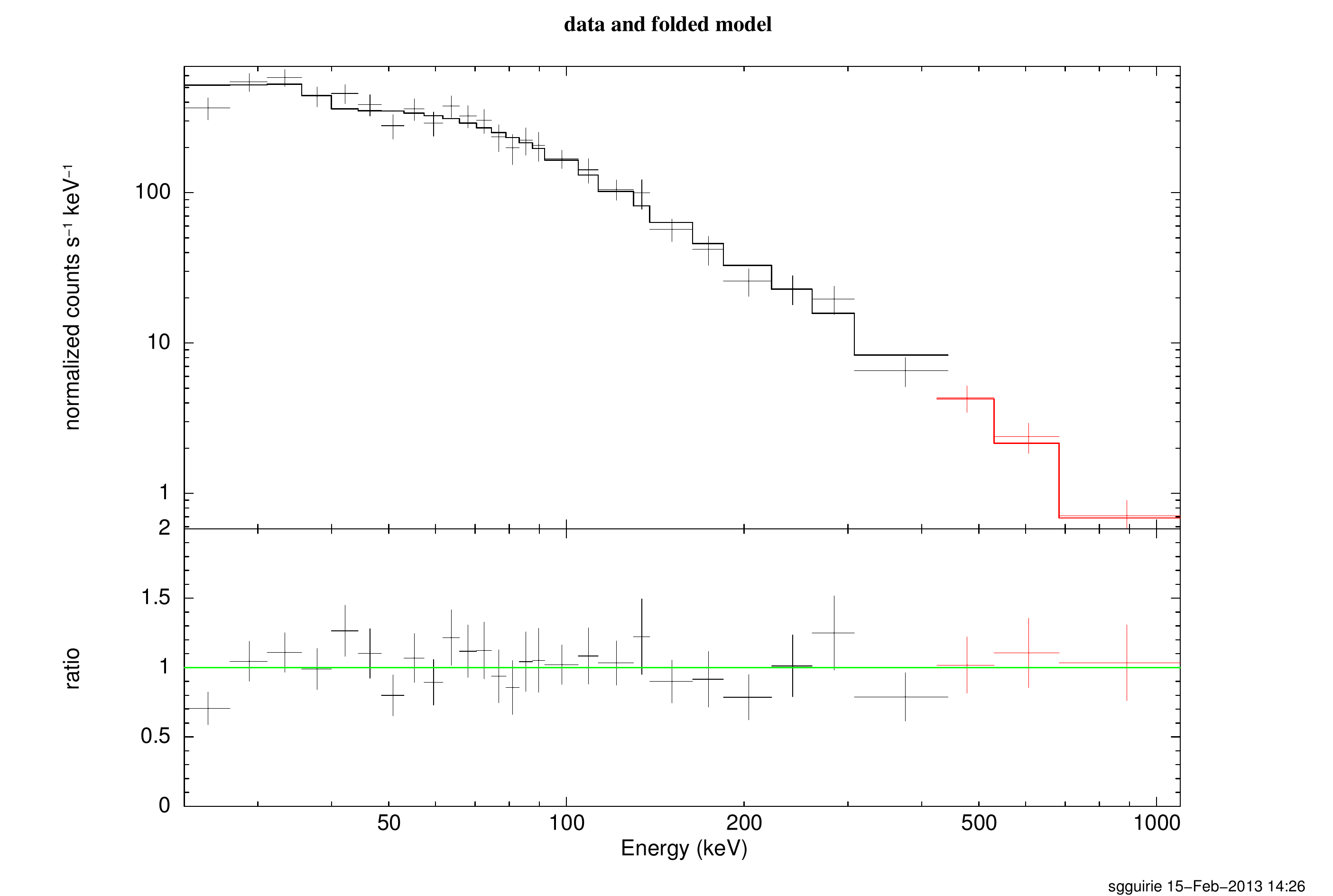}
\includegraphics[totalheight=0.245\textheight, clip, viewport=40 75 523 717,angle=270]{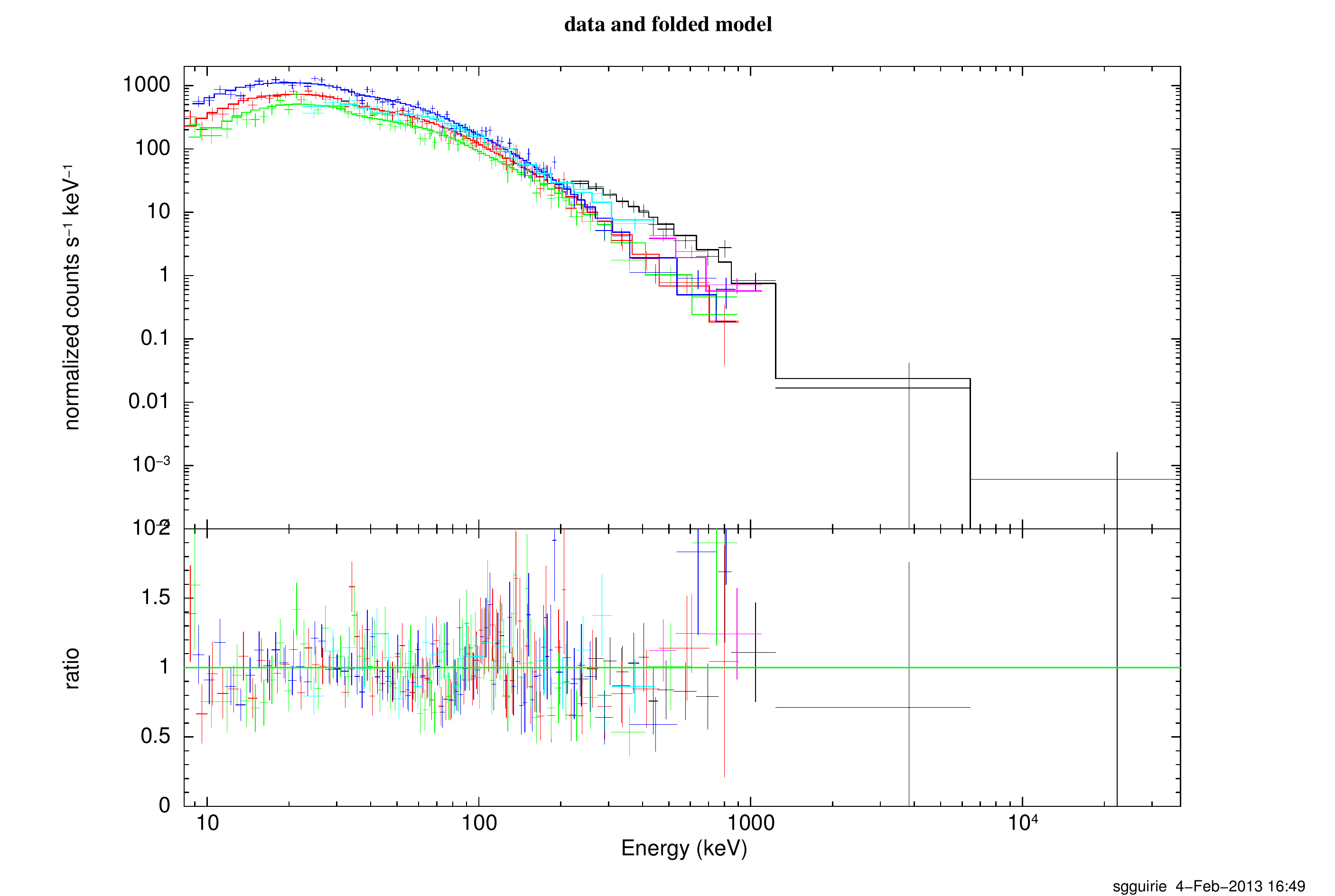}
\caption{\label{fig3}GBM, Konus and GBM+Konus count data (left, center and right columns, respectively) in time interval sp1 (T$_\mathrm{0}$ to T$_\mathrm{0}$+0.064 s) with the residuals resulting from the CPL, Band, B+BB and C+BB fits (lines 1, 2, 3-4 and 5 from top to bottom, respectively). GBM energy channels have been grouped in larger energy bins for display purpose only; this grouping does not affect the fitting process that uses the best energy resolution of the instrument.}
\end{center}
\end{figure*}

\newpage

\begin{figure*}
\begin{center}
\includegraphics[totalheight=0.26\textheight, clip, viewport=40 38 523 717,angle=270]{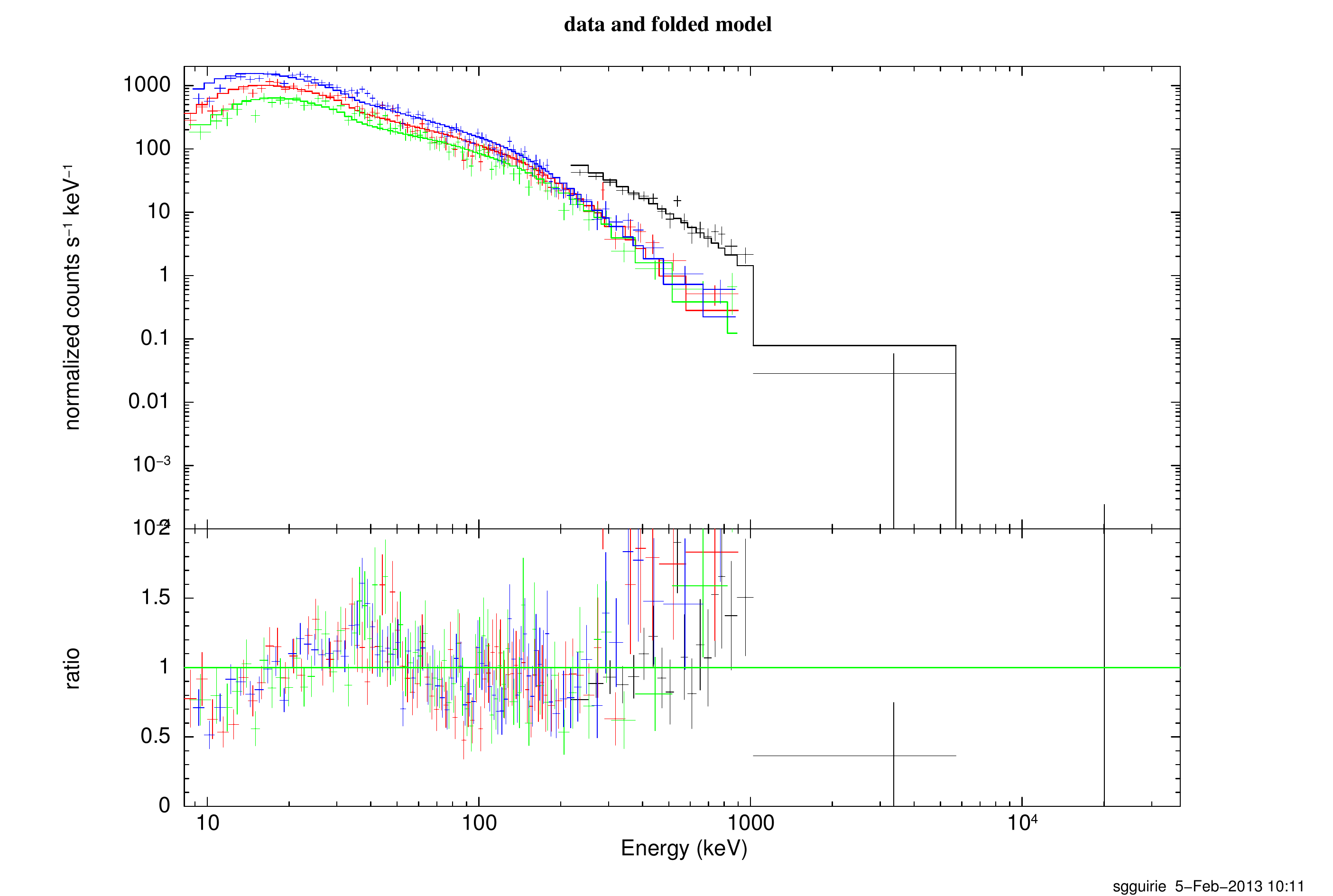}
\includegraphics[totalheight=0.245\textheight, clip, viewport=40 75 523 717,angle=270]{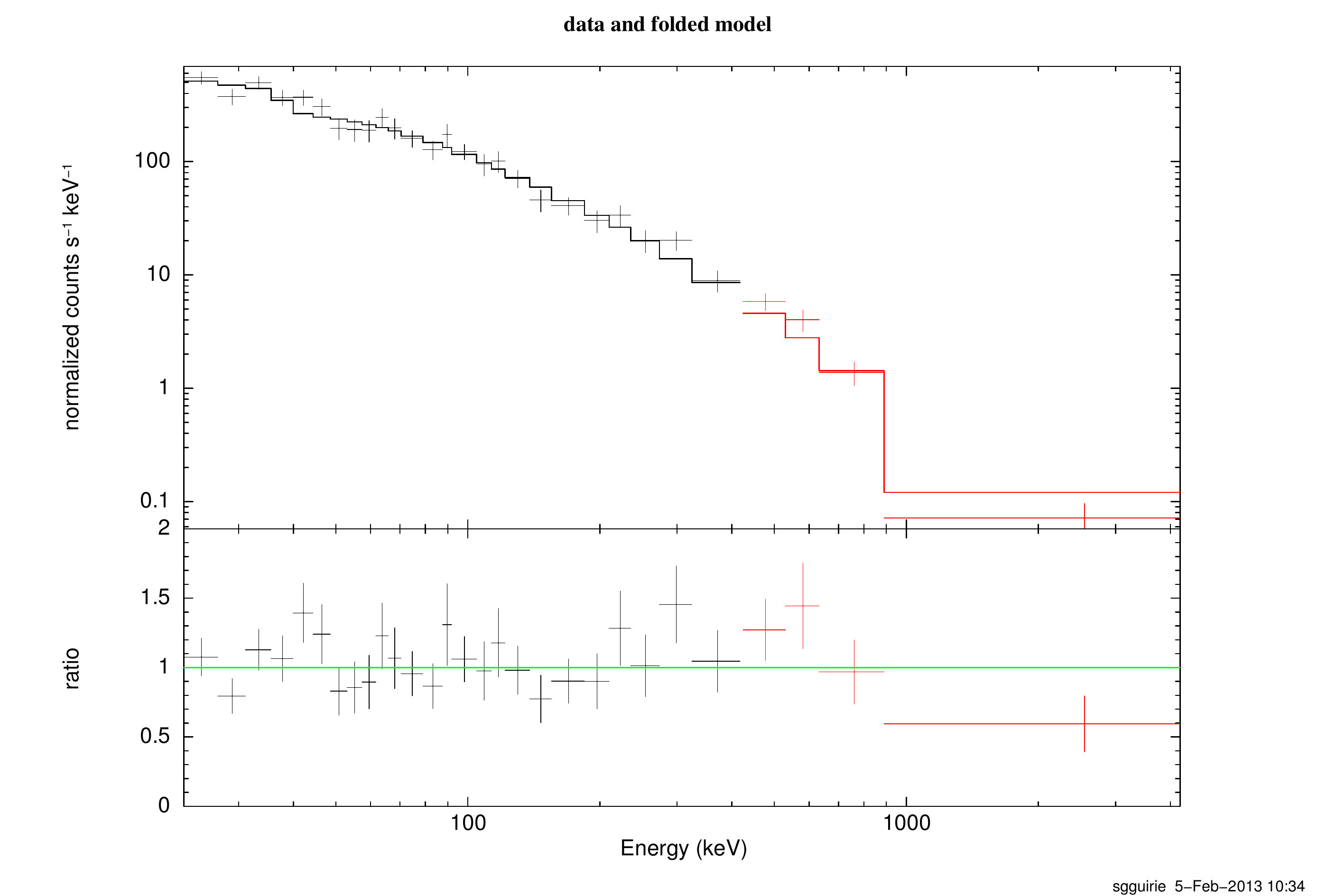}
\includegraphics[totalheight=0.245\textheight, clip, viewport=40 75 523 717,angle=270]{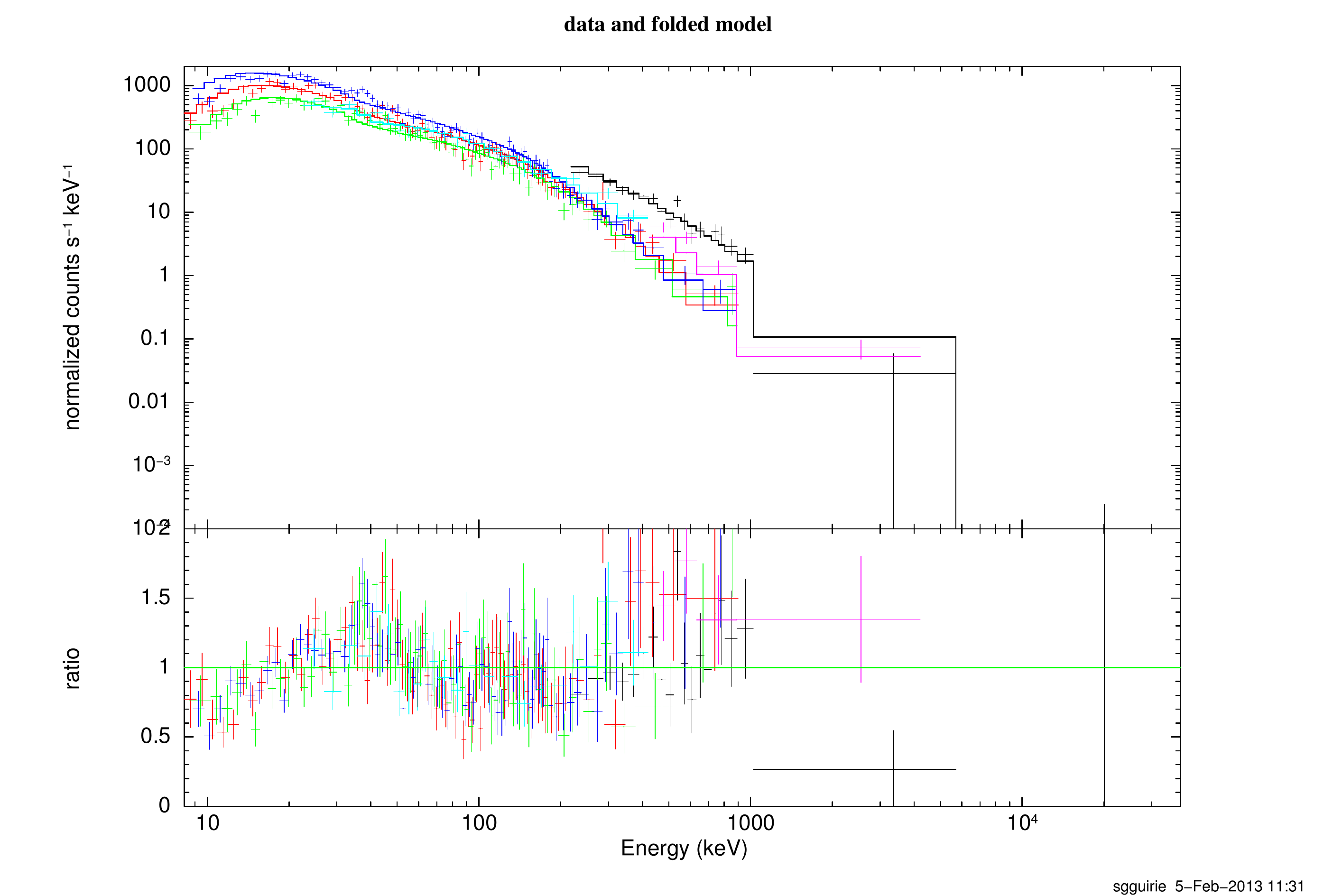}

\includegraphics[totalheight=0.26\textheight, clip, viewport=40 38 523 717,angle=270]{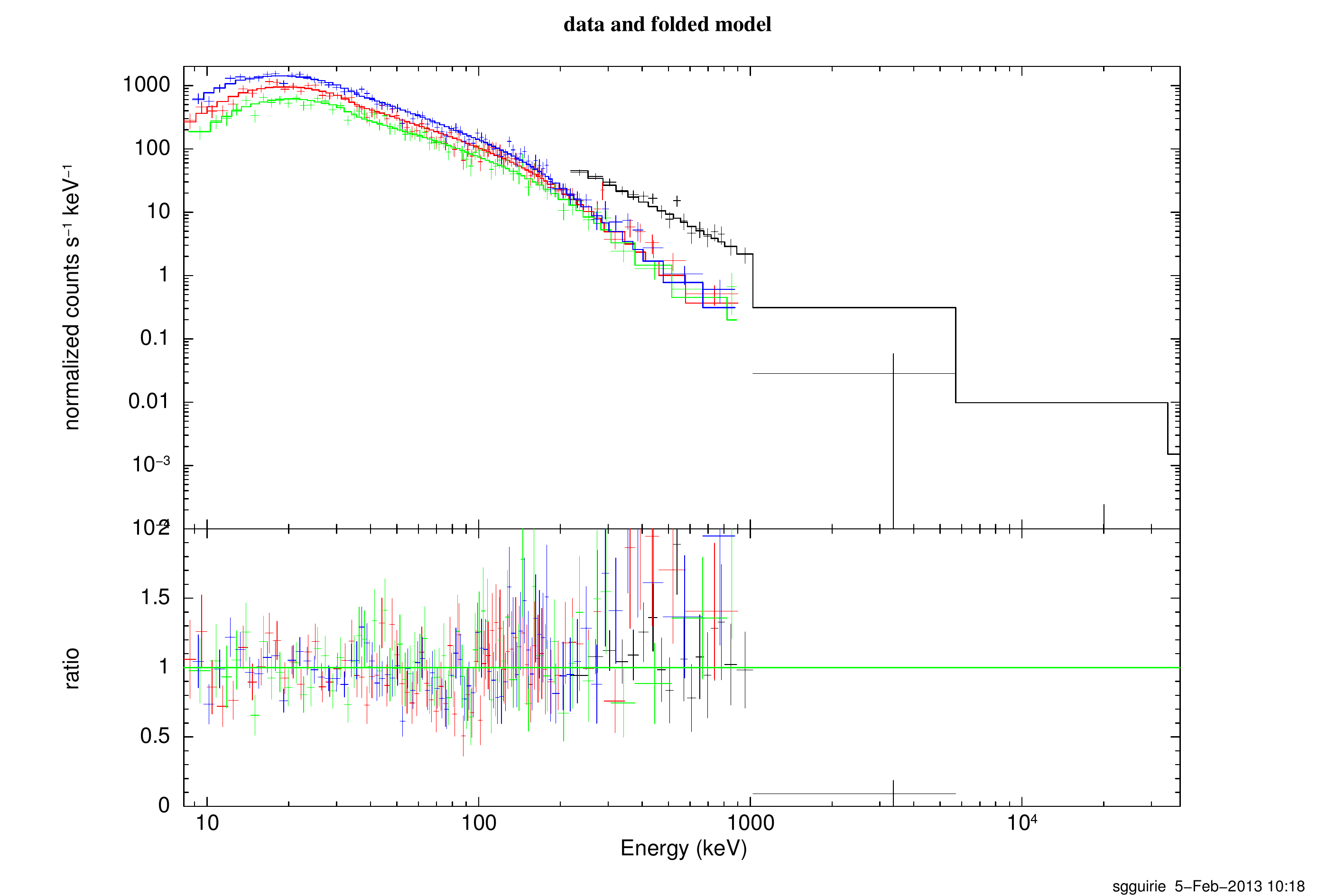}
\includegraphics[totalheight=0.245\textheight, clip, viewport=40 75 523 717,angle=270]{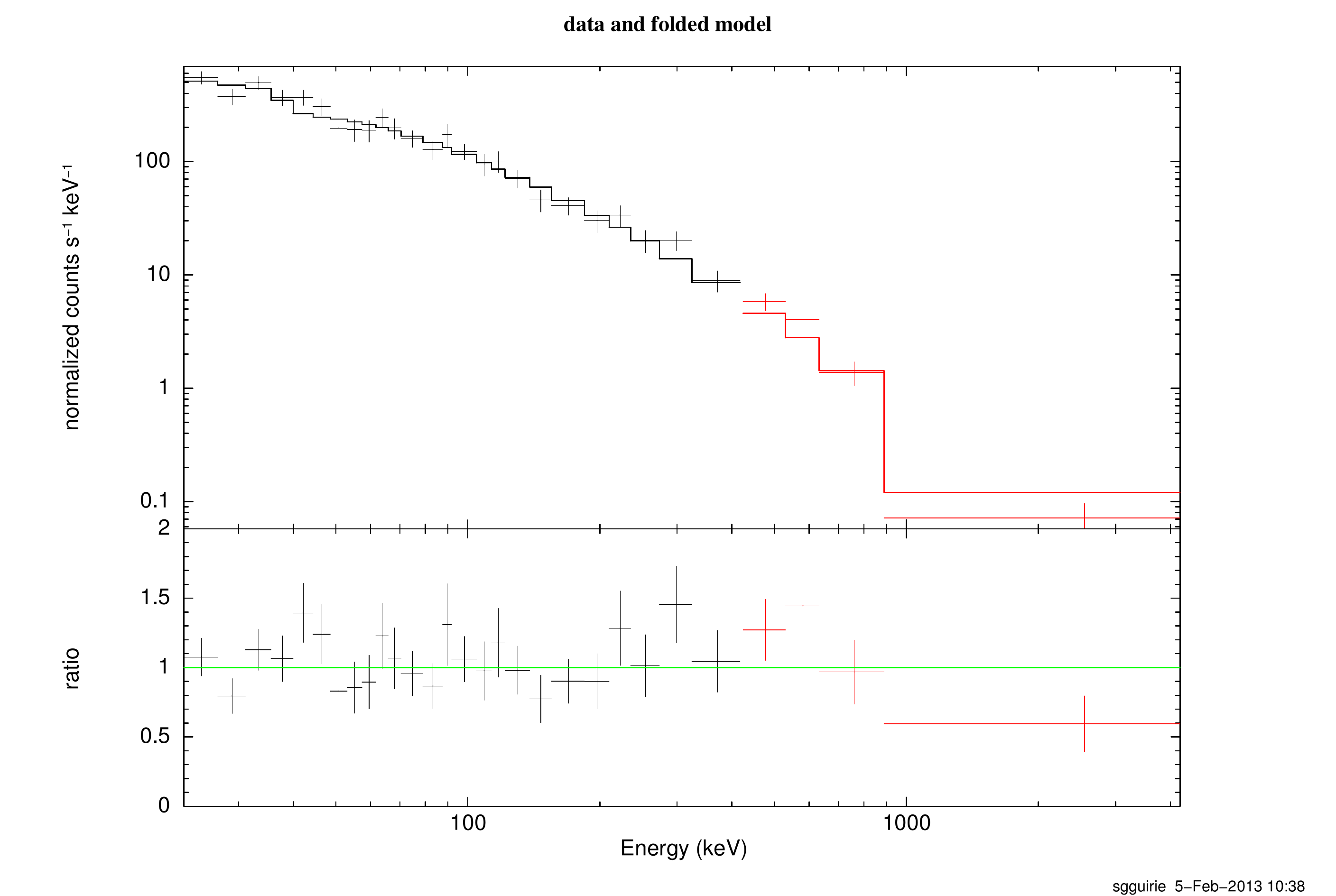}
\includegraphics[totalheight=0.245\textheight, clip, viewport=40 75 523 717,angle=270]{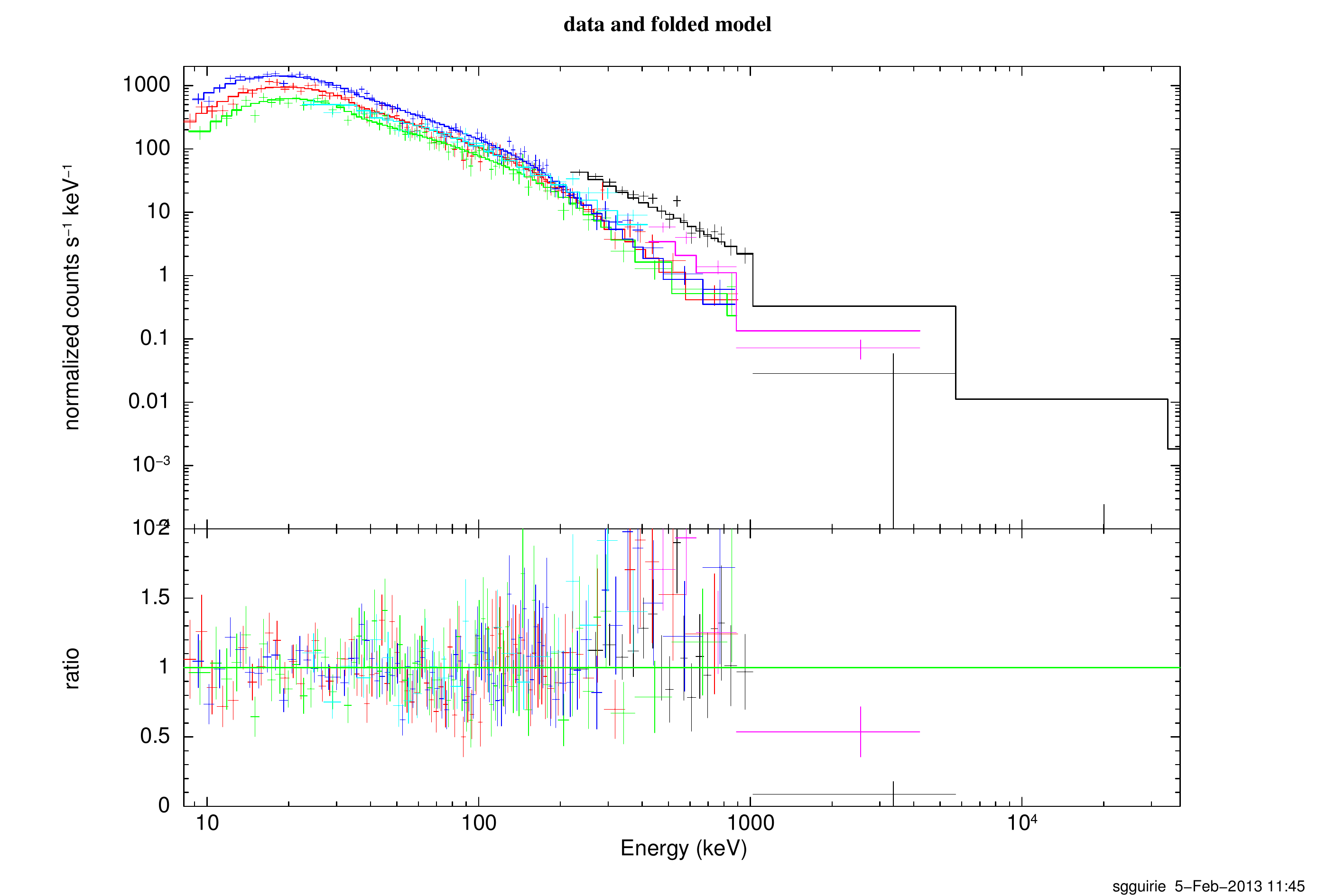}

\includegraphics[totalheight=0.26\textheight, clip, viewport=40 38 523 717,angle=270]{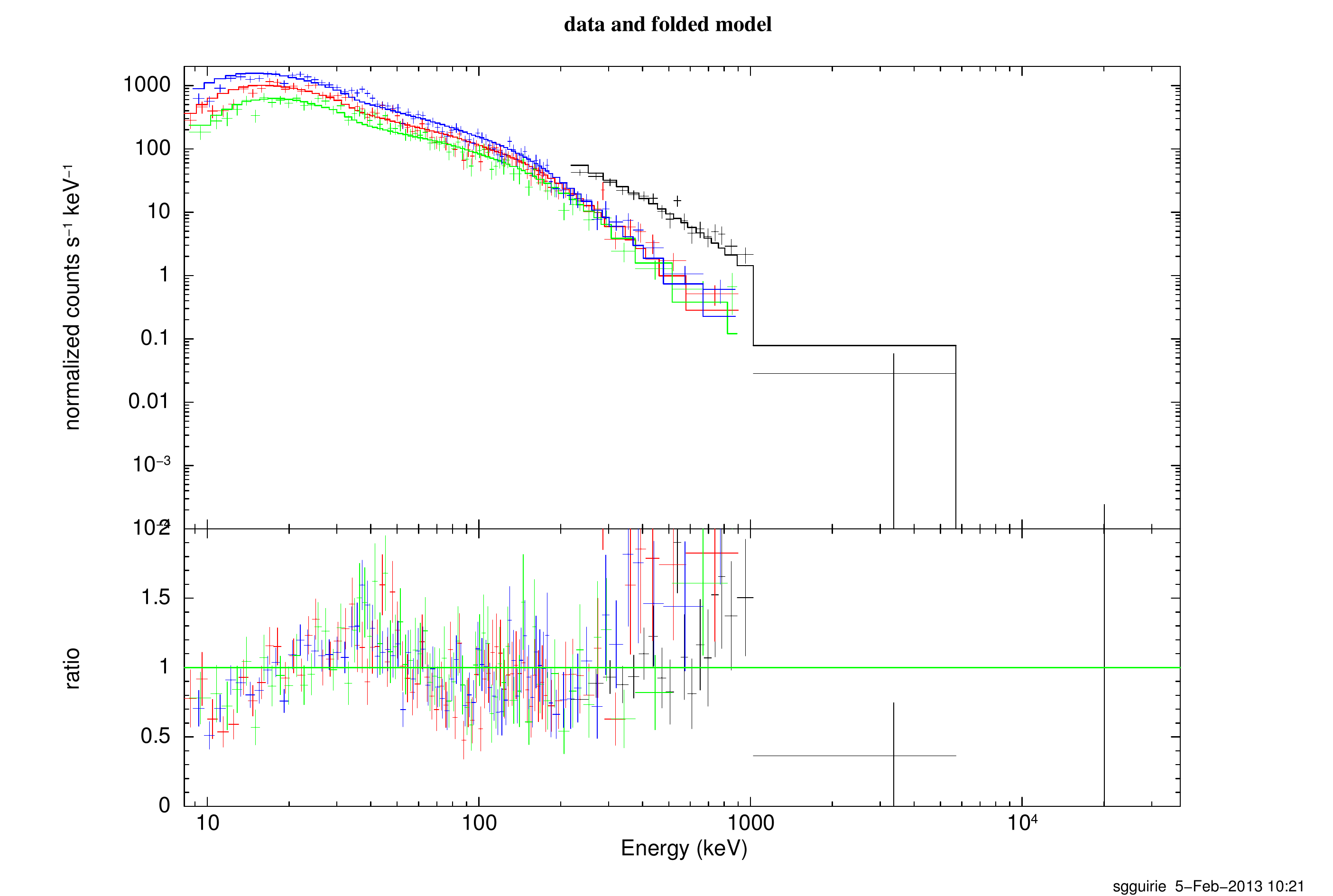}
\hspace{5.48cm}
\includegraphics[totalheight=0.26\textheight, clip, viewport=40 38 523 717,angle=270]{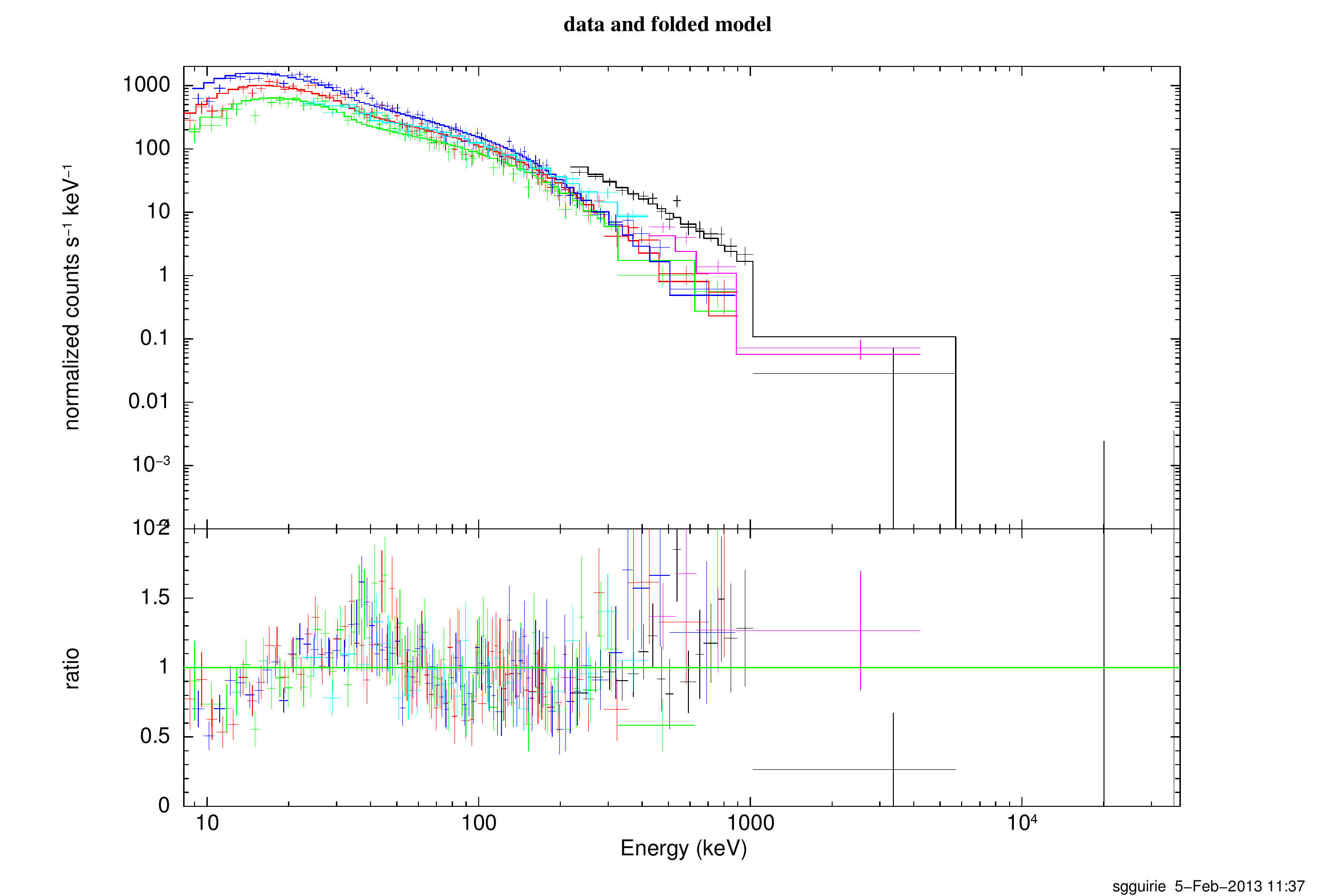}

\includegraphics[totalheight=0.26\textheight, clip, viewport=40 38 523 717,angle=270]{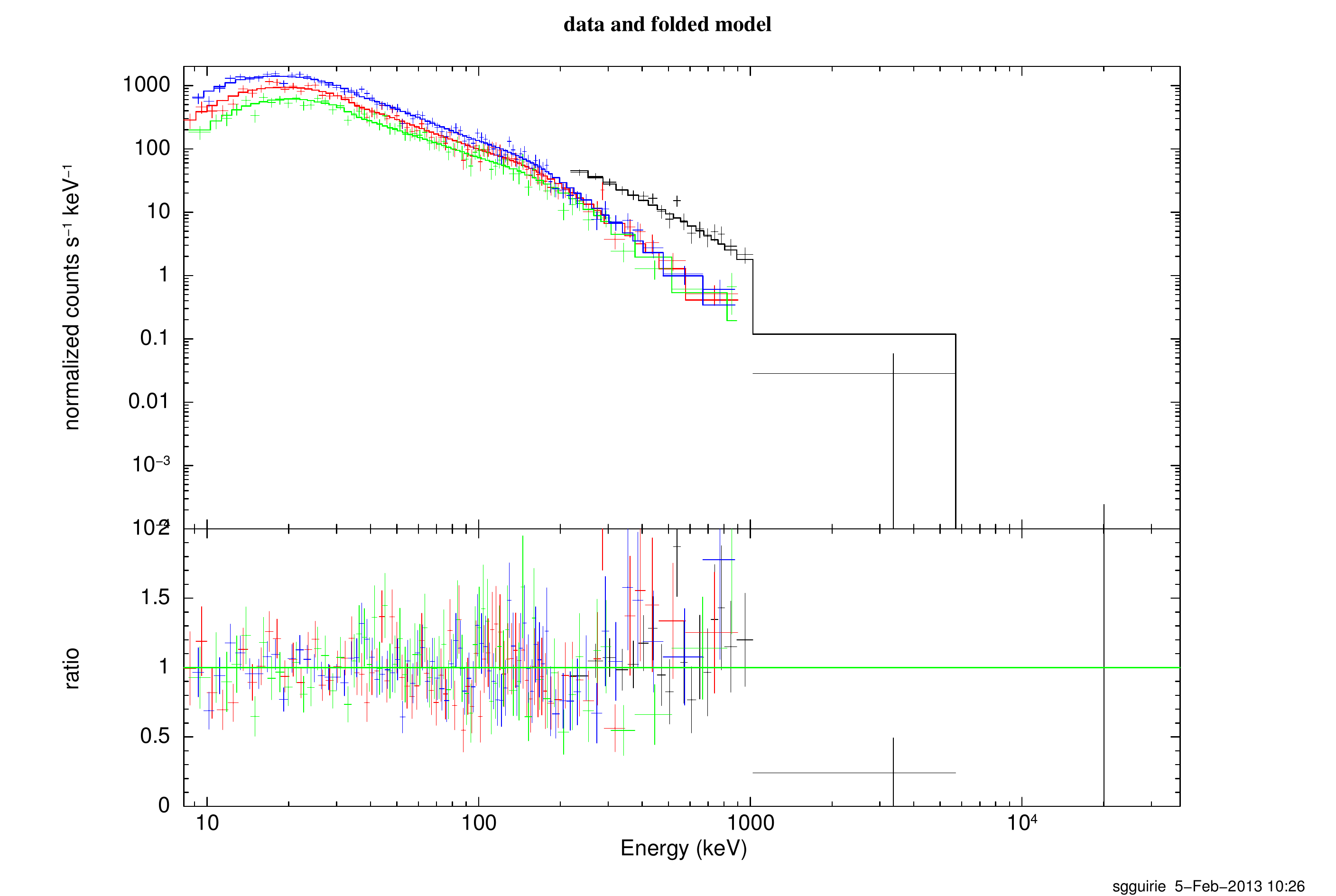}
\includegraphics[totalheight=0.245\textheight, clip, viewport=40 75 523 717,angle=270]{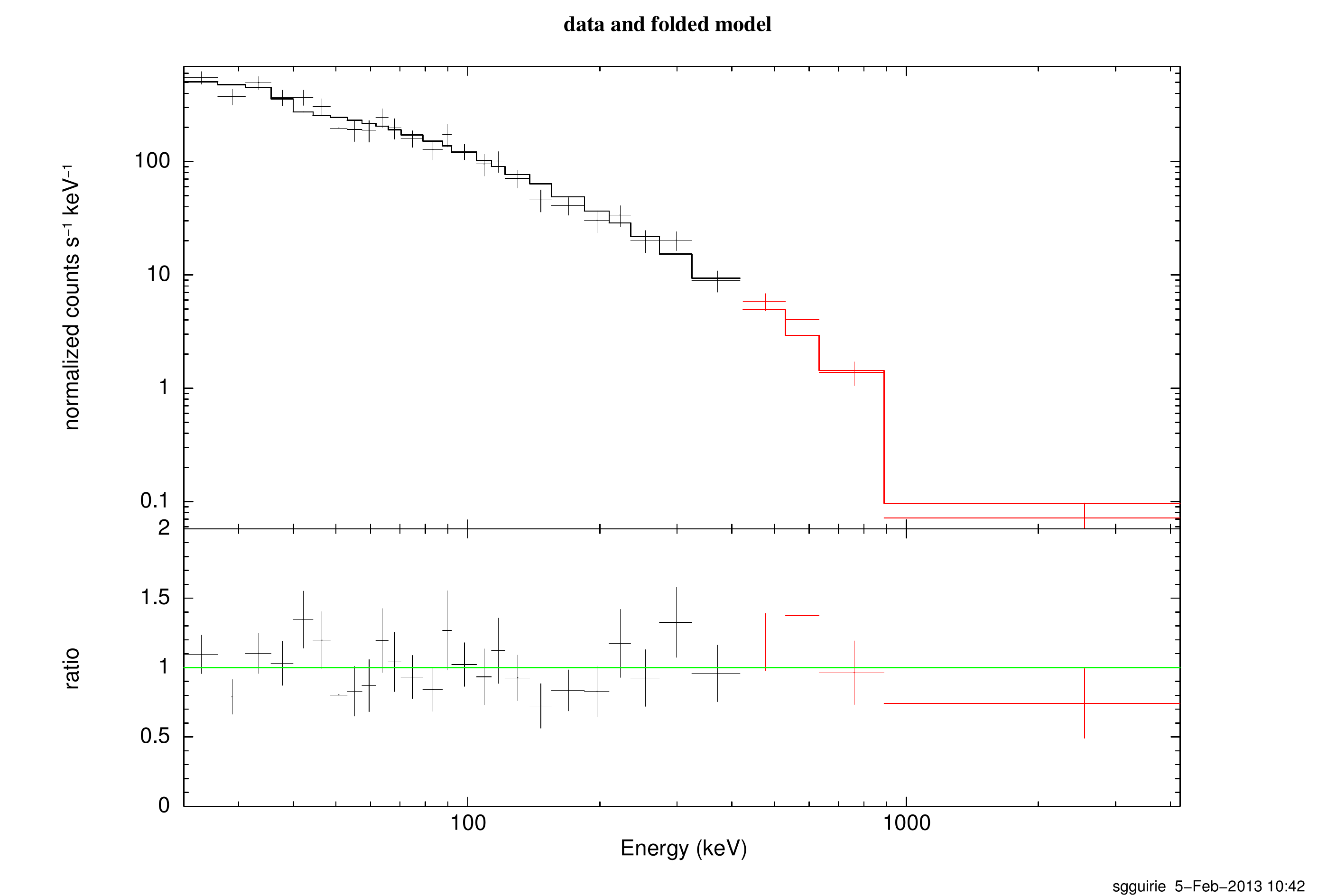}
\includegraphics[totalheight=0.245\textheight, clip, viewport=40 75 523 717,angle=270]{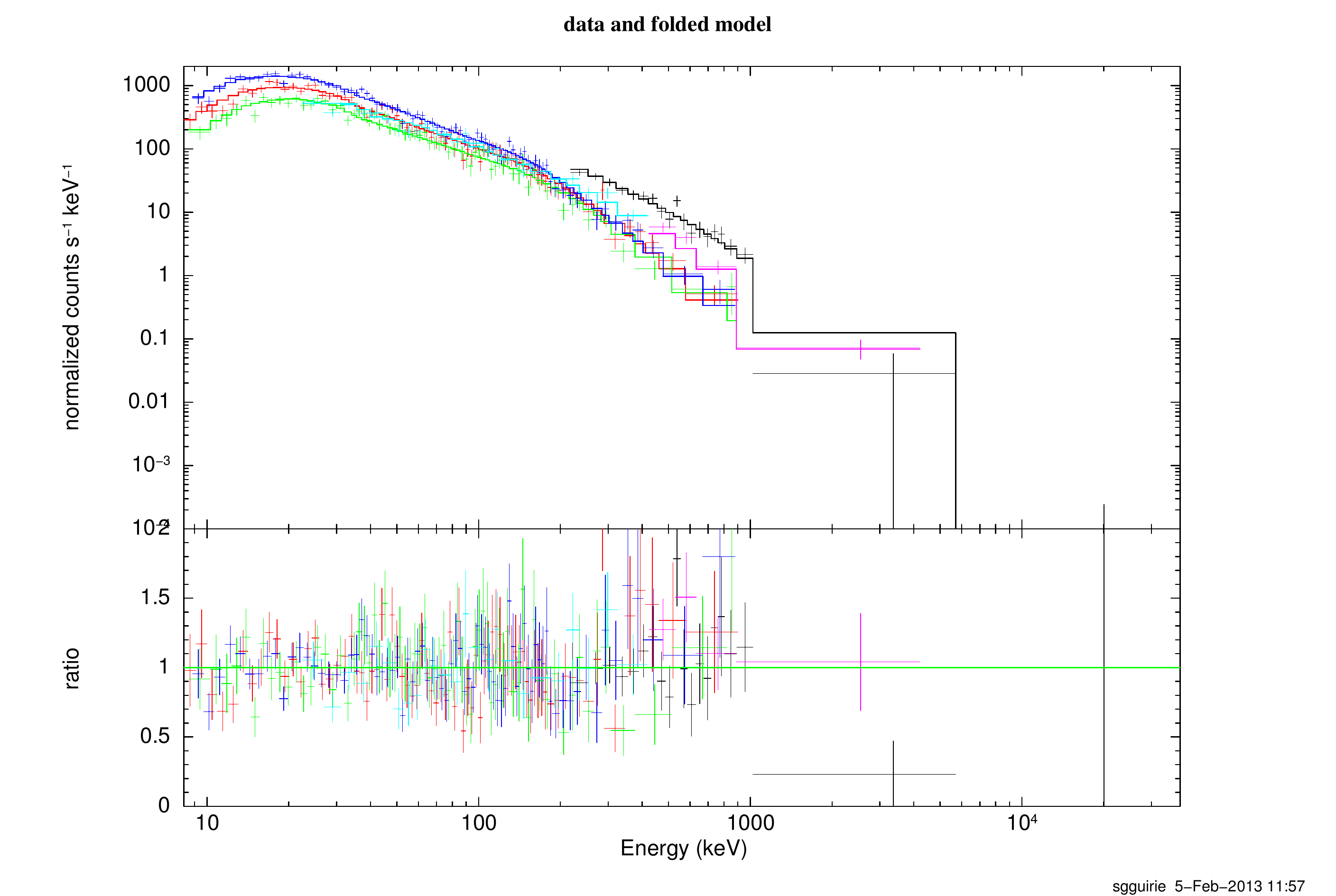}

\includegraphics[totalheight=0.26\textheight, clip, viewport=40 38 523 717,angle=270]{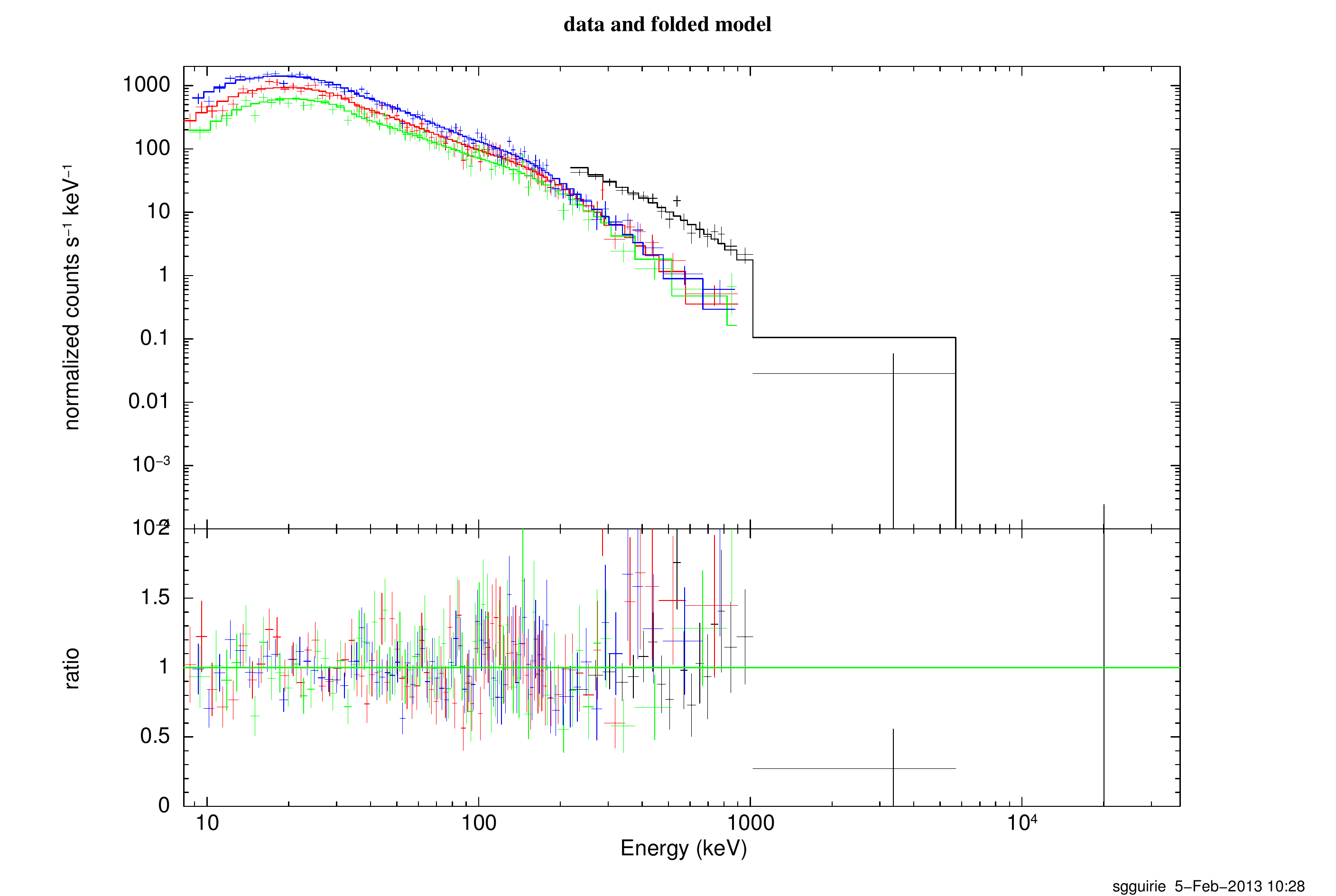}
\includegraphics[totalheight=0.245\textheight, clip, viewport=40 75 523 717,angle=270]{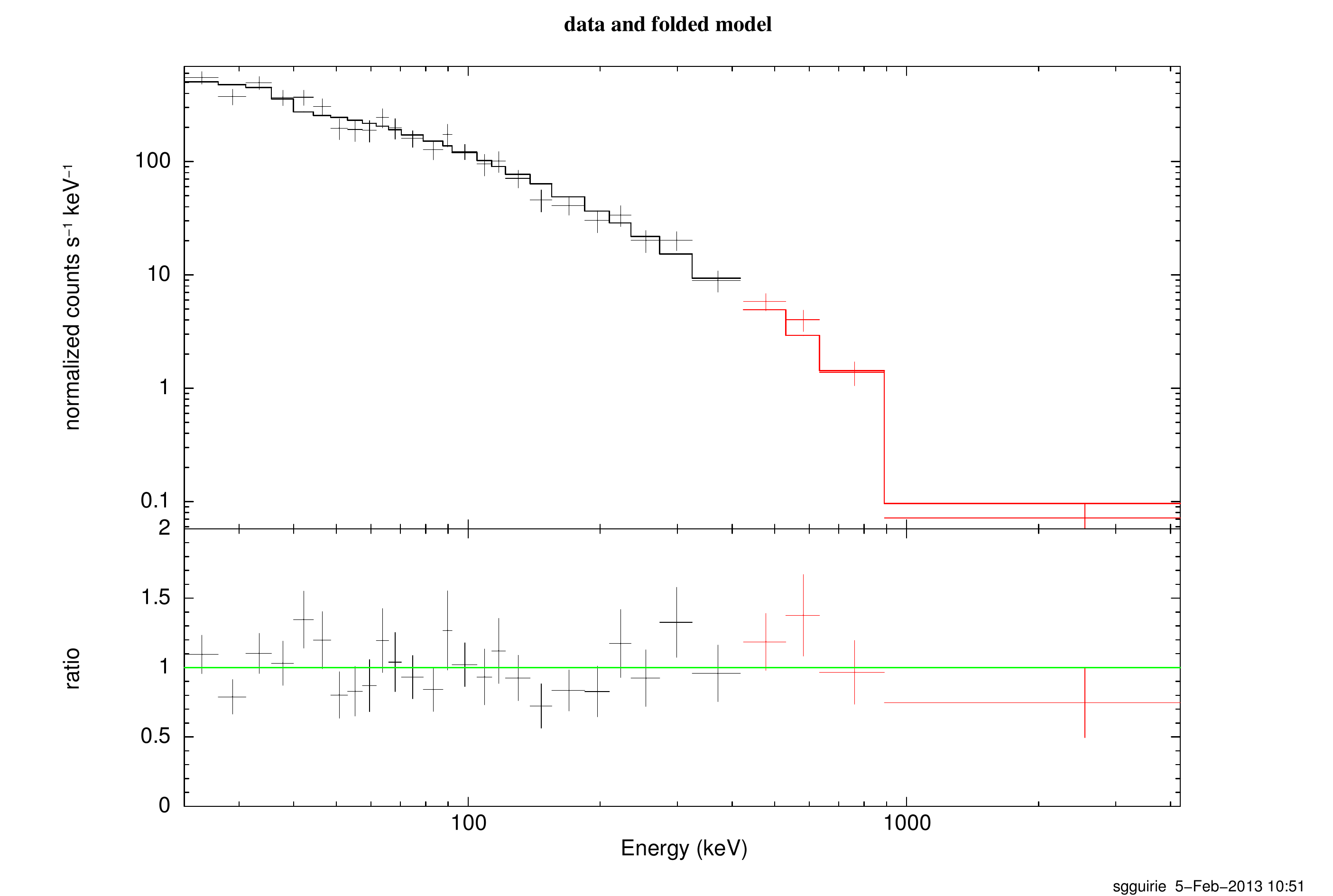}
\includegraphics[totalheight=0.245\textheight, clip, viewport=40 75 523 717,angle=270]{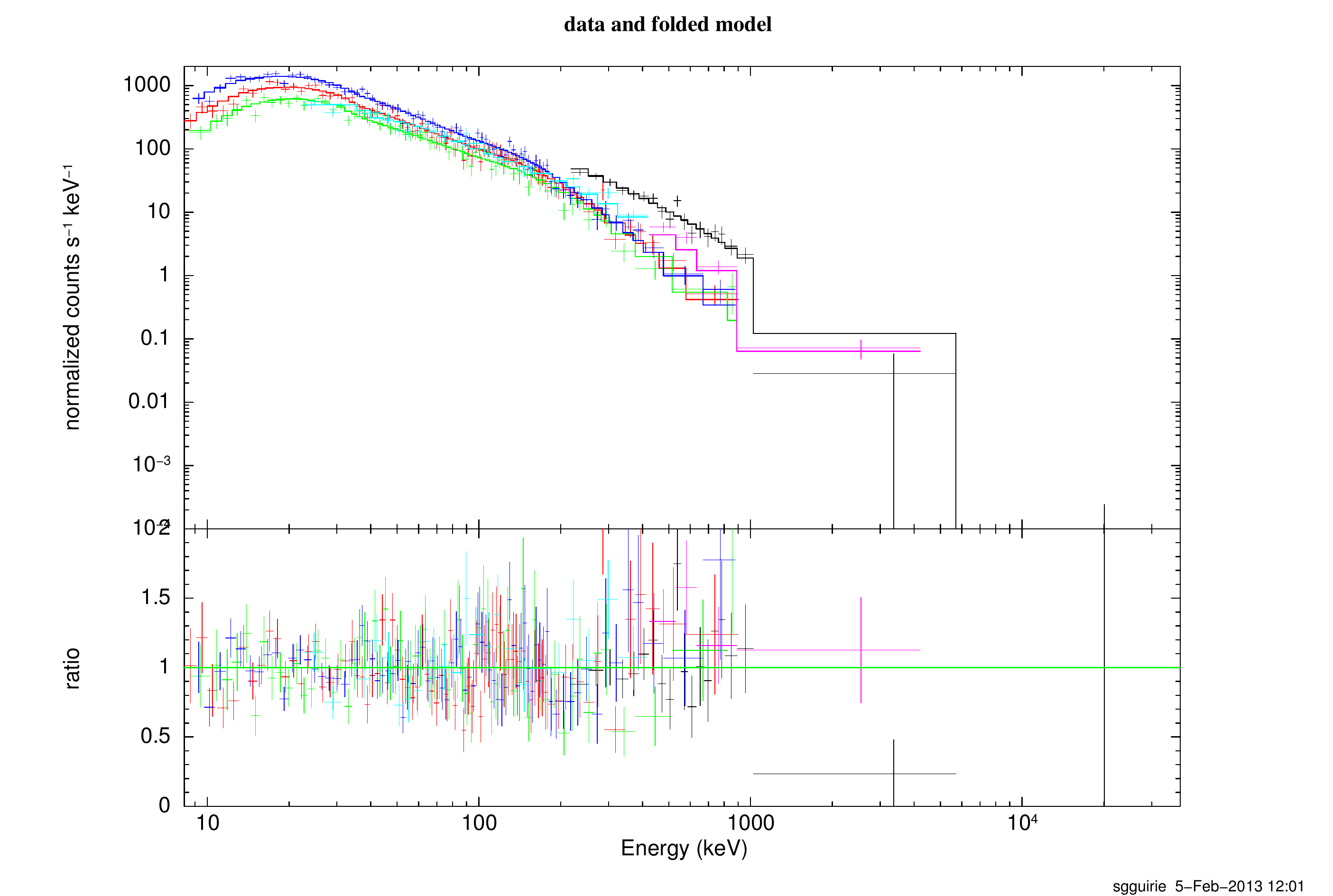}
\caption{\label{fig4}GBM, Konus and GBM+Konus count data (left, center and right columns, respectively) in time interval sp2 (T$_\mathrm{0}$+0.064 s to T$_\mathrm{0}$+0.128 s) with the residuals resulting from the CPL, Band, B+BB and C+BB fits (lines 1, 2-3, 4 and 5 from top to bottom, respectively). GBM energy channels have been grouped in larger energy bins for display purpose only; this grouping does not affect the fitting process that uses the best energy resolution of the instrument.}
\end{center}
\end{figure*}



\begin{thebibliography}{}





\bibitem[\protect\astroncite{{Band} et al.}{1993}]{Band:1993}
Band, D., Matteson, J., Ford, L., et al.: 1993,
\newblock{\apj}, 413, 281








\bibitem[\protect\astroncite{{Golenetskii} et al.}{2012a}]{Golenetskii:2012a}
Golenetskii, S., et al.: 2012a,
\newblock{GCN}, 13102, 1

\bibitem[\protect\astroncite{{Golenetskii} et al.}{2012b}]{Golenetskii:2012b}
Golenetskii, S., et al.: 2012b,
\newblock{GCN}, 13103, 1



\bibitem[\protect\astroncite{{Gruber} \& {Connaughton}}{2012}]{Gruber:2012}
Gruber, D., \& Connaughton, V.: 2012,
\newblock{GCN}, 13099, 1








\bibitem[Guiriec et al.(2010)]{Guiriec:2010} Guiriec, S., Briggs, M.~S., Connaugthon, V., et al.\ 2010, \apj, 725, 225

\bibitem[Guiriec et al.(2011)]{Guiriec:2011} Guiriec, S., Connaughton, V., Briggs, M.~S., et al.\ 2011, \apjl, 727, L33

\bibitem[Guiriec et al.(2013)]{Guiriec:2013} Guiriec, S., Daigne, F., Hasco{\"e}t, R., et al.\ 2013, \apj, 770, 32

\bibitem[Guiriec et al.(2015a)]{Guiriec:2015a} Guiriec, S., Kouveliotou, C., Daigne, F., et al.\ 2015a, \apj, 807, 148

\bibitem[Guiriec et al.(2015b)]{Guiriec:2015b} Guiriec, S., Mochkovitch, R., Piran, T., et al.\ 2015b, \apj, 814, 10

\bibitem[Guiriec et al.(2016a)]{Guiriec:2016a} Guiriec, S., Gonzalez, M.~M., Sacahui, J.~R., et al.\ 2016a, \apj, 819, 79

\bibitem[Guiriec et al.(2016b)]{Guiriec:2016b} Guiriec, S., Kouveliotou, C., Hartmann, D.~H., et al.\ 2016b, \apjl, 831, L8





\bibitem[\protect\astroncite{{Kouveliotou} et al.}{1993}]{Kouveliotou:1993}
Kouveliotou, C., et al.: 1993,
\newblock{\apjl}, 413, L101 












\bibitem[Svinkin et al.(2016)]{Svinkin:2016} Svinkin, D.~S., Frederiks, D.~D., Aptekar, R.~L., et al.\ 2016, \apjs, 224, 10


\end{thebibliography}
\end{document}